\documentclass[reprint, twocolumn]{revtex4}
\usepackage{latexsym, amsmath, amssymb,amsthm,amsopn,amsfonts, graphicx, epstopdf, multirow}

\def\simlt{\lower.5ex\hbox{$\; \buildrel < \over \sim \;$}}
\def\simgt{\lower.5ex\hbox{$\; \buildrel > \over \sim \;$}}
\def\simgtalt{\lower.5ex\hbox{$\buildrel > \over \sim \;$}}

\def\bd#1{{\bf #1}}
\def\l#1{\left #1}
\def\r#1{\right #1}

\def\eref#1{(\ref{#1})}

\begin{document}

\title{Framework for performance forecasting and optimization of CMB $B$-mode observations in presence of astrophysical foregrounds}

\author{J. Errard\footnote{E-mail: josquin.errard@apc.univ-paris-diderot.fr}}
\affiliation{Laboratoire Astroparticule et Cosmologie, Universit{\'e} Paris-Diderot, 10 rue A.Domon et L. Duquet, 75205 Paris Cedex 13, France}
\author{F. Stivoli\footnote{E-mail: federico.stivoli@gmail.com}}
\affiliation{INRIA, Laboratoire de Recherche en Informatique, Universit\'e Paris-Sud 11, B\^atiment 490, 91405 Orsay Cedex, France} 
\author{R. Stompor\footnote{E-mail: radek@apc.univ-paris-diderot.fr}}
\affiliation{CNRS, Laboratoire Astroparticule et Cosmologie, Universit{\'e} Paris-Diderot, 10 rue A.Domon et L. Duquet, 75205 Paris Cedex 13, France}

\begin{abstract}
We present a formalism for performance forecasting and optimization of future cosmic microwave background (CMB) experiments. We implement it in the context of nearly full sky, multifrequency, $B$-mode polarization observations, incorporating statistical uncertainties due to the CMB sky statistics, instrumental noise, as well as the presence of the foreground signals. We model the effects of a subtraction of these using a parametric maximum likelihood technique and optimize the instrumental configuration with predefined or arbitrary observational frequency channels, constraining either a total number of detectors or a focal plane area. We showcase the proposed formalism by applying it to two cases of experimental setups based on the CMBpol and COrE mission concepts looked at as dedicated $B$-mode experiments. We find that, if the models of the foregrounds available at the time of the optimization are sufficiently precise, the procedure can help to either improve the potential scientific outcome of the experiment by a factor of a few, while allowing one to avoid excessive hardware complexity, or simplify the instrument design without compromising its science goals. However, our analysis also shows that even if the available foreground models are not considered to be sufficiently reliable, the proposed procedure can guide a design of more robust experimental setups. While better suited to cope with a plausible, greater complexity of the foregrounds than that foreseen by the models, these setups could ensure science results close to the best achievable, should the models be found to be correct.\\
\\
%{\bf Version \draftVersion (\today)}
\end{abstract}

\maketitle

\section{Introduction}

\label{sect:intro}

Cosmic Microwave Background (CMB) $B$-mode polarization offers some of the most exciting goals for the next stage of observational and experimental
effort in cosmology. These goals are already aimed at by an entire
slew of current, forthcoming, and planned CMB observations, e.g.,~\cite{2010SPIE.7741E..39A,2010SPIE.7741E..37R, 2011arXiv1102.2181T, 2009arXiv0903.0902A, 2011arXiv1105.2044K}. Probably most importantly, CMB $B$-mode measurements could open up a window, as direct
as likely ever possible, onto the physics of the very early Universe, giving us unique insights on the physical laws governing at the highest
energies. Such outstanding, anticipated consequences seem 
to be however matched by difficulties, which need to be overcome, first, to deliver an incontrovertible, reliable detection and
sufficiently precise characterization of the primordial $B$-mode signal, and later to interpret it. The obstacles are of fundamental and instrumental
origins and stem from the
fact that the anticipated $B$-mode amplitudes are expected to be nearly 2 orders of magnitude below those of CMB E-mode 
polarization and up to 10 times lower than the $B$-mode signal generated by the Galactic foregrounds.
To meet and successfully address such a challenge progressively more sophisticated and advanced observatories have to be devised
and built. Their complexity results in a number of design choices and decisions instrumental teams have to make
in the course of their development. 
As those have often a direct impact on the science output of the instrument, these are science goals which should drive the decision-making process. Though such a situation is not new, the sheer size, complexity, and precision of the modern instruments and data sets call for novel, more robust ways of addressing the instrumental optimization problem. 

In this paper we propose a general, methodological framework for the experiment optimization and then apply it in specific cases
of CMB $B$-mode observatories. We note that, however sophisticated an adopted optimization procedure may be, it is likely to always come up short in doing justice to all the complexity of an instrument under consideration. The goal of such a procedure, as we pursue here, is therefore not just to find a single best (in some sense) instrumental configuration. Rather, the goal is to provide, on the one hand, a reference against which to judge actual hardware designs and, on the other, guidelines of, first, how to propose, given some science goals, a suitable and viable experimental design and, later, how to modify it to implement inevitable, real-life limitations and constraints in a way which will have a minimal impact on its scientific performance.

Though the discussed formalism lends itself straightforwardly to a number of generalizations, in this paper we demonstrate it
in the context of the $B$-mode detection by multifrequency observatories taking
into account the presence of the astrophysical (diffuse) foregrounds, leaving a study of some of the most common instrumental effects to a future work.
We note that even in this limited context a result of the instrument optimization problem will depend on a number of factors: scientific goals as set for the experiment in question; models of the physical effects, e.g., foregrounds; specific techniques and assumptions they require, selected to be used for the component separations step. This emphasizes the need for using the state-of-the-art physical models of the foregrounds and the separation techniques in this kind of problem, as well as for continuing effort aiming at better, more reliable understanding of the foreground physics.

As the optimization requires a capability to predict the performance of an instrument given its characteristics, it is very closely
connected with performance forecasting. In fact, in most of the similar work to date, the problem of selecting the most suitable experimental
configurations is typically treated as a performance forecasting problem applied to some predefined, and limited, set of potential candidate 
experimental setups, the relative merits of which are subsequently evaluated and compared, e.g.,~\cite{2005PhRvD..72l3006A, 2006JCAP...01..019V, 2009A&A...503..691B, 2009AIPC.1141..222D,
2010MNRAS.408.2319S, 2011JCAP...08..001F}. This is in contrast with this paper, which employs an actual optimization procedure.
In this respect our approach is most similar to the one by~\citet{2007PhRvD..75h3508A}. Here we generalize and extend the latter work on both methodological and implementation levels. 
We consider broader parameter space and optimization
strategies, search for families of acceptable configurations, and by adopting the parametric 
component separation approach
 as the component separation technique of the choice, we manage to propagate  
 realistic ensemble-averaged errors to our selected figure of merit indicators in a statistically sound manner.

The paper is organized as follows. In the next section we first describe a general framework of our approach and then specialize
it to our specific science case of  CMB $B$-mode observations. In that section we show how the parametric component separation
technique can be used to assess the performance of CMB experiments in the presence of galactic foregrounds, developing 
the approach to a performance forecasting in such cases.
In Sec.~\ref{sect:foregrounds} we detail the foreground model we use in this work. Section~\ref{sect:applications} describes
applications of the proposed formalism to two fiducial satellite experiments, based on the CMBpol~\cite{2009arXiv0903.0902A} and COrE~\cite{2011arXiv1102.2181T} proposals.
In Sec.~\ref{sect:conclusions} we present our conclusions. Some of the lengthy calculations are collected in Appendixes~\ref{sect:algebra} and~\ref{sect:algebra_2}.

\section{Method}

\label{sect:method}

Our approach is as follows. We start off from expressing our science goals in terms of acceptable ranges of values of some 
proposed figures of merit (Sec.~\ref{subsect:fom}), which are chosen to reflect the physical context of the considered experiment.
We then first treat all figures of merit (FOMs) separately and for each of them perform a strict optimization procedure (Sec.~\ref{subsect:minProc}), i.e.,
minimize or maximize it over a set of considered instrumental parameters. This is 
usually done in the presence of some external 
constraints arising for instance due to some hardware requirements but also some other science-driven restrictions, (Sec.~\ref{subsect:minProc}).
 This first step aims at determining the best possible instrument performance from the perspective of the considered FOMs and their
 corresponding configurations.  If for any of the FOMs the best performance value does not fulfill our science goals, the procedure
 halts and either the set of instrumental parameters have to be enlarged or the science goals/FOMs rethought. Otherwise, 
for each FOM, but one, we select a threshold value, which need to be attained by any acceptable configuration and perform the optimization of
the one left-over FOM over the parameter space under additional constraints, requiring that all or some of the 
remaining FOMs are not worse than their established
thresholds. If the optimization fails, we may need to adjust some of the thresholds and repeat the procedure again. This may be also the
case if the solution found does not ensure an acceptable value for the  FOM, which is used in the optimization. If the tuning of
the thresholds succeeds, the solution obtained via the above procedure is used as a starting point for further 
post-processing and the corresponding set of values of all FOMs used as a reference to compare any other configuration against.
The post-optimization processing is used to implement some additional constraints and/or simplifications, which for some reason could
not have been imposed on the formal optimization procedure.

 Below we present a specific implementation of this general framework in the context  of primordial CMB $B$-mode 
 observations by multifrequency multidetector observatories in the presence of Galactic foregrounds. In this case our FOMs need
 to account for some effects arising due to the component separation procedure, which has to be applied to data to recover a
  genuine CMB signal.
We therefore start below by discussing a specific component separation approach, the so-called parametric maximum likelihood
technique, and its impact on a CMB $B$-mode detection.
 
\subsection{Effects of foreground separation}

An estimation of the presence of the foregrounds involves two main steps. On the first step, we estimate the error incurred while constraining the spectral 
parameter values from the data. On the second, we translate that error into some figures of merit expressing the overall quality of the separation process
and which are then used in our optimization procedure.
 
\subsubsection{Formalism}

\label{sect:formalism}

Hereafter we use the parametric maximum likelihood component separation approach implemented as  in \cite{2009MNRAS.392..216S}. We thus 
assume a linear data model, where a signal measured in each pixel $p$ is given by
\begin{eqnarray}
\bd{d}_p = \bd{A} \, \bd{s}_p \, + \, \bd{n}_p,
\label{eqn:dataModelFull}
\end{eqnarray}
where for each pixel $p$,
\begin{itemize}
\item $\bd{d}_p$ is a multifrequency data vector with each entry corresponding to a different frequency channel;
\item $\bd{s}_p$ is a multicomponent sky signal vector each entry of which corresponds to a different sky component and which is to be estimated from the data;
\item $\bd{A}$ is a mixing matrix defining how the components need to be combined to give a signal for each of the considered frequency channels; and
\item $\bd{n}_p$ is a vector containing the instrumental noise and assumed to be Gaussian and uncorrelated with a dispersion given by $\bd{N}$.
\end{itemize}
Here both $\bd{A}$ and $\bd{N}$ are assumed to be pixel independent for simplicity, with an exception of Sec.~\ref{subsect:robustmessForeModels}.

In the parametric approach, one assumes that $\bd{A}$ is parametrized by a set of spectral parameters, $\bd{\beta}$, which need to be determined together with the sky signal
estimates. The noise level per channel, number of frequency channels, etc., are all dependent on instrument properties, which thus will affect the results of the
component separation process and could therefore serve as optimization parameters.  Some other effects such as calibration, beam sizes,  and bandwidths are also typically 
relevant and may need to be included in the modeling. We leave a thorough evaluation of those effects for future work, while neglecting them in this paper.

Given values of $\bd{\beta}$ and defined instrumental parameters we can estimate the component signal using a standard maximum likelihood solutions,
\begin{eqnarray}
\bd{s}_p & \equiv& \l( \bd{A}^t\,\bd{N}^{-1}\,\bd{A}\r)^{-1}\,\bd{A}\,\bd{N}^{-1}\,\bd{d}_p.
\label{eqn:mlMapDef}
\end{eqnarray}
To estimate the spectral parameters  we will use a pseudo (or profile) likelihood \citep{2009MNRAS.392..216S} given as,
\begin{eqnarray}
-2\,\ln\,{\cal L}  =  -\,\sum_{p}\, \l(\bd{A}\,\bd{N}^{-1}\,\bd{d}_p\r)^t\,\l( \bd{A}^t\,\bd{N}^{-1}\,\bd{A}\r)^{-1}\,\bd{A}\,\bd{N}^{-1}\,\bd{d}_p.
\label{eqn:profileLikeDef}
\end{eqnarray} 
We will refer to this likelihood as the spectral likelihood and will identify its peak value with the best estimate of the spectral indices and the curvature matrix at its peak 
as the measure of the uncertainties expected for the spectral parameter estimation. These will be used to construct our figures of merit.
\\
\\
\subsubsection{Spectral parameter uncertainty}
\label{subsubsect:specParErr}
The profile likelihood derivatives with respect to the spectral parameters can be readily computed and the relevant formulas are collected in Appendix~\ref{sect:algebra}.
As our purpose is to gain some insight in the constraining power of different plausible experimental setups rather than analyze any specific data set we will average over the
possible noise realization assuming that the noise correlation matrix, $\bd{N}$, is known. Using Eq.~\eref{eqn:firstDer} from the Appendix we then arrive at
\begin{eqnarray}
\l\langle\frac{\partial \ln {\cal L}}{\partial\beta}\r\rangle_{noise} = \sum_p\, \l(\bd{A}_{,\beta}\,\bd{\bar{s}}_p\r)^t \, \bd{N}^{-1}\,
\l(\hat{\bd{A}}\,\hat{\bd{s}}_p-\bd{A} \,\bd{\bar{s}}_p\r) \label{eqn:firstDerAver}
\end{eqnarray}
for the first derivative. In this equation, as well as everywhere hereafter, we will use a hat over a quantity to mark that we refer to its true, rather than just an estimated,
value. $\bd{\bar{s}}$ is a sky signal estimate in the case of the noiseless data and it is defined in Eq.~\eref{eqn:mpDef}. If the data model in Eq.~\eref{eqn:dataModelFull} is 
correct both in terms of assumed scaling laws but also a number of components, the first derivative in Eq.~\eref{eqn:firstDerAver}  vanishes for the true values of the parameters, 
$\bd{\beta}\equiv\bd{\hat{\beta}}$, emphasizing that the estimator is on average unbiased. Indeed in such a case we have $\bd{\hat {A}} = \bd{A}$ and $\bd{\hat{s}} = \bd{\bar{s}}$.
Under the same assumptions the second order derivatives  taken at the true values of the parameters can be then written as [see Eq.~\eref{eqn:secDervGen0}]
\begin{widetext}
\begin{eqnarray}
\l.\l\langle\frac{\partial^2 \ln {\cal L}}{\partial\beta\,\partial\beta'}\r\rangle_{noise}\r|_{\bd{\beta}=\hat{\bd{\beta}} }
& = & \, {\rm tr}\,\l\{\l[ \bd{A}_{,\beta}^t\, \bd{N}^{-1}\,  \bd{A}\, \l(\bd{A}^t\bd{N}^{-1}\bd{A}\r)^{-1} \,  \bd{A}^t\bd{N}^{-1}\bd{A}_{,\beta'}\, 
- \,  \bd{A}_{,\beta}^t \, \bd{N}^{-1} \,  \bd{A}_{,\beta'}\,\r]
 \sum_p\,\hat{\bd{s}}_p\,\hat{\bd{s}}_p^t \r\}. 
 \label{eqn:secDervSat}
\end{eqnarray}
\end{widetext}
 Hereafter we will use the inverse of this matrix to approximate the error matrix, $\bd{\Sigma}$,  for the recovered scaling parameters, i.e.,
\begin{eqnarray}
\l[\bd{\Sigma} ^{-1}\r]_{\beta\beta'} \simeq - \l.\l\langle\frac{\partial^2 \ln {\cal L}}{\partial\beta\,\partial\beta'}\r\rangle_{noise}\r|_{\bd{\beta}=\bd{\hat{\beta}}}.
\label{eqn:errMatDef}
\end{eqnarray}

\begin{figure}
	 \includegraphics[width=8cm]{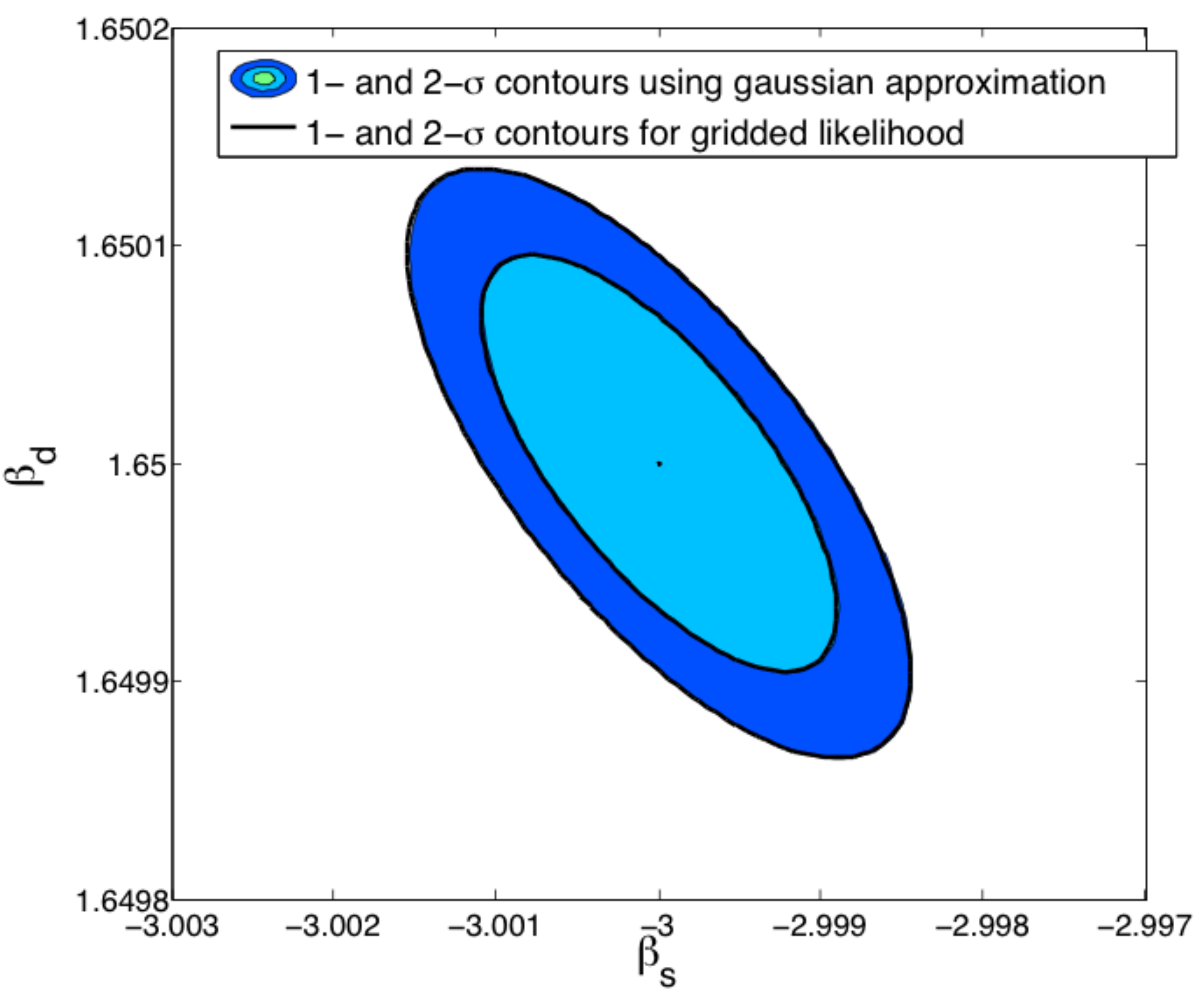}
\caption{1-$\sigma$ and 2-$\sigma$ contours in the $\beta_{dust}$ - $\beta_{sync}$ space of the spectral likelihood, ${\cal L}$,  calculated using Eq.~\eref{eqn:profileLikeDef} for a random realization of the CMB and noise contributions, shaded areas, and compared against the Gaussian approximation with a dispersion as given by Eq.~\eref{eqn:secDervSat}, solid lines. The former likelihood has been recentered at the true values of the parameters.}
\label{fig:likelihoods}
\end{figure}

We note that the spatial morphology of the sky components enter the calculation of the errors only in a form of pixel averaged correlations, 
\begin{eqnarray}
\hat{\bd{F}} \equiv \frac{1}{N_{pix}}\sum_p\,\hat{\bd{s}}_p\hat{\bd{s}}_p^t.
\label{eqn:FmatDef}
\end{eqnarray}
Moreover, only those of the columns and rows of this correlation matrix matter, which correspond to sky components characterized by the scaling laws including some 
unknown parameters.
Mathematically, this just follows from the fact that only columns of the derivatives of the mixing matrix, $\bd{A}_{,\beta}$, corresponding to such components do not vanish. 
Physically,
this indicates that the components for which the scaling laws are known unambiguously, e.g., CMB, are subtracted cleanly during the separation process
and do not affect the result of spectral indices estimations. An immediate consequence of this is that the resulting expressions are indeed equivalent to those obtained while averaging over an ensemble of realization of noise and CMB signal.

We note that though our conclusion about the impact of different components on the spectral parameter estimation is general, a simple form of the dependence of the latter
on the foreground signal morphology is due to our simplifying assumption of a pixel-independent noise level. In general, the relation is more complex, with noise levels selective (de)emphasizing the contributions of some of the pixels on the sky.  
Though the formalism developed here is  general and can be straightforwardly adopted to a case of arbitrary and correlated noise it can  quickly become 
computationally heavy.  We will therefore leave a discussion to more complex and realistic cases for future work.

In Fig.~\ref{fig:likelihoods} we show examples of the contours likelihoods, Eq.~\eref{eqn:profileLikeDef}, computed for dust and synchrotron spectral indices for simulated data  as described in Sec.~\ref{sect:foregrounds} and for some fiducial nearly full sky experiment.
They are compared with a Gaussian approximation based on the variance derived with help of the error matrix, Eq.~\eref{eqn:errMatDef}. Generally we find
a very good agreement. This may breakdown somewhat in cases with very few pixels when the actual spectral likelihoods typically become
somewhat skewed \citep{2009MNRAS.392..216S}. Nevertheless, we find that even in those cases though the Gaussian approximation may fail to reproduce properly the tails of 
the distributions, its overall performance is still rather good. In applications of interest for this paper a sufficient number of pixels is always granted.

An interesting question is then how the precision of the spectral parameter estimation depends on the matrix $\hat{\bd{F}}$. 
The short answer is that given the noise levels the higher density contrast of the components, i.e., larger diagonal elements of $\hat{\bd{F}}$, the better precision of estimated $\bd{\beta}$,
while large cross-correlation terms tend to increase the error.

\subsubsection{Residuals}

\label{subsect:residuals}

 From the discussion in the previous section it is clear that the precision of the spectral parameters determination though relevant is clearly not a single factor
important in quantifying the component separation effects on the $B$-mode science. This is due to the fact that  better 
precision is usually related to a higher foreground contrast and vice versa making it not straightforward to infer an effective foreground contribution
left over in the CMB map after the separation process, given just the spectral indices errors. 
However, given the estimated value of the spectral parameters, $\bd{\beta}$, we can always calculate the level of the foreground residuals, i.e., a mismatch between the estimated
and true sky components. This can be expressed as follows \citep{2010MNRAS.408.2319S},
\begin{eqnarray}
\bd{\Delta} = \bd{s}\, - \hat{\bd{s}} = \l(\bd{Z}\l(\bd{\beta}\r)-\bd{I}\r)\,\hat{\bd{s}},
\end{eqnarray}
where
\begin{eqnarray}
\bd{Z\l(\bd{\beta}\r)} \equiv \l(\bd{A}^t\l(\bd{\beta}\r)\,\bd{N}^{-1}\,\bd{A}\l(\bd{\beta}\r)\r)^{-1}\,\bd{A}^t\l(\bd{\beta}\r)\,\bd{N}^{-1}\,\bd{A}(\hat{\bd{\beta}}),
\end{eqnarray}
 $\bd{I}$ is a unit matrix and, as usual a hat over a quantity denotes its  true underlying value.

The foreground residuals left in the CMB map are just one component of the vector, $\bd{\Delta}$, which for definiteness is assumed to be the zeroth one.
 We will now restrict ourselves to the CMB component and linearize the problem,
assuming that the errors in spectral parameter determination are small. We thus obtain
\begin{eqnarray}
\Delta^{\rm CMB} = \sum_{k,j}\,\delta\bd{\beta}_k\,\bd{\alpha}^{0j}_k \, \hat{\bd{s}}^j,
\label{eqn:deltaCMBlin}
\end{eqnarray}
where
\begin{eqnarray}
\bd{\alpha}^{ij}_k \equiv \frac{\partial\,\bd{Z}_{ij}(\hat{\bd{\beta}})}{\partial \bd {\beta}_k},
\end{eqnarray}
and we assumed that the CMB component is stored as first (i.e., with an index equal to 0) in the component vector, $\bd{s}$.
We can now characterize the level of the residuals either simply by its {\em rms} value or, in a more informative way
 we can estimate the noise average (though noiseless) foreground residual power spectrum, which reads
\begin{eqnarray}
{\bd C}_\ell^{\Delta} \equiv
\sum_{k,k'}\sum_{j,j'}\,\bd{\Sigma}_{kk'}\,\bd{\alpha}^{0j}_k\,\bd{\alpha}^{0j'}_{k'} \, \hat{\bd{C}}^{jj'}_{\ell}.
\label{eqn:resSpec0}
\end{eqnarray}
Given that as mentioned before (see also, \citep{2010MNRAS.408.2319S}) no CMB signal is left in the CMB map residuals, which combine just
the foreground signals, the noise ensemble averages coincide with those made over a full CMB + noise set of realizations.
Clearly to compute
the residual spectra we need to make assumptions concerning the spatial morphology of the considered foregrounds,
i.e., the knowledge beyond the $\hat{\bd{F}}$ matrix defined earlier. This is reflected in Eq.~\eref{eqn:resSpec0} by the presence of
true auto- and cross- spectra for each considered foregrounds, $\hat{\bd{C}}^{jj'}$.
However, the $\hat{\bd{F}}$ matrix provides a sufficient description necessary to calculate the rms value of the residuals. This can be seen noting that
\begin{eqnarray}
{\Delta^{\rm CMB}_{rms}}^2 = \sum_{k,k'}\sum_{j,j'\ne 0}\,\bd{\Sigma}_{kk'}\,\bd{\alpha}^{0j}_k\,\bd{\alpha}^{0j'}_{k'} \, \hat{\bd{F}}_{jj'}.
\label{eqn:rmsCMBdef}
\end{eqnarray}
In the following we will use the $C_{\ell}^{\Delta}$ quantity to construct our FOMs making some specific assumptions about the foregrounds
spatial properties as described in Sec.~\ref{sect:foregrounds}. We point out that the formulas presented above are just a special case
of those already studied in \citep{2010MNRAS.408.2319S}. The important difference is however that the spectral indices uncertainties used 
in this work are computed  effectively as the full CMB + noise, ensemble averages rather than derived in a 
single, particular study case as in that previous work.

\subsection{Figures of merit}

\label{subsect:fom}

Given the estimates of the foreground residuals provided in the previous Section, we can now define our figures of merit.
Hereafter, we will use three FOMs: two referring to the effects of the foreground residuals found in the recovered CMB
map as a consequence of the separation process, and the third related to the noise level of that map. As our scientific goals
here are related to the primordial $B$-mode signal two of the proposed FOMs express the effects of the foreground residuals
on a tensor-to-scalar ratio (of the respective CMB spectra), $r$. The third one is more generic and is just to ensure that the least-noisy
map of the sky is produced.

\paragraph{{\bf FOM\#1}: $r_{stat}$ -- an $r$ value detectable on $95$\% confidence level incorporating the component separation uncertainties.}
\ \\
This FOM is computed in two steps. First,
we use a generalized Fisher matrix expression to estimate the uncertainty of estimating the tensor-to-scalar ratio, $r$, for any given assumed $r$ value,
and subsequently we determine a value of $r \equiv r_{stat}$, which is detectable on $95$\% confidence level. 
This limiting value is defined as
\begin{eqnarray}
r_{stat} \simeq 2\, F_{rr}^{-1/2}\l(r_{stat}\r).
\label{eqn:rLimStat}
\end{eqnarray}

The Fisher matrix we propose to use here accounts for usual cosmic, sampling, and noise variance, but also for an extra error resulting from the shortcomings of
the foreground component separation, which is presumed to be applied to the maps beforehand.
We model the separation residuals following the formalism introduced in Sec.~\ref{subsect:residuals} and which treats
the map-level residuals  as a linear combination of the foreground templates with Gaussian distributed amplitudes. 

The detailed derivation of the Fisher formula is presented in Appendix~\ref{sect:algebra_2}. Recalling that $C_{\ell}^{\Delta}$ denotes the power spectrum of the residuals, 
the final expression for the Fisher matrix, $F_{rr}$, reads then
\begin{widetext}
\begin{eqnarray}
F_{rr} = \sum_{\ell,\ell'}^{\ell_{max}} \,   \frac{\partial C_{\ell}}{\partial r} \,  \left\{ \frac{\left(2\ell+1\right) \delta_{\ell\ell'}}{2\,f_{sky}^{-1}\,C_{\ell}^{2}} \,   - 
 \, \frac{\left(2\ell+1\right)C_{\ell}^{-3}C_{\ell}^{\Delta}\, \delta_{\ell\ell' }}{\left( 1 +{\displaystyle \sum_{\ell''}^{\ell_{max}}\; (2\ell''+1)\,\frac{C_{\ell''}^{\Delta}}{C_{\ell''}}}\right)}
 +  \frac{ \left(2\ell+1\right)\,\left(2\ell'+1\right)\,C_{\ell}^{\Delta} \, C_{\ell'}^{\Delta}}{2C_{\ell}^2C_{\ell'}^2\left( 1 +{\displaystyle \sum_{\ell''}^{\ell_{max}}\; (2\ell''+1)\,\frac{C_{\ell''}^{\Delta}}{C_{\ell''}}}\right)^2}
\right\} \,\frac{\partial C_{\ell'}}{\partial r} 
\label{eq:exact_fisher_result}
\end{eqnarray}
\end{widetext}
where for shortness we set $C_{\ell} \equiv C_{\ell}^{CMB}+C_{\ell}^{noise}$.

A choice of experimental parameters will in general affect both the noise level as quantified by $C_\ell^{noise}$ but also the level of residuals resulting in different
$r_{stat}$ values derived for different proposal configurations.

We note that if the level of residuals is very high as a result of the errors on spectral parameters being large then the first order expansion
used to obtain Eqs.~\eref{eqn:deltaCMBlin}~and~\eref{eqn:resSpec0}  may not be any more sufficient. Likewise, if the foreground contributions are 
large so their residuals are comparable 
to the CMB signal, sufficiently precise knowledge of the foregrounds would become necessary to ensure that the above formulas produce 
reliable results. As one
may not be completely comfortable with such a presumption, we will introduce another FOM designed to penalize such configurations.

\paragraph{{\bf FOM\#2}: $r_{eff}$ -- an effective $r$ value of the foreground residuals.}
\ \\
We use a proposal of  \cite{2007PhRvD..75h3508A} and we characterize any obtained foreground residuals using its effective value of $r$ defined as
\begin{eqnarray}
s(r_{eff}) \simeq u,
\label{eq:amblard}
\end{eqnarray}
where
\begin{eqnarray}
s\l(r\r) &\equiv& \sum_\ell^{\ell_{max}} C_\ell^{cmb}(r)- C_\ell^{cmb}(0),\nonumber
\\
u &\equiv& \sum_\ell^{\ell_{max}} C_\ell^{\Delta}.\nonumber
\end{eqnarray}
We note that due to a missing factor of $2\ell+1$ this criterion does not compare power contained in the primordial $B$ spectrum with that of the residuals (up to $\ell_{max}$),
and  in contrast to the latter it gives more weight to low multipoles.

\paragraph{{\bf FOM\#3}: $\sigma^{noise}_{\rm CMB}$ - noise level of the recovered CMB map.}
\ \\
When the true values of the spectral parameters are available the only uncertainty of the recovered component maps, Eq.~\eref{eqn:mlMapDef}, is due to the instrumental noise and reads
\begin{eqnarray}
{\cal N} & = & \l(\bd{A}^t\,\bd{N}^{-1}\bd{A}\r)^{-1},
\label{eqn:noiseProp}
\end{eqnarray}
and therefore
 depends on the number of detectors and frequency channels. With our focus on the CMB
we will therefore use the diagonal element of ${\cal N}$ corresponding to the CMB component as one of our criteria, which we would like to keep as low as only possible. We thus have
\begin{eqnarray}
\l(\sigma^{noise}_{\rm CMB}\r)^2 \equiv {\cal N}_{00}.
\end{eqnarray}
 We note that only when $\bd{A}$ is a unit matrix the above formulas corresponds to a standard, inverse-noise-coaddition. This
in turn can only happen if no sky components are mixed together, implying no foregrounds. In any other case the final noise of the CMB map is higher than the 
inverse noise weighting would imply~\cite{2011MNRAS.tmp..449B} and
its exact value will depend on the details of the component scalings and experimental set up. We note that unlike two other FOMs implemented here this applies on
a map rather than a power spectrum level. Moreover, as the spectral parameters, $\bd{\beta}$, are assumed to be known ahead of the computation, this FOM may
lead to configurations in which the estimation of those is not feasible and thus rendering the residuals effectively arbitrary and unknown. Nevertheless, though it 
needs to be used with a care, it provides a meaningful reference against which to gauge other configurations.

\subsection{Optimization procedure}

\label{subsect:minProc}

\subsubsection{Parameters and optimization approaches}

In this work typically we will optimize a number of detectors in each of the pre-defined frequency channels. This is clearly one of the most basic hardware parameters
one would like to know designing a $B$-mode experiment. Though the central frequency of the channels is often constrained from the onset by some hardware constraints,
we will also consider more general optimization problems in which a number of frequency channels, their central frequencies, and a number of detectors per channel
are all to be optimized with respect to.

In the former case we perform a single global optimization operation. Our numerical codes  use a minimization algorithm for constrained nonlinear multivariate function, as implemented in {\sc matlab}, which is based on a line-search algorithm with constraints introduced via a quadratic approximation to the Lagrangian function.

\begin{figure}
 	\includegraphics[width=8cm]{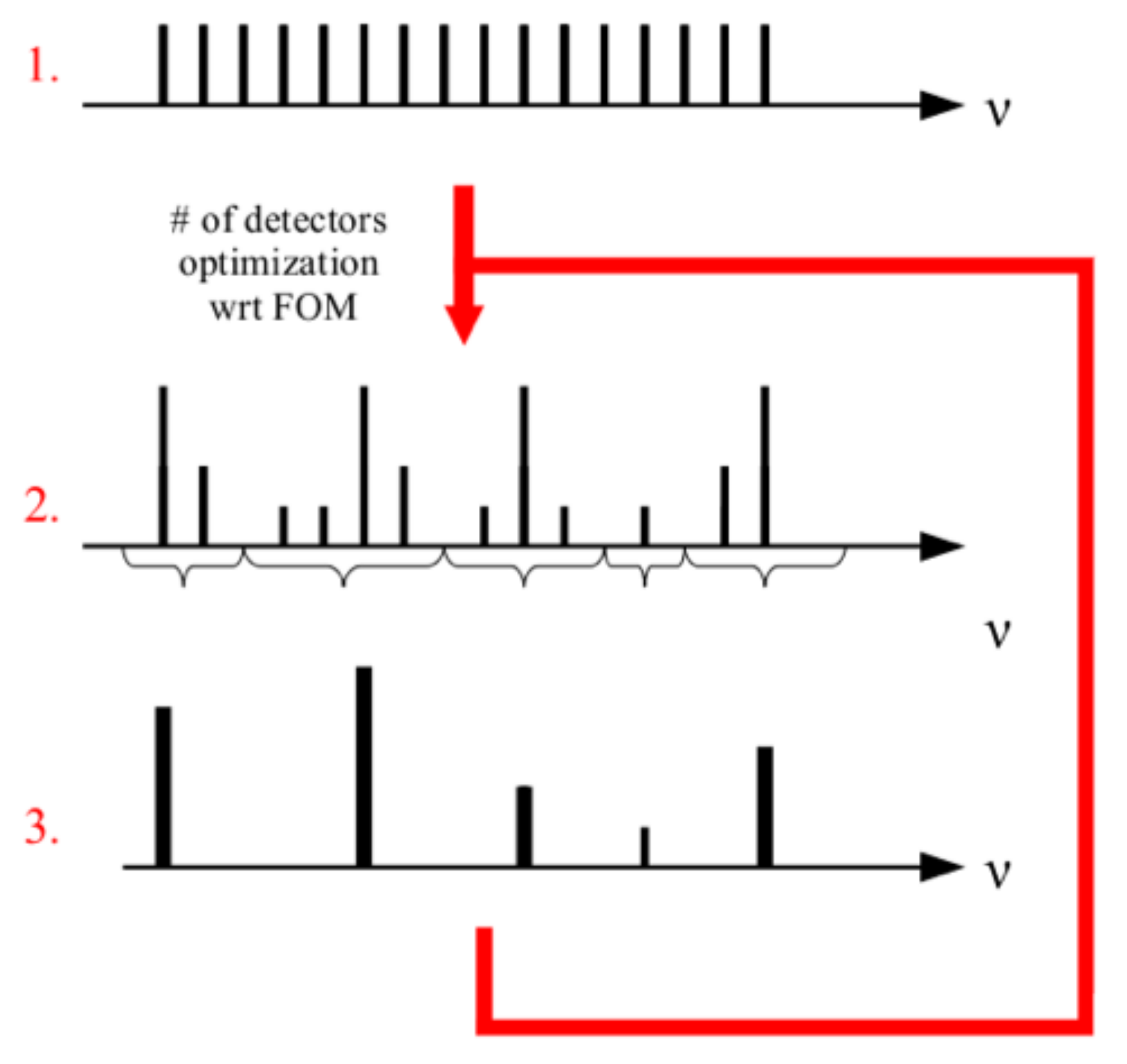}
\caption{Schematic illustration of our optimization procedure in a case of an adjustable number of channels, a number of detectors per channel, and their central frequencies.}
\label{fig:scheme_freq_and_opt_explanation}
\end{figure}

In the second type of the optimization problems we have found that attempts of performing a global optimization are often frustrated by numerical issues and the results are consequently not very
reliable. Instead we have devised a multi-step approach which is shown schematically in Fig.~\ref{fig:scheme_freq_and_opt_explanation}.
In the proposed method we start from a configuration consisting of a focal plane  overpopulated with a large number of mock channels uniformly covering the requested interval of frequencies. 
Each of these channels is assigned the same number of detectors or a fraction of the focal plane area, depending on which hardware constraint we use (step 1). We then optimize the number of detectors or focal plane area as in the
standard case with the fixed frequency channels  with respect to a given FOM (step 2). As the obtained detector distribution is typically rather inhomogeneous we then merge the channels with close central frequencies, e.g., closer than the expected 
band-width of the anticipated channels. In the process of merging we replace some subsets of channels by a new channel, centered at the barycenter of the previous  frequencies as weighted either by a number of
detectors or focal plane  assigned to each of the merged channels, and assign to it either their detectors or the corresponding 
focal plane area (step 3).
We optimize this new configuration again with respect to numbers of detectors per channel, and go back to step 2 whenever the resulting configuration is found very inhomogeneous. Then we repeat this process
again. We find however that usually a single pass over the optimization sequence produces satisfactory results. 

\subsubsection{Constraints}

The constraints can be imposed straightforwardly via Lagrangian multipliers
therefore permitting a wide variety of those, which can, and sometimes have to, be introduced. 

These include some trivial constraints stemming from the physical interpretation of the optimized parameters, e.g.,  ensuring non-negative values for detector numbers or 
focal plane area, which have to be usually  included explicitly. 

There are also some fundamental constraints without each the convergence could not be reached at all. These 
typically followed from the hardware restrictions. As an example of the hardware constraint, hereafter we will use either a constraint on a total area of the focal plane or on a total 
number of detectors, corresponding to cases where we have full freedom to fill in the entire focal plane as densely as only needed or when such freedom is restricted, for instance,
by capability of our read-out systems.

Yet another type of constraints invoked in the optimizations studied here includes those driven by the science goals rather than hardware requirements. For instance, we could require that some specific 
frequency
channel map has a noise level better than some pre-set level in order to make such a map good enough to investigate some sky objects or features of interest. 
These kinds of constraints are often needed in the post-processing phase described later.

In addition, while considering multiple FOMs simultaneously we will typically use some of them as constraints restricting the optimization to such configurations for which the required 
values of these FOMs is better than some suitable threshold.

\subsection{Post-optimization processing}

\label{subsect:postProc}

The optimized solution formally determined as described here in most of the cases will require further adjustments and tuning,
before it could become a basis for an actual instrument design and later its potential development.

Specific instances of such post-optimization processing, which we consider hereafter include:
\begin{itemize}
\item {\bf design simplification} -- including either rounding of numbers of detector per channels and/or removing some channels 
altogether, in particular those assigned a small number of detectors. 
\item {\bf addition of some \textit{ad hoc} frequency channels} -- for instance, either to improve the overall robustness of the derived
configuration with respect to potential surprises concerning physical properties of the foregrounds, or to extend the science
goals beyond  what is already encoded in the FOMs.
\end{itemize}

In all these cases a crucial question is how significant modifications from the initial optimized setup are allowed before the science goals, 
as expressed by the FOMs, are compromised too significantly to be acceptable. Below we outline a general approach devised  to answer such questions in some specific cases relevant to the applications considered here, leaving a more detailed description of its practical implementation in our study 
cases to Sec.~\ref{sect:applications}.

\subsubsection{Detector number rounding}

Let us consider only channels for which the optimization procedure has assigned a nonzero number of detectors. Moreover we start from the channels for which we want to decrease a number of detectors, as a result of the rounding procedure, and postpone the treatment of the remaining ones
 for later.  For the time being we also relax all the constraints imposed on the optimization, with an exception of the ones ensuring
positivity of a number of detectors or focal plane area. Removing some of the detectors decreases the instrument sensitivity and thus will  affect 
our science goals, unavoidable rendering the experiment less competitive. For any specific configuration we can always calculate exactly its performance in terms of the adopted FOMs. However, on the experiment designing stage, when many such configurations may need to be considered and often quickly discarded, the need for the case-by-case computation may be a hinderance. In such a context a fast, even if rough and approximate, approach could be therefore a handy substitute permitting one, on the one hand, to  zoom quickly on an interesting family of potential solutions, and, on the other, to reject configurations which are clearly of no interest.
One way to address such a need could be to construct, for each FOM,  a series 
of hyper-volumes, ${\cal V}_k,\ (k=0,..., n_{\cal V}-1)$, centered on the optimized configuration and such that  ${\cal V}_0 \subset {\cal V}_1 \subset \dots \subset {\cal V}_{n_{\cal V}-1}$. To each volume, ${\cal V}_k$, we can assign uniquely a value, $\tilde{v}_k$, such as,
\begin{eqnarray}
\tilde{v}_k \equiv {\min_{\l\{d_i\r\} \in {\cal V}\l(v_k\r)}} \Bigl\{{\rm FOM}\bigl(\{d_i\}\bigr)\Bigr\},
\label{eqn:vtildeDef}
\end{eqnarray}
i.e., which defines the worst performance plausible within the volume.
The values $\tilde{v}_k$ are directly arranged in a descending order given that any volume contains all the previous ones. If now a configuration of our interest belongs to the 
$k$-th volume
and does not to the $(k-1)$-th one we immediately can infer that its performance, $\tilde{v}$, expressed in terms of the given FOM, is bracketed by the two values corresponding
to these two hyper-volumes, i.e., $\tilde{v}_{k-1} \le \tilde{v} \le \tilde{v}_{k}$.

Two features are essential to make such a scheme useful. First, we have to have an easy way to identify whether a given configuration is or is not contained in a given 
hypervolume. Second, the volumes have to be defined in such a way that the values of $\tilde{v}_k$ assigned to them span a range of interesting values and do so sufficiently
densely. Given potential high-dimensionality of the parameter space we consider here, none of these two requirements is straightforward to satisfy. To address the first of them we propose to use as the volumes hyperellipsoids defined as
\begin{eqnarray}
 {\cal V}_k \equiv \Bigl\{ \bigl\{d_i\bigr\} \Big{|} \sum_i\,\frac{(d_i-d_i^{opt})^2}{{\sigma_i^{\l(k\r)}}^2} \le1,\,  d_i < d_i^{opt} \Bigr\},
\label{eqn:hyper_ellipses}
\end{eqnarray}
where the last condition on the right hand side narrows the volume to the cases of our interest here. The semiaxes of the ellipsoid, $\sigma_i^{\l(k\r)}$, need to reflect the fact
that the rate at which the given FOM changes will be in general different in different directions in the parameter space. We therefore determine them for every direction corresponding to varying detector numbers in a single channel separately and we do it for each channel of relevance here,  i.e., for which  $d_i^{opt} \ne 0$. The procedure
here involves two steps. First, we select a grid of values of the considered FOM, $v_k$,
which covers the range of its values of our interest and does that with a sufficient density. This grid is used consistently  for all directions and channels.
Subsequently, for every channel, $i$, we find numerically a dependence between a value of FOM and a distance from the optimized solution along $i$-th axis of the 
parameter space
and use this relation to determine $\sigma_i^{\l(k\r)}$ so  $FOM\bigl( \sigma_i^{\l(k\r)}\bigr) = v_k$. Typically, the grid point values, $v_k$, will provide a good approximation to 
the worst case values, $\tilde{v}_k$, defined earlier. The latter are therefore expected to be automatically well-spaced and to span a sufficient interval of FOM values.
In actual applications, we compute more precise estimates of $\tilde{v}_k$ than those provided by $v_k$. This is done by using Eq.~\eref{eqn:vtildeDef} and randomly sampling the volume of the corresponding hyperellipsoid.

The proposed construction therefore obeys the two requirements we defined earlier and provides a quick and easy way to find out how far the configuration can be tweaked,
without compromising the science goals. The parameters $\sigma_i^{\l(k\r)}$ and $\tilde{v}_k$  constitute an additional and important piece of information, 
which should be determined and provided alongside any optimized configuration to render the optimization process helpful. We demonstrate this in actual applications in 
Sec.~\ref{subsect:postProcApp}.

So far we have neglected the hardware constraints. Those would require that any subtraction of the detectors from some of the channels
 needs to be accompanied by adding detectors somewhere else. However, as adding detectors can only improve our FOMs, the procedure outlined above
is conservative as the final outcome of the rounding with the constraints fulfilled can be only better than what the procedure implies. 

We can now get back to the channels for which we might have wanted to round up the number of channels. This can be done but only by appropriately distributing 
the detectors we have removed earlier, as the overall hardware constraint has to be fulfilled.
If we do not have however strong preferences regarding their distribution we may try to
perform a second round of the optimization to find out how it can be done in an optimized way.
This could be done by solving the optimization problem as the initial one but adding extra constraints fixing the number of detectors to their rounded
value in all the channels, where the rounding has been applied.

\subsubsection{Low-populated channels}

The formal optimization procedure proposed here may result in configurations, which include a number of channels with a relatively low
number of detectors. As extra frequency channels contribute to an overall complexity of the instrument, it could be advantageous to remove those
if there is no strong science driver behind them. Removing entire channels is more delicate than a removal of  some fraction of the detectors as discussed above.
This is because it can render the separation process singular or nearly so with separation errors growing rapidly. The singularities however can be
usually avoided by keeping track of a number of channels needed to separate some specific number of components, each described by a well-defined number
of parameters. We will therefore assume throughout that this is indeed the case. We then proceed as follows with the underpopulated channels. We remove such a
channel or contiguous group of those and  either redistribute the extra detectors between the adjacent channels or create a new channel with a central
frequency computed as a detector (or focal plane area) weighted average of the frequencies of the channels to be replaced. 
We then test the change in the FOM values. If either of the options is not satisfactory, we can try to further to improve on it by performing formal optimization but now 
using only channels which contain a nonzero number of detectors. If that still turns out to be much worse than the optimized values of the FOM, we subsequently need
to identify, which of the low-populated channels are crucial from the performance point of view and retain them in our final configuration, while 
removing or merging the others.

\subsubsection{Ad hoc extra channels}

Clearly our optimized configuration is only as good as the foreground model assumed in the optimization process. The impact of some of the uncertainties in the foreground
modeling can be  discussed directly within the formalism presented here as, for example, that of details of the foreground correlation matrix and/or shape of their power spectra. 
It is more difficult however to investigate the role of our assumptions about a number of spectral parameters and/or a number of foreground components.
In that respect one may feel more at ease with the configurations, which have the entire frequency range accessible to the instrument sufficiently populated, as they, at least on the intuitive level,
may appear more robust with regard to the unknown.

If the optimization does not  lead to a configuration, which satisfies such a condition on its own, one may want to impose it by adding one or more \textit{ad hoc} frequency channels in the areas they
are missing. This can be done straightforwardly by adding a constraint requiring at least some predefined and nonzero number of detectors in those channels. If this
number is fixed exactly, it will be obviously not  anymore a parameter of the optimization, however the channel will still take part in the optimization process as it will be taken into
account in the FOM computation. We use this approach to answer an important question, i.e., how close such a new configuration would perform as compared to
the original, optimized one. In other words, should the foreground model used turn out to be correct, would we lose much by trying to make the configuration
more robust~? Ideally, the loss of performance will not be significant, permitting us to reach both these goals simultaneously: near optimality whenever our modeling
is correct, and ability to meet the surprises. In Sec.~\ref{subsect:postProcApp} we discuss how the parameters of such \textit{ad hoc} channels can be proposed in a specific application.

\subsection{Design robustness}

\label{subsect:robustness}

A problem closely related to the one discussed above is that of the robustness of the final configuration.
Given some unavoidable failure rates in a technological process involved in the instrument design and development,
a final version of the instrument typically comes short of the actual design target. An important and valid question then
is how robust the science goals posed for the experiment are assuming that the target has been defined using the 
procedure described here.
We address this problem in a specific case in which we admit some failure rate for the detector production process, $\varepsilon$.
For a set of realistic values of $\varepsilon$ we perform a random sampling of the parameter space randomly drawing a
number of failed detectors. We then evaluate the full set of FOMs for each of the samples and find what is an average, likely
on $95$\% confidence level impact of the considered failure rates on the FOM values.

\section{Foreground modelling}

\label{sect:foregrounds}

As discussed earlier in our formalism there are two key quantities needed 
to describe completely the effect of foregrounds. These are the auto- and cross- spectra
characterizing the spatial distribution of the foreground components and the component
correlation matrix, $\bd{\hat{F}}$. To calculate these we will rely on a specific model of the
Galaxy and since we are interested in the $B$-modes, we will consider only diffuse foregrounds,
synchrotron and dust, with known and non-negligible polarization emission. 

To simulate these emissions in polarization we implement the same recipe as in
\cite{2010MNRAS.408.2319S}, which starts off from deriving reliable total intensity
templates from the available data  (the Haslam map \cite{1982S&T....63..230H} for the
synchrotron and the combined COBE-DIRBE and IRAS for the dust
\cite{1998ApJ...500..525S}), rescales them using some constant overall polarization
efficiency factor, fixed to $10$\% in order to match the large scale $E$ and $B$ spectra of
\cite{2007ApJS..170..335P}, therefore producing polarization intensity templates. The polarization
angles on the largest scales are then determined using a combination of the WMAP data and three-dimensional modeling of the Galactic magnetic
field as in \cite{2007ApJS..170..335P},  while on the small angular scales ($\simlt 1^\circ$), by randomly simulating those using their angular power spectra as derived from the data \cite{2002A&A...387...82G}.
 
We assume spatially constant frequency 
scalings: a power law with index $\beta_s=-3$ for the synchrotron, i.e.,
\begin{eqnarray}
A_{\rm sync}\l(\nu,\nu_{ref}\r) = \l(\frac{\nu}{\nu_{ref}}\r)^{\beta_{s}}\ 
\end{eqnarray}
and a uniform greybody scaling law, as in Model 3 of \citep{1999ApJ...524..867F}, 
\begin{eqnarray}
\label{eq:dust_scal}
A_{\rm dust}\l(\nu,\nu_{ref}\r) = \l(\frac{\nu}{\nu_{ref}}\r)^{\beta_{d}+1}\ 
\frac{ \exp  \frac{ h\nu_{ref}}{kT_d}-1} { \exp \frac {h\nu}{kT_d}-1}\ ,
\end{eqnarray}
where $T_d=18.0$ K and $\beta_d=1.65$ for the dust.

As pointed out in \cite{2010MNRAS.408.2319S}, by
adopting this model a large amount of correlation is expected
between dust and synchrotron both because the Galactic magnetic field
is a common ingredient and because of the lack of high resolution data
that forces us to extend the correlation to small scales. This is
reflected in the fact that the off-diagonal terms of $\hat{\bd{F}}$ are of
the same order of the diagonal terms. However, as we discuss in Sec.~\ref{subsubsect:specParErr}
large off-diagonal terms inflate the errors on spectral parameters, so from the
perspective of foreground residuals the employed model can be considered 
conservative.

\begin{figure}
 	\includegraphics[width=7.5cm]{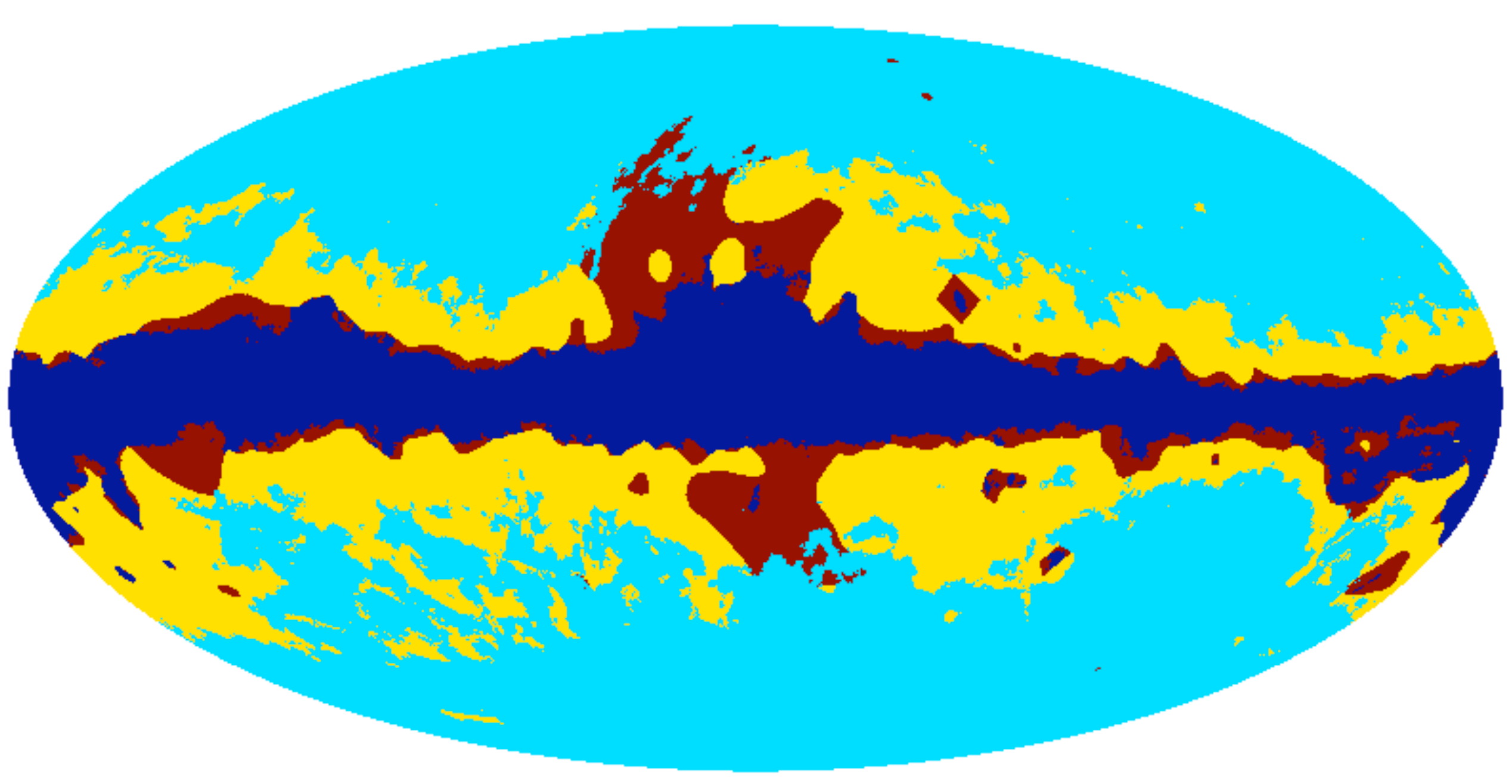}
	\caption{Three foreground masks as used in this work. Yellow (largest mask), dark red (large mask round the galactic bulge), and dark blue (narrowest mask around the galactic plane) mark sky areas excluded from the masks: Mask II, P06, and Mask I, respectively.}
	\label{fig:masks}
\end{figure}

\begin{figure}
 		\includegraphics[width=8.5cm]{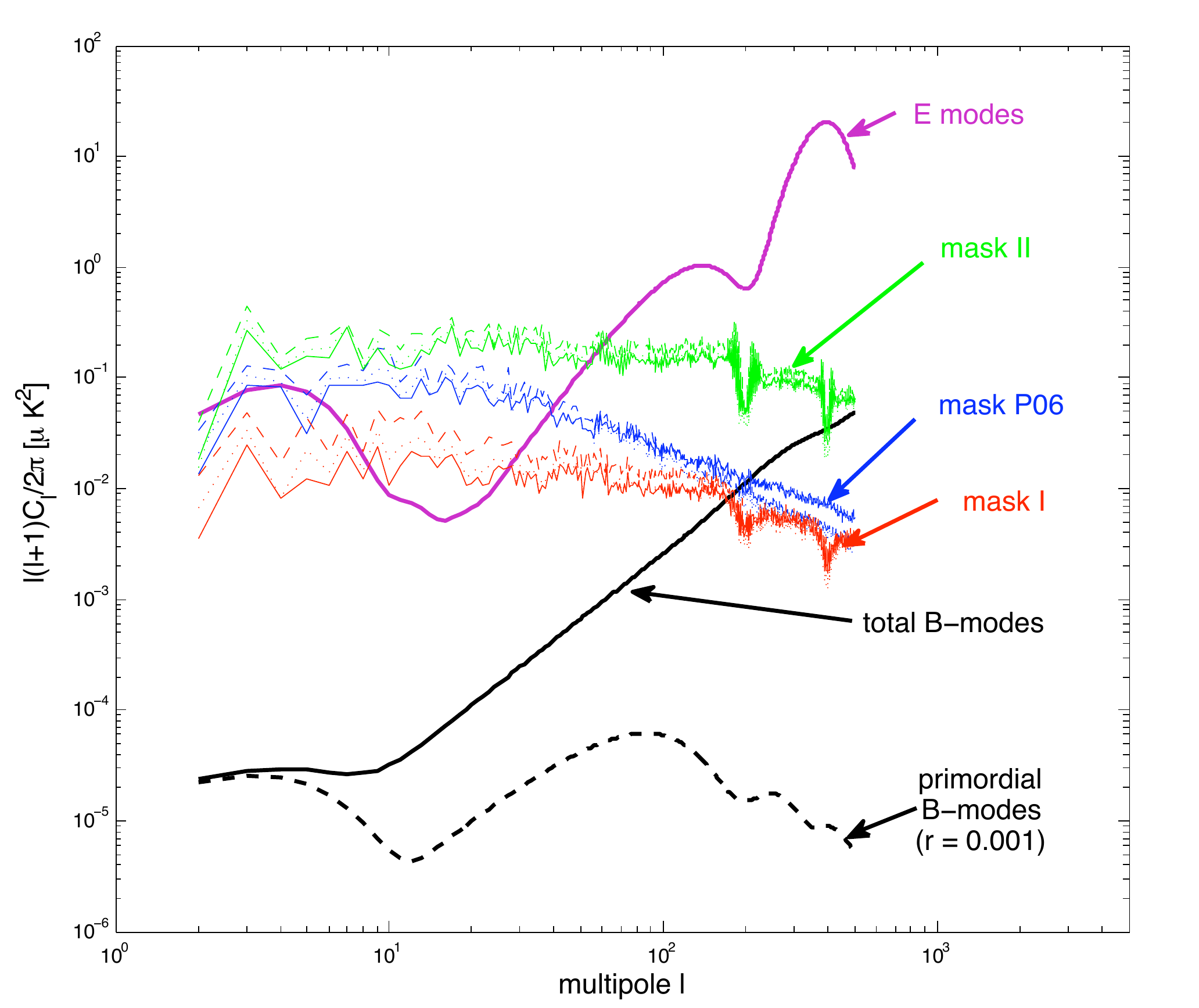}
		\caption{Pseudopower spectra of the foreground templates for the three different masks considered in this work and contrasted with the CMB $B$-mode power spectrum. For each mask the three lines show dust (solid line), synchrotron (dashed line), and their cross-correlation (dotted line). The foreground signals are computed at the $65$ GHz. All the spectra used in this work are computed from HEALPIX-pixelized maps with nside = 512.}
	\label{fig:fig4}
\end{figure}

To investigate the effects of different foreground contrasts and morphology we
consider here three different sky masks. Mask I and Mask II are tailored in such a way that they
have the possible total polarized foreground contrast (synchrotron
plus dust) lower than a predefined threshold equal to $0.86$ and $0.36\,\mu$K, respectively. We also employ more standard the P06 mask from the WMAP team,
which is optimized for the low frequency coverage of WMAP, i.e. it is skewed toward cutting out more
the synchrotron than the dust emission. All three masks are shown in Fig.~\ref{fig:masks} and their corresponding foreground (pseudo) power spectra are displayed in Fig.~\ref{fig:fig4}. In addition, in Table~\ref{table:Ff_components} we list the elements of the matrix $\hat{\bd{F}}$ for each of them.

\begin{table}
\begin{tabular}{|c|cccc|}
\hline 
Mask & $f_{sky}$ & $\hat{\bd{F}}_{\rm dust-dust}$ & $\hat{\bd{F}}_{\rm dust-sync}$ & $\hat{\bd{F}}_{\rm sync-sync}$ \\
\hline
P06 mask & 0.73 & 3.20 &  0.082 & 0.0025 \\
Mask I & 0.82 & 1.12 &  0.029 & 0.00084 \\
Mask II & 0.51 & 1.74 & 0.053 & 0.0019\\
\hline
\end{tabular}
\caption{$\hat{\bd{F}}$ matrix elements computed for two foreground components, dust and synchrotron,  at the fiducial frequency of $70$ GHz  for the three masks used in this work and  all pixelized using \textsc{healpix} scheme with ${\rm nside} = 128$.}
\label{table:Ff_components}
\end{table}

These masks are thought to be applied  \textit{a posteriori} to the full sky map, assumed to be homogeneously observed by the experiments. This means that 
the noise level per pixel, described in Sec.~\ref{noise_levels}, will be the same for each of them and thus the results of the FOM\#3 optimization will
be the same in all three cases.

\section{Applications}

\label{sect:applications}

\begin{figure}
\begin{center}
	 \includegraphics[width=4.05cm]{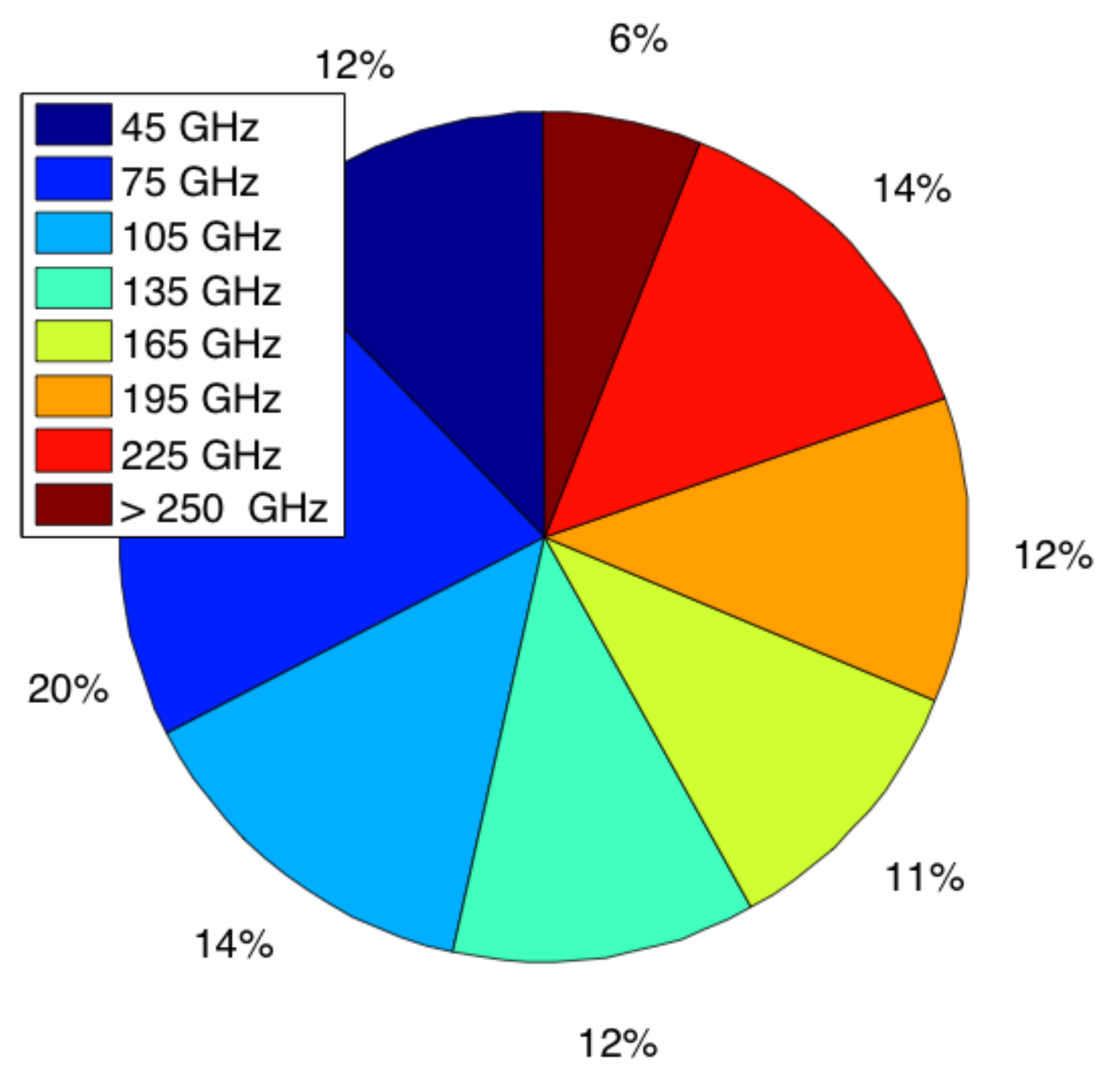}
	 \includegraphics[width=4cm]{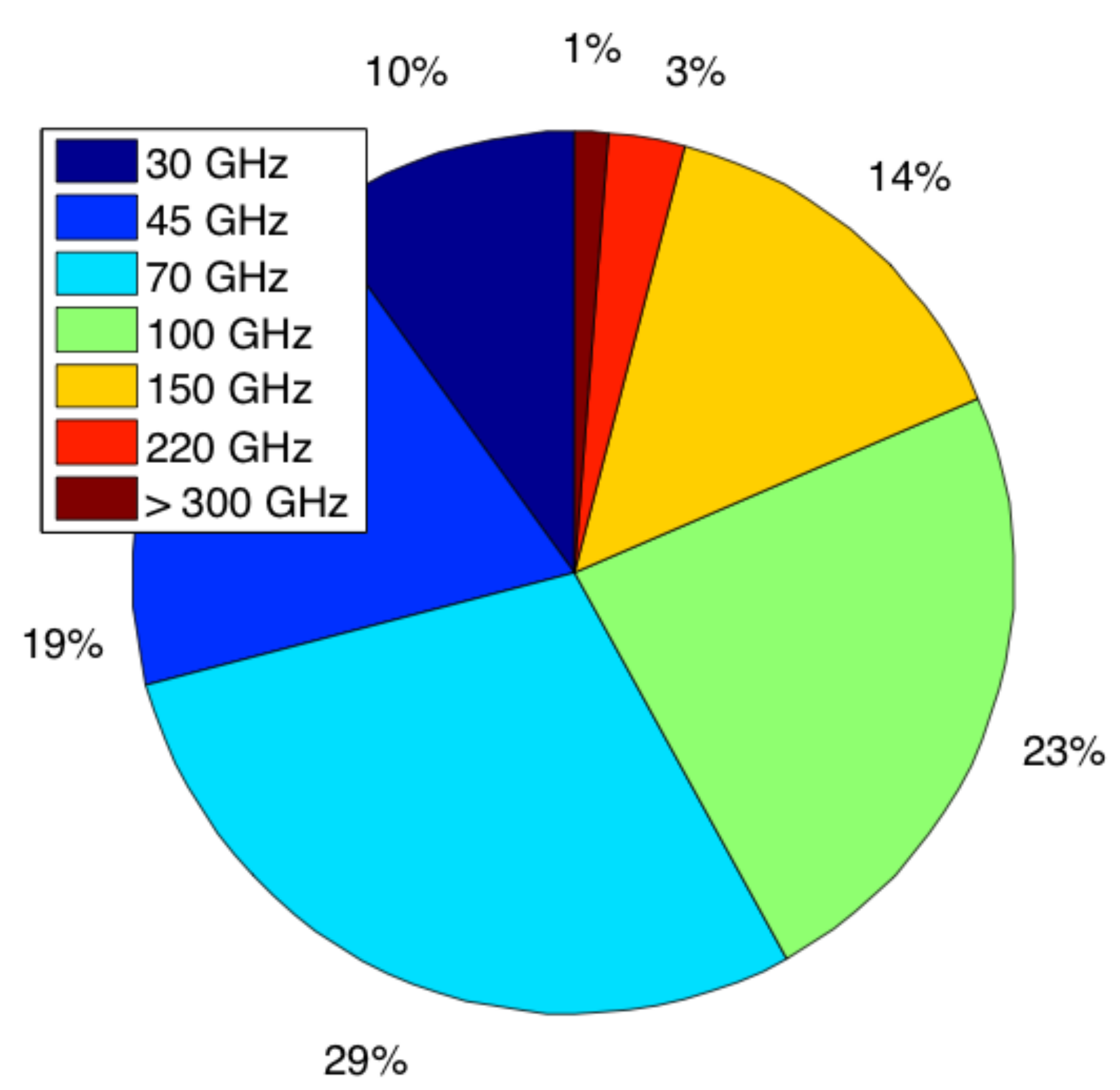}
	 \caption{Breakdown of the focal plane area between the frequency channels as originally proposed for the COrE, left, and CMBpol, right, satellites. In the case of COrE all the channels with frequencies larger than $250$ GHz represent less than 10\% of the total focal plane area.} 
	 \label{fig:frac_area_full_core}
\end{center}
\end{figure}

As an illustration of the method detailed in the previous sections, we will consider the optimization of two different full sky satellite designs:  Cosmic Origins Explorer (COrE)
proposed in response to the European Space Agency Cosmic Vision 2015-2025 Call~ \cite{2011arXiv1102.2181T}, and CMBpol~\cite{2009arXiv0903.0902A, 2009astro2010S..67D}, proposed as part of the NASA mission concept study. The respective frequency channels and a number of detectors per channel corresponding to the original designs are summarized in Table~\ref{table:cmbpol_focal_plane} for CMBpol and in Table~\ref{table:core_focal_plane} for COrE. 
\begin{table*}
\begin{tabular}{c|ccccccccc} 
Frequency [GHz]  & 30 & 45 & 70 & 100 & 150 & 220 & 340 & 500 & 850 \\
\hline
Number of detectors & 84 & 364 & 1332 & 196 & 3048 & 1296 & 744 & 938 & 1092
\end{tabular}
\caption{CMBpol distribution of detectors among the different channels.}
\label{table:cmbpol_focal_plane}
\end{table*}
\begin{table*}
\begin{tabular}{c|ccccccccccccccc} 
Frequency [GHz]  & 45 & 75 & 105 & 135 & 165 & 195 & 225 & 255 & 285 & 315 & 375 & 435 & 555 & 675 & 795\\
\hline
Number of detectors & 64 & 300 & 400 & 550 & 750 & 1150 & 1800 & 575 & 375 & 100 & 64 & 64 & 64 & 64 & 64
\end{tabular}
\caption{COrE distribution of detectors among the different channels.}
\label{table:core_focal_plane}
\end{table*}
In our analysis we will assume the same noise levels per detector for each of the experiments, Sec.~\ref{noise_levels}, and that they scan the sky homogeneously with all the detectors observing simultaneously over the course of $4$ years. Everywhere in this paper, but in Sec.~\ref{subsect:varying_number_channels}, we will aim at optimizing  a number of detectors per channel, assuming that the latter are fixed and known, and 
keep either the effective area of the focal plane or total number of detectors constant. The assumed values for the two constraints are derived given the proposed configurations 
of COrE (Table \ref{table:core_focal_plane}) and CMBpol (Table \ref{table:cmbpol_focal_plane}). 
In the case of the focal plane area we assume that an area of the focal plane occupied by a single, diffraction-limited detector operating at frequency, $\nu$, can be expressed as
\begin{eqnarray}
	\mathcal{A} \sim \frac{c^{2}}{\nu^{2}}.
	\label{eq:area_expression}
\end{eqnarray}
The total focal plane area is then obtained by summing over the contribution coming from all the detectors. We note that this gives at the best some
 \emph{effective} area because we do not take into account any kind of filling factor, which is usually driven by technical constraints such as the shape of the detectors, the wiring, etc. 
 Fig.~\ref{fig:frac_area_full_core} shows the fractional area as occupied by each channel in the case of the  proposed versions.

Hereafter we neglect the effects of the $E$-$B$ leakage, e.g., \cite{2009PhRvD..79l3515G}, both in the calculations of the foreground spectra as well as the CMB variance. In the former case this is justified
given the fact that $E$ and $B$ spectra for foregrounds are on comparable levels and the leakage is usually harmless. For the CMB variance we assume that the effects of such a leakage can be largely removed using one of the methods proposed in the literature.
Though corrections of this sort usually lead to some extra precision loss, this is typically only a fraction of the standard cosmic variance and, at least for experiments 
with a sufficiently large sky coverage, small enough not to change our results in a significant way. For small-scale observations the effect may not be negligible and should be taken into account, e.g., \cite{2009PhRvD..79l3515G, 2010MNRAS.408.2319S}.

For some alternative analyses of performance of these two experiments see, e.g., \cite{2009AIPC.1141..222D, 2009A&A...503..691B, 2011MNRAS.tmp..449B}.

\subsection{Mixing matrix}

To define the mixing matrix, Eq.~\eref{eqn:dataModelFull} relevant for the problem at hand, we will use the component frequency scaling laws as defined in Sec.~\ref{sect:foregrounds}.
We set the reference frequency, i.e., frequency at which all the component maps are recovered as equal to $150$ GHz. We also account for frequency
band-shapes. For this we will assume that they are top-hat-like with a width equal to $1/3$ of the central value. Therefore, an element, $\bd{A}_{ij}$ of the mixing matrix will be given as
\begin{eqnarray}
\bd{A}_{ij} \equiv \int \, d\nu\, \Phi_j\l( \nu, \nu_{ref}\r) W_{TH}\l(\l|\nu-\nu_i\r|, \frac{1}{3}\nu_i\r),
\end{eqnarray}
where $\nu_i$ is a frequency of the $i$-th channel, $\Phi_j\l( \nu, \nu_{ref}\r)$ is a photon flux as measured
at frequency $\nu$ relatively to $\nu_{ref}$, and $W_{TH}(\cdot,\, \sigma_{TH})$ is a top hat window centered at 0 and
with a width $\sigma_{TH}$.  As mentioned earlier we assume hereafter that the scaling laws adopted on this stage coincide with
the true ones modulo the unknown parameters.
Nonetheless we will limit the frequency range of the
channels included in our discussion below to between $30$ and $400$ GHz, to, on the one hand, avoid channels where the
CMB is completely swamped by the foregrounds and, on the other,  not to stretch the adequacy of the frequency scaling model 
of the dust over a too broad interval.

\subsection{Noise levels}
\label{noise_levels}

We assume sky-noise limited  detectors. Their noise level, in antenna units, is taken to be independent on a detectors operating frequency and 
set to be equal to $\sigma_t\,\sim\, 30\, \mu{\rm K}\sqrt{s}$~ \cite{2011arXiv1102.2181T}. A single detector noise level per pixel will 
then be given by an observation total length, $T_{obs}$ and pixel area. The detector 
noise per channel will also depend on a number of detectors operating at a given frequency. The numbers of detectors for each channel, $\l\{d_i\r\}_{(i=0,\dots,n_{f}-1)}$, are the parameters we will
be most frequently trying to optimize in the reminder of this paper. The noise correlation matrix will be then assumed to be diagonal and the diagonal elements will be given by
\begin{eqnarray}
\bd{N}_{ii} = \frac{4\,\sigma_t^2\,N_{pix}^{tot}}{T_{obs}\,d_i}.
\label{eqn:noiseMat}
\end{eqnarray}
Here, $N_{pix}^{tot}$ is a total number of observed pixels (to be distinguished from $N_{pix}$ a number of pixels included in the analysis ($N_{pix}$ will depend on the mask we will consider; see Sec.~\ref{sect:foregrounds}).

\subsection{Resolution}

So far we have ignored completely the fact that detectors operating at different frequencies will likely have a different resolution, in particular if they are diffraction-limited. Because the parametric maximum likelihood
component separation approach adopted here is pixel-based all the channel maps will have to be however smoothed to some common resolution before the separation can be accomplished. 
The extra smoothing required here is not generally lossless and may introduce
noise correlation between the pixels. Hereafter we will ignore such effects and keep using Eq.~\eref{eqn:noiseMat} to compute the noise levels with only the pixel size, and thus a number of pixels, adjusted accordingly.
As far as the sky signals are concerned, given that our science
goals are mostly constrained by the large angular scales, we will mimic the common resolution by setting a hard limit on the  considered value of $\ell$ to be $\ell_{max} = 500$, as we have found that 
for the considered noise levels there is no information beyond that range. We note that in a more refined approach one may want to introduce the resolution as an optimization parameter and constraint it by
requiring that the gain due to its decrease is larger than some threshold.  All the power spectra used in this work have been derived using \textsc{healpix} pixelized maps with the \textsc{healpix} resolution level,  ${\rm nside} = 512$. This is clearly sufficient given the hard $\ell$-space cut off we have adopted here. We stress that this
resolution is higher than the one used in Sec.~\ref{sect:foregrounds} for the determination of the matrix $\bd{\hat{F}}$. This is because in the latter calculation only  pixel-domain quantities are involved, which are overwhelmingly dominated by the large scale fluctuations for which ${\rm nside} = 128$ maps are entirely sufficient.

\subsection{Fixed number of channels with pre-defined, fixed frequencies}
\label{subsec:fixed_num_det_and_fixed_ch}

\begin{figure*}
	\begin{center}
		 \includegraphics[width=7.5cm]{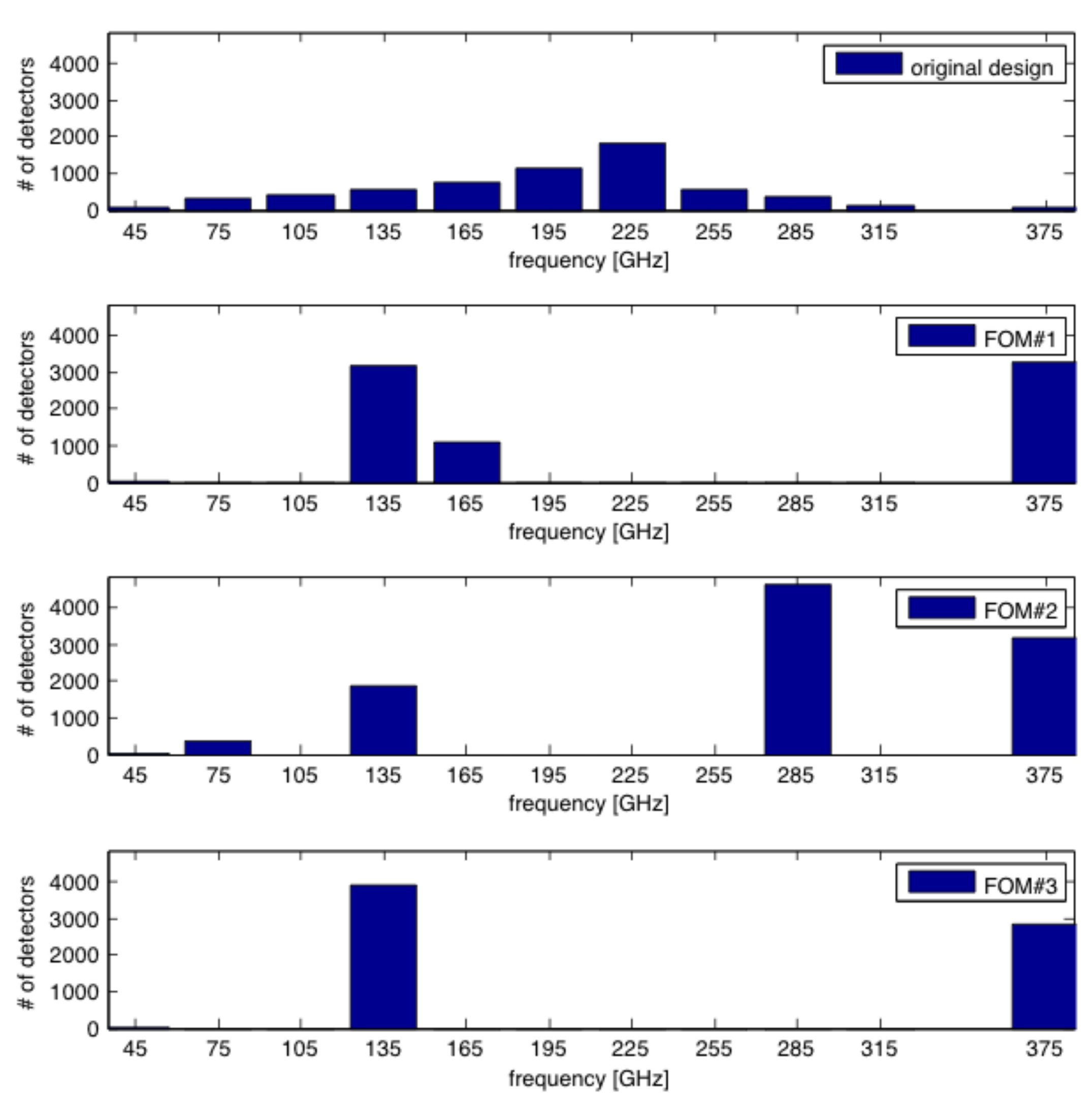}
		 \includegraphics[width=9cm]{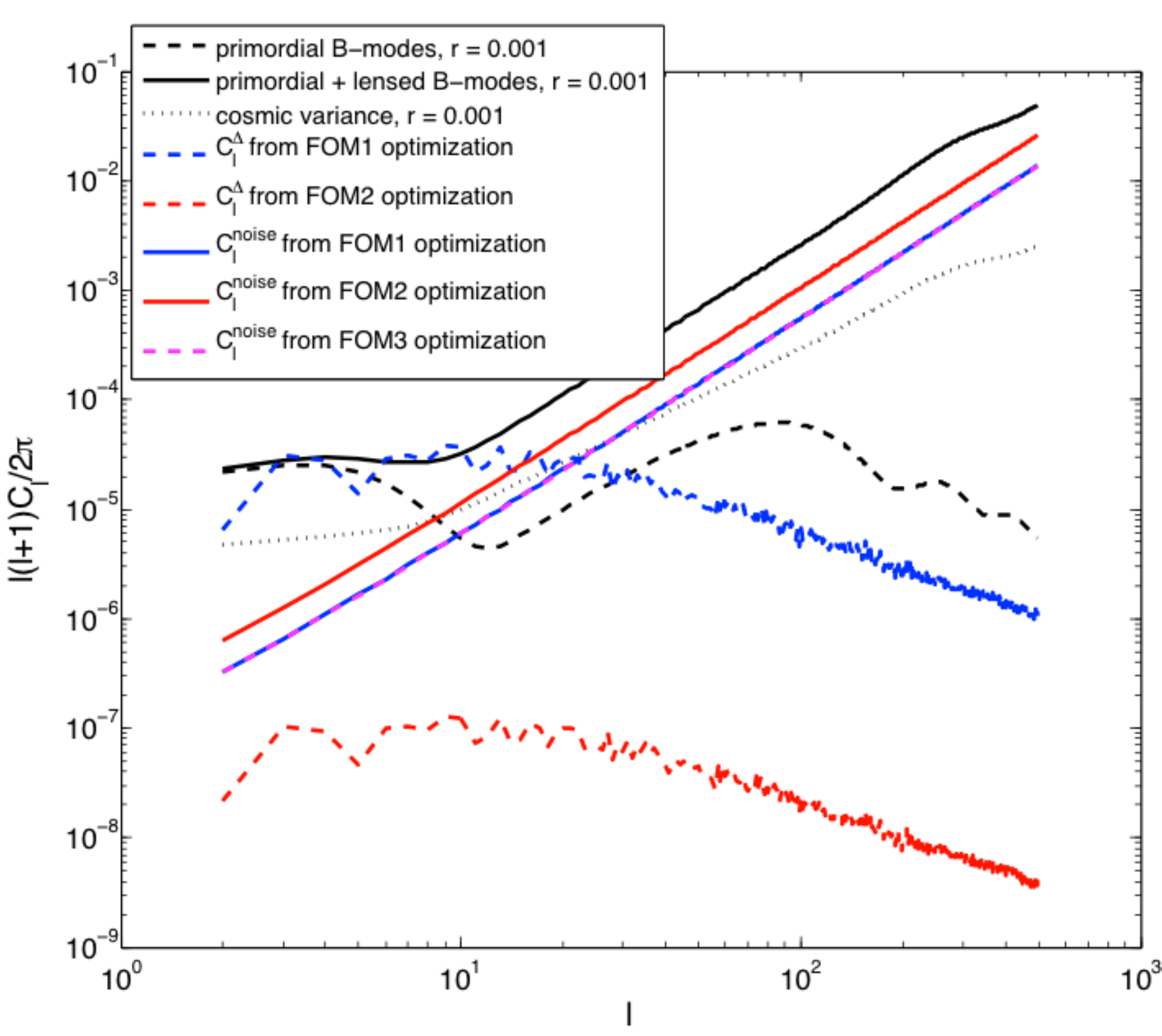}
	 \caption{\textit{Left}: Optimized distributions of  numbers of detectors per channel derived under the total focal plane area constraint for the COrE satellite, including only channels below $400$ GHz. From  top to bottom we show first the original distribution followed by the three optimized ones derived using FOM\#1 to \#3, respectively. \textit{Right}: Corresponding power spectra of the residuals  and the noise computed for the optimized configurations shown on the left and compared against the spectrum of the CMB $B$-modes with $r = 0.001$.} 
	 \label{fig:opt_results_core_area_constr}
\end{center}
\end{figure*}

\begin{figure*}
	\begin{center}
 		 \includegraphics[width=7.5cm]{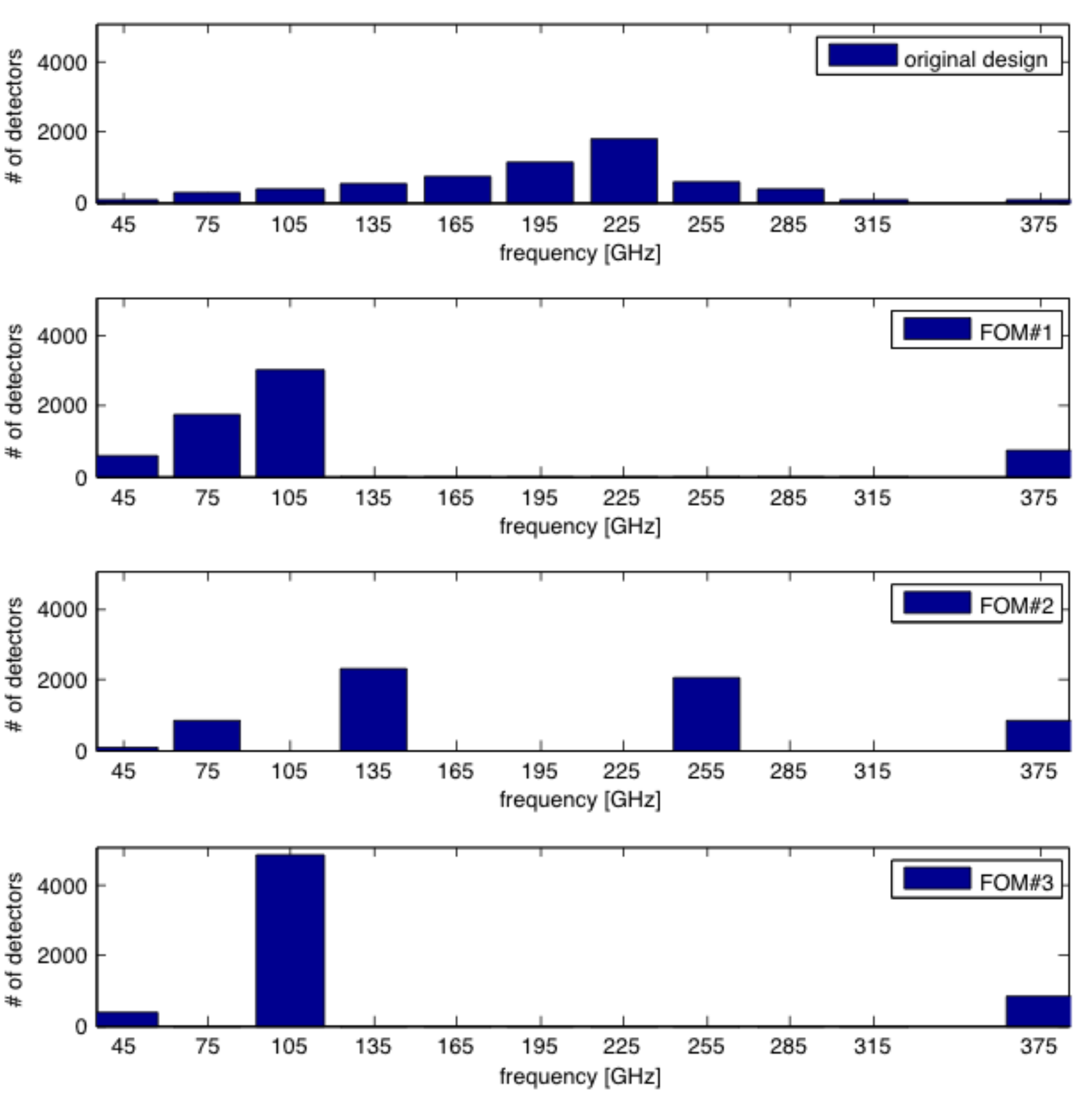}
 		 \includegraphics[width=9cm]{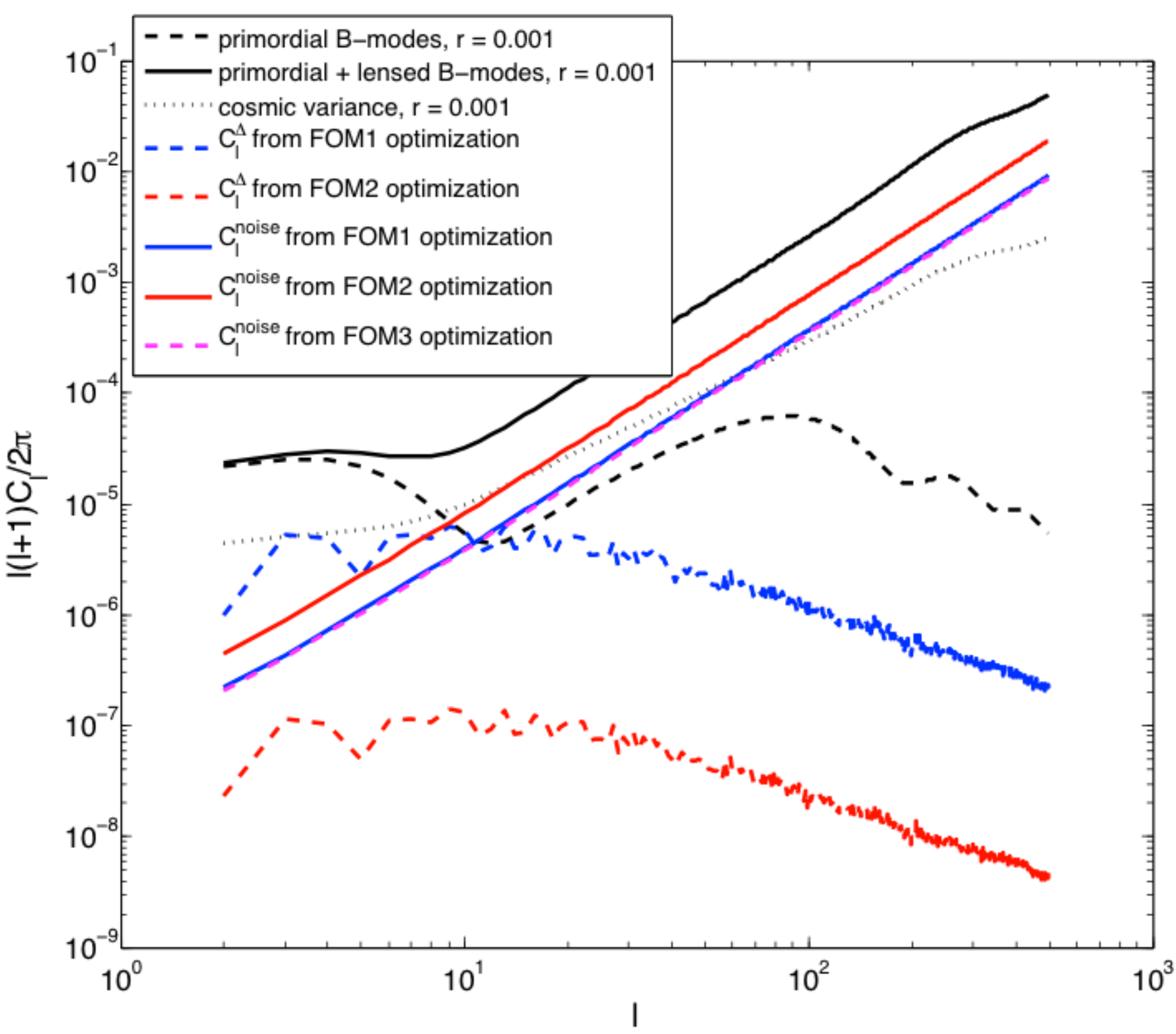}
	 \caption{As in Fig.~\ref{fig:opt_results_core_area_constr} but imposing the constraint on the total number of detectors.} 
	 \label{fig:opt_results_core_num_det_constr}
\end{center}
\end{figure*}

\begin{figure*}
	\begin{center}
		 \includegraphics[width=7.5cm]{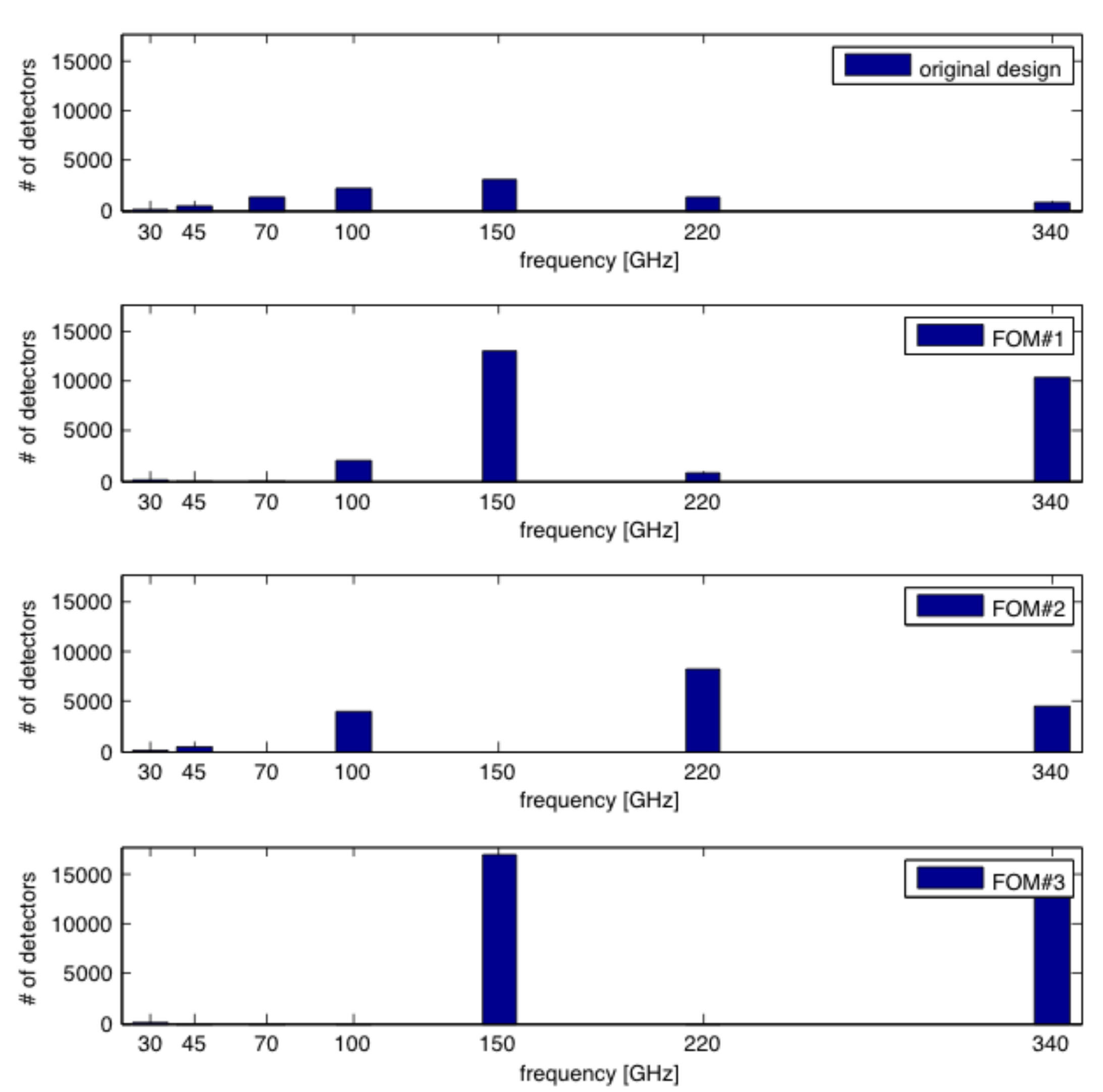}
		 \includegraphics[width=8.5cm]{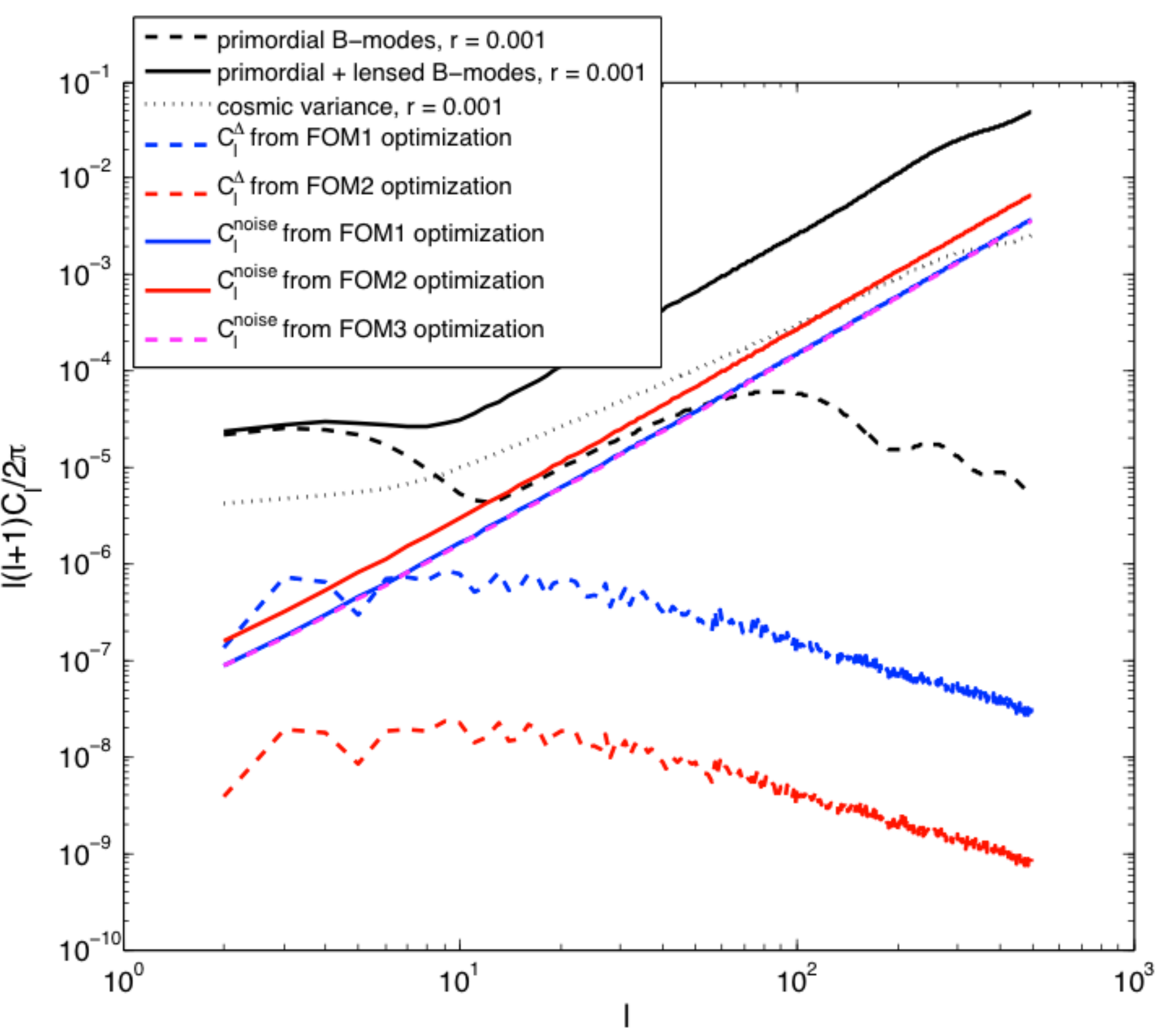}
	 \caption{As in Fig.~\ref{fig:opt_results_core_area_constr} but for the CMBpol satellite.} 
	 \label{fig:opt_results_cmbpol_area_constr}
\end{center}
\end{figure*}

\begin{figure*}
	\begin{center}
 		 \includegraphics[width=7.5cm]{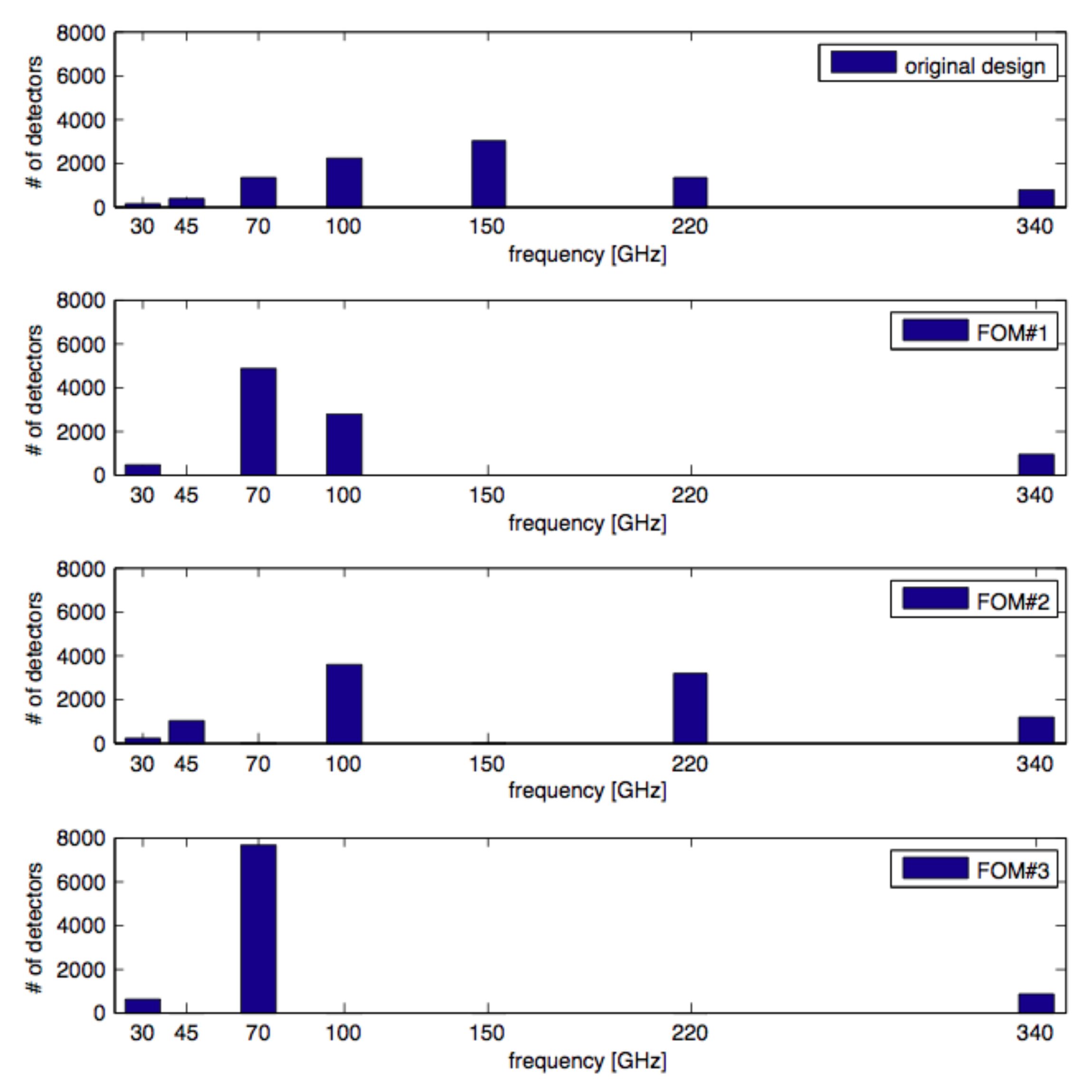}
 		 \includegraphics[width=8.5cm]{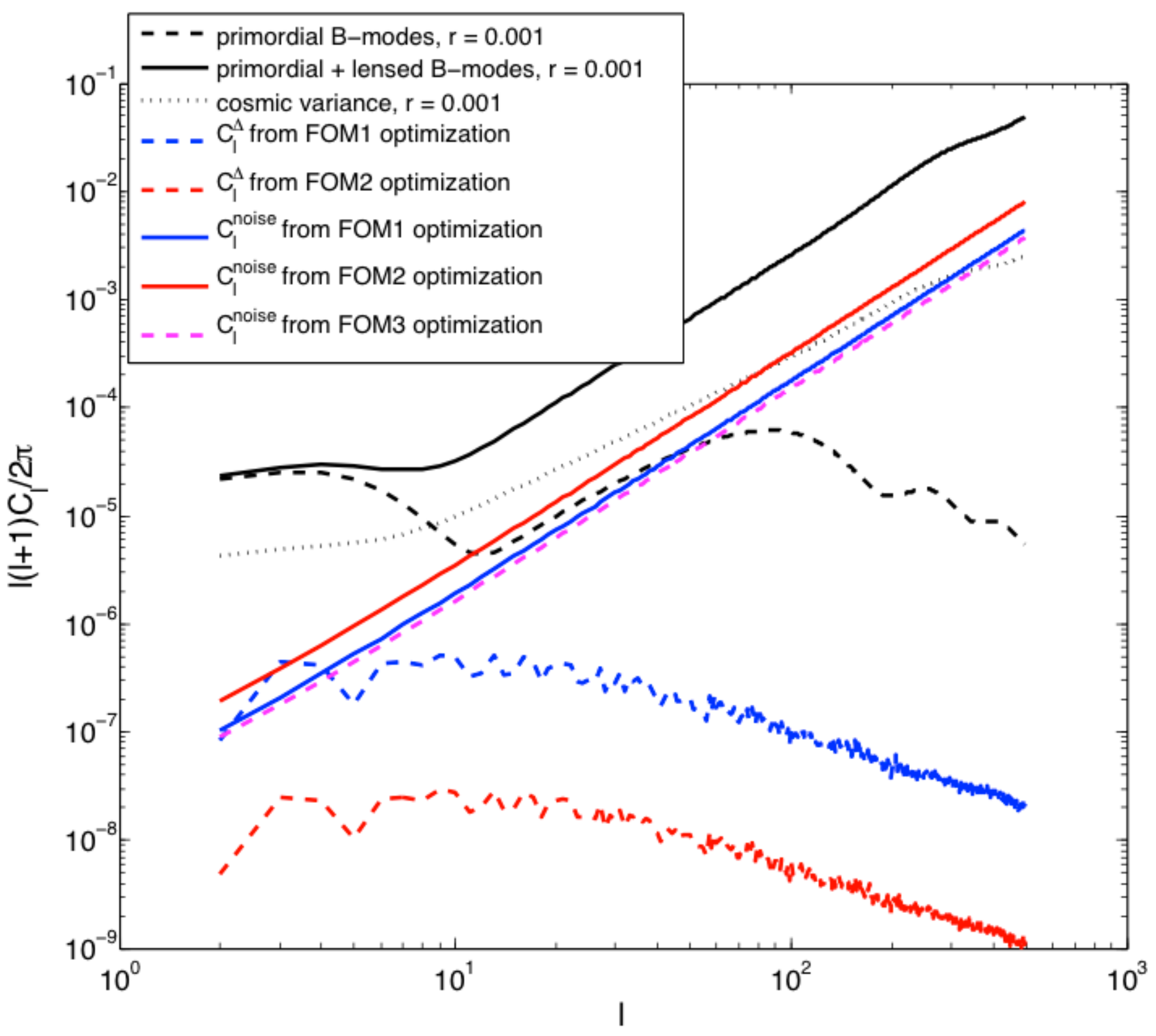}
	 \caption{As in Fig.~\ref{fig:opt_results_core_num_det_constr} but for the CMBpol satellite.} 
	 \label{fig:opt_results_cmbpol_num_det_constr}
\end{center}
\end{figure*}

In this Section, we describe the optimization of the two experiments, assuming that the frequency channels are fixed ahead of the procedure.
The results are summarized in
Tables~\ref{table:optimization_numbers_core} and~\ref{table:optimization_numbers_cmbpol} for COrE and CMBpol, respectively, and for each FOM (called there 
for shortness as F1, F2 or F3), three considered sky masks (P06, Mask I or Mask II), and two hardware constraints (total area or total number of detectors), 
and are contrasted with results obtained for the original designs of the experiments, as shown in the rightmost columns of the Tables. We note that though the latter 
configurations are mask-independent, the corresponding FOMs values differ somewhat  from mask to mask due to differences of the sky included in the analysis. 
For each of the optimized configurations the tables show a corresponding total number of detectors, focal plane area, effective noise levels, spectral index determination precision, and 
values of the three FOMs.  
A selection of these results is also depicted in  Figs.~\ref{fig:opt_results_core_area_constr}-\ref{fig:opt_results_cmbpol_num_det_constr}, showing, as bars, a number of detectors for each of the considered channels, left panels, and power spectra of the residuals corresponding to each configuration, right panels. The visualized cases are those based on the P06 mask, however  the other cases would look similar. In each Figure the upper left panel shows a corresponding original configuration followed by three panels displaying
configurations optimized with respect to each of the three FOMs. Four general observations are in order here.
\begin{enumerate}
\item The optimized configurations depend on the FOM used for the optimization.
\item The constraints imposed on the problem affect the results. Constraining the focal plane area gives preference to the high frequency channels with detectors occupying a small area and
thus leads to a worse determination of the synchrotron signal, which in turn leads to a higher level of residuals, if these are left unconstrained, i.e. in cases of FOM\#1 and FOM\#3. Also the overall noise, FOM\#3, tends to be higher. 
\item The final configurations obtained for each of the three masks are essentially identical, though the actual values of FOMs do differ mostly due to a different number of pixels with Mask II containing the fewest of those.
\item The optimized configuration contain significantly fewer frequency channels than allowed for in the optimization and therefore fewer than proposed in the original versions of the both these experiments.
\end{enumerate}
Below we comment on some of the result in more detail and leaving a general discussion for the conclusions, Sec.~\ref{sect:conclusions}.

\begin{table*}
{\tiny
\begin{tabular}{|c|c||ccc|ccc||cc|cc||cc|cc||c|c|c|}
\hline 
&channels& \multicolumn{6}{c||}{P06 mask} & \multicolumn{4}{c||}{mask {\rm I}} & \multicolumn{4}{|c||}{mask {\rm II}}  & \multicolumn{3}{c|}{proposed version}\\
Constraint& & \multicolumn{3}{c}{area} & \multicolumn{3}{|c||}{total \#}& \multicolumn{2}{c|}{area} &  \multicolumn{2}{c||}{total \#} & \multicolumn{2}{|c|}{area} &  \multicolumn{2}{c||}{total \#} & \multicolumn{3}{c|}{}\\
& (GHz)  & F1 & F2 & F3 & F1 & F2 & F3 & F1 & F2 & F1 & F2 &  F1 & F2 &  F1 & F2 & P06 mask & mask {\rm I} & mask {\rm II}\\
\hline
&45& 45&22&48&610&87&382&45&21&610&72&45&22&607&88&64&-&-\\
&75& -&370&-&1775&827&-&-&366&1775&778&-&37&1759&832&300&-&-\\
&105&-&-&-&3027&-&4876&-&-&3026&-&-&-&3042&-&400&-&-\\
&135&3160&1872&3918&-&2313&-&3161&1886&-&2322&3124&1871&-&2315&550&-&-\\
&165&1092&-&-&-&-&-&1091&-&-&-&1146&0&-&-&750&-&-\\
Number of &195&-&-&-&-&-&-&-&-&-&-&-&-&-&-&1150&-&-\\
detectors &225&-&-&-&-&-&-&-&-&-&-&-&-&-&-&1800&-&-\\
&255&-&-&-&-&2081&-&-&-&-&2141&-&-&-&2073&575&-&-\\
&285&-&4623&-&-&-&-&-&4669&-&-&-&4610&-&-&375&-&-\\
&315&-&-&-&-&-&-&-&-&-&-&-&-&-&-&100&-&-\\
&375&3281&3186&2859&717&820&870&3281&3156&717&816&3294&3188&719&820&64&-&-\\
\hline
Total area&&0.023&0.023&0.023&0.081&0.032&0.057&0.023&0.023&0.081&0.031&0.023&0.023&0.080&0.032&0.023&-&-\\
$\sum$ [number of detectors]$_{{}_{\,}}$&&7579&10073&6824&6128&6128&6128&7577&10099&6128&6128&7608&10062&6128&6128&6128&-&-\\
\hline
&45&0.085&0.042&0.091&0.34&0.12&0.30&0.085&0.040&0.34&0.10&0.085&0.042&0.36&0.12&0.12&-&-\\
&75&-&0.25&-&0.35&0.42&-&-&0.25&0.35&0.41&-&0.25&0.35&0.42&0.21&-&-\\
&105&-&-&-&0.31&-&0.69&-&-&0.31&-&-&-&0.31&-&0.14&-&-\\
&135&0.67&0.40&0.83&-&0.36&-&0.67&0.40&-&0.37&0.66&0.40&-&0.36&0.12&-&-\\
Fractional &165&0.15&-&-&-&-&-&0.15&-&-&-&0.16&-&-&-&0.11&-&-\\
area &195&-&-&-&-&-&-&-&-&-&-&-&-&-&-&0.12&-&-\\
&225&-&-&-&-&-&-&-&-&-&-&-&-&-&-&0.14&-&-\\
&255&-&-&-&-&0.090&-&-&-&-&0.097&-&-&-&0.090&0.034&-&-\\
&285&-&0.22&-&-&-&-&-&0.22&-&-&-&0.22&-&-&0.018&-&-\\
&315&-&-&-&-&-&-&-&-&-&-&-&-&-&-&0.0039&-&-\\
&375&0.090&0.088&0.079&0.006&0.016&0.010&0.090&0.087&0.0057&0.017&0.090&0.088&0.0057&0.016&0.0018&-&-\\
\hline
&45&0.37&0.52&0.35&0.099&0.27&0.13&0.37&0.53&0.099&0.30&0.37&0.52&0.099&0.26&0.31&-&-\\
&75&-&0.13&-&0.058&0.085&-&-&0.13&0.058&0.088&-&0.13&0.058&0.085&0.14&-&-\\
&105&-&-&-&0.044&-&0.035&-&-&0.044&-&-&-&0.044&-&0.12&-&-\\
&135&0.044&0.057&0.039&-&0.051&-&0.044&0.056&-&0.051&0.044&0.036&-&0.051&0.10&-&-\\
Noise &165&0.074&-&-&-&-&-&0.074&-&-&-&0.072&-&-&-&0.089&-&-\\
per &195&-&-&-&-&-&-&-&-&-&-&-&-&-&-&0.072&-&-\\
channel&225&-&-&-&-&-&-&-&-&-&-&-&-&-&-&0.058&-&-\\
$[\mu K_{antenna}]$&255&-&-&-&-&0.054&-&-&-&-&0.053&-&-&-&-&0.10&-&-\\
&285&-&0.036&-&-&-&-&-&0.036&-&-&-&0.036&-&-&0.13&-&-\\
&315&-&-&-&-&-&-&-&-&-&-&-&-&-&-&0.24&-&-\\
&375&0.043&0.043&0.046&0.091&0.085&0.083&0.043&0.044&0.091&0.086&0.043&0.043&0.091&0.085&0.31&-&-\\
\hline
$\delta \beta_{d}\,[10^{-3}]$&&0.96&0.12&-&0.95&0.16&-&0.83&0.074&0.82&0.10&1.47&0.19&1.48&0.25&0.28&0.18&0.45\\
$\delta \beta_{s} \,[10^{-3}]_{{}_{\,}}$&&30&2.9&-&4.3&2.2&-&26&1.9&3.7&1.4&38&3.9&5.6&2.9&3.4&2.2&4.5\\
$\frac{\delta \beta_{d}\delta \beta_{s}}{\delta \beta_{d}\times \delta \beta_{s}}_{{}_{\,}}$&&-0.92&-0.44&-&-0.92&-0.57&-&-0.96&-0.46&-0.96&-0.58&-0.91&-0.44&-0.91&-0.57&-0.67&-0.70&-0.67\\
\hline
F1 $\l[10^{-3}\r]$&&0.22&0.26&-&0.21&0.24&-&0.20&0.23&0.19&0.21&0.31&0.37&0.29&0.34&0.28&0.25&0.40\\
F2 $\l[10^{-3}\r]$&&0.95&0.0097&-&0.16&0.011&-&1.1&0.0057&0.18&0.0065&0.79&0.086&0.14&0.0094&0.028&0.018&0.025\\
F3 $\l[{\rm nK}_{{\rm cmb}}\r]_{{}_{\,}}$&&5.4&10&5.3&3.6&7.4&3.4&5.4&10&3.6&7.7&5.4&10&3.6&7.4&14&14&14\\
\hline
\end{tabular}
}
\caption{Summary of the optimization results in the case of COrE considering channels only below 400 GHz. For each of the three masks, we present results
for each of the three FOMs optimized under one of the two constraints, either fixing the focal plane area or the total number of detectors. The results for FOM\#3 are quoted only
once as they do not depend on the choice of the mask.The rightmost columns show
the results computed using the original version of COrE as proposed in~\cite{2011arXiv1102.2181T}. In the latter case the configuration is always the same, whatever the choice of the mask. }
\label{table:optimization_numbers_core}
\end{table*}

\begin{table*}
{\tiny
\begin{tabular}{|c|c||ccc|ccc||cc|cc||cc|cc||c|c|c|}
\hline 
&channels & \multicolumn{6}{|c||}{P06 mask} & \multicolumn{4}{|c||}{mask {\rm I}} & \multicolumn{4}{|c||}{mask {\rm II}}  & \multicolumn{3}{|c|}{proposed version}\\
Constraint& & \multicolumn{3}{|c|}{area} & \multicolumn{3}{|c||}{tot \#}& \multicolumn{2}{c|}{area} &  \multicolumn{2}{c||}{tot \#} & \multicolumn{2}{|c|}{area} &  \multicolumn{2}{c||}{tot \#} & \multicolumn{3}{c|}{}\\
& (GHz)  & F1 & F2 & F3 & F1 & F2 & F3 & F1 & F2 & F1 & F2 &  F1 & F2 &  F1 & F2 & P06 mask & mask {\rm I} & mask {\rm II}\\
\hline
&30&35&62&52&472&185&601&33&61&448&168&56&62&672&187&84&-&-\\
&45&-&491&-&-&1016&-&-&493&-&975&10&491&1240&1021&364&-&-\\
Number of &70&-&-&-&4861&-&7646&-&-&4935&-&-&-&3643&-&1332&-&-\\
detectors &100&1970&4056&-&2776&3546&-&1400&4101&2579&3567&6311&4049&2583&3544&2196&-&-\\
&150&13159&-&16995&-&-&-&14639&-&-&-&3518&-&-&-&3048&-&-\\
&220&823&8328&-&-&3164&-&-&8228&-&3207&178&8340&-&3157&1296&-&-\\
&340&10364&4525&13210&954&1154&817&11586&4259&1102&1148&7988&4566&926&1154&744&-&-\\
\hline
Total area&&0.084&0.084&0.084&0.16&0.10&0.20&0.084&0.084&0.16&0.099&0.084&0.084&0.21&0.10&0.084&-&-\\
$\sum$ [number of detectors]$_{{}_{\,}}$&&26352&17462&30258&9064&9064&9064&27658&17143&9064&9064&18061&17508&9064&9064&9064&-&-\\
\hline
&30&0.042&0.074&0.063&0.29&0.18&0.30&0.040&0.073&0.28&0.17&0.068&0.074&0.32&0.18&0.10&-&-\\
&45&-&0.26&-&-&0.44&-&-&0.26&-&0.44&0.0054&0.26&0.26&0.44&0.19&-&-\\
Fractional&70&-&-&-&0.55&-&0.70&-&-&0.57&-&-&-&0.31&-&0.29&-&-\\
area&100&0.21&0.44&-&0.15&0.31&-&0.15&0.44&0.15&0.32&0.68&0.44&0.11&0.31&0.24&-&-\\
&150&0.63&-&0.81&-&-&-&0.70&-&-&-&0.17&-&-&-&0.15&-&-\\
&220&0.018&0.19&-&-&0.057&-&-&0.18&-&0.060&0.0040&0.19&-&0.057&0.029&-&-\\
&340&0.097&0.042&0.12&0.0046&0.0088&0.0032&0.11&0.040&0.0054&0.0090&0.074&0.043&0.0034&0.0087&0.0069&-&-\\
\hline
&30&0.41&0.31&0.34&0.11&0.18&0.010&0.42&0.31&0.12&0.19&0.33&0.31&0.094&0.18&0.27&-&-\\
&45&-&0.11&-&-&0.077&-&-&0.11&-&0.078&0.77&0.11&0.070&0.077&0.13&-&-\\
Noise &70&-&-&-&0.035&-&0.028&-&-&0.035&-&-&-&0.041&-&0.067&-&-\\
per &100&0.055&0.038&-&0.046&0.041&-&0.065&0.038&0.048&0.040&0.031&0.038&0.048&0.041&0.052&-&-\\
channel&150&0.021&-&0.019&-&-&-&0.020&-&-&-&0.041&-&-&-&0.044&-&-\\
$[\mu K_{antenna}]$&220&0.085&0.027&-&-&0.044&-&-&0.027&-&0.043&0.18&0.027&-&0.044&0.068&-&-\\
&340&0.024&0.036&0.021&0.079&0.072&0.086&0.023&0.038&0.074&0.072&0.027&0.036&0.080&0.072&0.090&-&-\\
\hline
$\delta \beta_{d}\,[10^{-3}]$&&0.25&0.086&-&0.71&0.13&-&0.37&0.055&0.62&0.055&0.41&0.14&0.66&0.21&0.16&0.10&0.25\\
$\delta \beta_{s}\,[10^{-3}]_{{}_{\,}}$&&2.39&0.51&-&1.5&0.38&-&3.2&0.33&1.4&0.33&2.7&0.68&0.67&0.50&0.55&0.36&0.73\\
$\frac{\delta \beta_{d}\delta \beta_{s}}{\delta \beta_{d}\times \delta \beta_{s}}_{{}_{\,}}$&&-0.66&-0.46&-&-0.96&-0.48&-&-0.10&-0.48&-0.88&-0.49&-0.88&-0.46&-0.54&-0.48&-0.63&-0.65&-0.62\\
\hline
F1 $[10^{-3}]$&&0.19&0.20&-&0.19&0.20&-&0.17&0.18&0.17&0.18&0.27&0.28&0.27&0.29&0.20&0.18&0.29\\
F2 $[10^{-3}]$&&0.024&0.0018&-&0.059&0.0023&-&0.076&0.0011&0.069&0.0014&0.020&0.0016&0.012&0.0020&0.0041&0.0026&0.0036\\
F3 $\l[{\rm nK}_{{\rm cmb}}\r]_{{}_{\,}}$&&1.5&2.7&1.4&1.6&3.1&1.5&1.4&2.7&1.6&3.2&1.6&2.7&1.7&3.1&3.0&3.0&3.0\\
\hline
\end{tabular}
}
\caption{As in Table~\ref{table:optimization_numbers_core} but for CMBpol~\cite{2009arXiv0903.0902A}.}
\label{table:optimization_numbers_cmbpol}
\end{table*}

\subsubsection{FOM\#1 optimization - $r_{min}$}

For all  configurations shown in Tables~\ref{table:optimization_numbers_core} and~\ref{table:optimization_numbers_cmbpol} for which FOM\#1 could be computed, i.e., 
those containing more than just 3 channels,  $r_{min}$ is found to be on the order of  $10^{-4}$ and varying from case to case by no more than a factor of $2$.
This is also the case for the original designs of the COrE and CMBpol satellites.
The values of FOM\#1  optimized under the constraint of the total number of detectors tend to be somewhat better (worse) than those derived under the total focal 
plane area constraint for COrE (CMBpol). The differences are however small across the board and probably irrelevant in practice.

In both the COrE and CMBpol cases, the optimization of FOM\#1 leads to configurations for which also FOM\#3 is close to the optimum, as the latter is found to be within $5$-$10$\% of its best value for the respective hardware constraints.  This suggests that this is the variance due to the noise rather than the foreground residual, which contributes to the recovered value of the FOM\#1 more significantly (see also~\cite{2010MNRAS.408.2319S}). Conversely, as a consequence in such cases the level of  the foreground residuals is not tightly controlled and therefore the FOM\#1-optimized configurations result in  values of FOM\#2, which are at least 1 order of magnitude above the best achievable $r_{eff}$, and worse than 
the values derived for the proposed designs.
As  we normally would prefer to avoid too high residuals we conclude that FOM\#1 is not sufficient as a stand-alone optimization criterion and preferably should be combined with some other indicator,
efficient in enforcing the low value of the residuals. We will get back to this issue later on in this Section.

\subsubsection{FOM\#2 optimization - $r_{eff}$}

From Eqs.~\eref{eqn:deltaCMBlin}--\eref{eqn:resSpec0} it follows that a good determination of the spectral parameters $\beta_{dust}$ and $\beta_{sync}$ is necessary and sufficient to ensure a low level of the foreground residuals. 
We therefore expect (see also  \cite{2007PhRvD..75h3508A}) that in the FOM\#2-optimized configuration the detectors should populate predominantly low frequency bands, which are dominated by the synchrotron signal, the CMB band, and high frequency bands, dominated by the dust. As we require at least 4 channels in the case at hand to avoid problem singularity and impose
the hardware constraint the actual answer is somewhat more complex, nevertheless the overall detector distribution conforms with the above intuition.
Indeed the FOM\#2-optimized configurations include channels below $50$ GHz,  around $100-130$ GHz,  and above $250$ GHz. This applies for both the experiments and for every mask. 
The details of the distribution depend on a type of the constraint. As the high frequency detectors have smaller area we find that the dust is better estimated ($\delta \beta_{dust}$ lower) under the total area constraint case
as more high frequency detectors can be had. The opposite can be seen for the synchrotron estimation. The resulting levels of the residuals are however essentially identical in both these cases. More
aggressive masking clearly helps, Mask I, but a balance has to be maintained between lowering the overall foreground level and the precision of the spectral index determination. The
latter, unlike the former, benefits from a larger number of pixels and higher foregrounds and, otherwise, can therefore start driving the effective residual up,  e.g., Mask II.

The FOM\#2-optimized configurations usually render good values for FOM\#1 (within $10-15$\% of the best achievable values), but result in the CMB map noise levels (FOM\#3) up to twice higher than 
the best ones.  The original versions of the considered experiments also yield the values of $r_{eff}$ close to the best ones.

\subsubsection{FOM\#3 optimization}

For this FOM, and in every considered case, the optimization of the focal plane with respect to the noise in the CMB map ends up with only three nonzero channels: two at  frequencies as extreme as only allowed for,  and one at an intermediate one contained in the CMB frequency band. The precise position of the latter is found again to be dependent on a type of the hardware constraint used. For the CMBpol satellite the values of the central frequencies are $70$ or $150$ GHz for the 
constraint on the total number of detectors and the area, respectively. For COrE they are $105$ and $135$ GHz, respectively. We recall that in the case of this FOM all the spectral 
indices are assumed to be known, otherwise the three channel configurations derived here would be singular and would not permit a determination of the spectral indices. The 
achieved noise levels are better when the total number of detectors is constrained, and are lower by a factor up to $\sim1.6.$
The original versions of the satellites result in quite high noise (higher by a factor of $2.5-4$) in comparison with the one derived for the optimized configurations.

\subsubsection{Consensus configuration}
Having postulated three different FOMs we have obtained three different, optimized configurations.  Moreover, as we have already mentioned, there is clearly tension between 
some of the considered FOMs. The issue now is therefore how to find a compromise between them in order to select a single configuration as a result of our procedure.
To do so we first recall that in our case the configurations preferred from the point of view of FOM\#1 fail to ensure a satisfactory level
of the residuals, as quantified by FOM\#2, while optimization of the latter yields a rather high level of noise, i.e., FOM\#3. Simultaneously however optimizing FOM\#1  
effectively ensures a near optimization of FOM\#3. Therefore we will retain the former as part of the optimization and  drop the latter, which from now on will be used only as a benchmark to compare against the obtained 
configurations. As FOM\#1 on its own is not
fully satisfactory we will therefore optimize it, while imposing a constraint based on a value of FOM\#2. 
Clearly if more FOMs are
used more constraints can be introduced in the same way. What values to choose for the thresholds is a somewhat  debatable question, an answer to which will depend on a specific application. 
In our case, we first note that
for the FOM\#2-optimized configuration the resulting $r_{eff}$ is an order of magnitude lower than the respective value of $r_{min}$. The latter is moreover typically $20$\% higher than its
corresponding best value. 

From the viewpoint of these two indicators the FOM\#2-optimized solution looks therefore quite satisfactory. This is particularly true for the CMBpol case for which this
solution can be accepted as indeed the final outcome of the procedure. 
For COrE the potential remaining problem could be the noise level. In search of the consensus configuration we may therefore want to let the residual grow, in particular, relatively to the value of $r_{min}$ and gain in terms of the noise. Clearly the more we compromise on $r_{eff}$ the more we can gain on $\sigma^2_{CMB}$. As for COrE the values of $r_{min}$ are close to  $2\times10^{-4}$ and we will allow $r_{min}$ to be as large as $10^{-4}$, and reoptimize the problem with respect to FOM\#1 with the constraint that $r_{eff} \le 10^{-4}$. This specific choice is in fact arguably rather  high.
In fact we find that imposing more strict limits of $r_{eff} \le 2.5\,\times\,10^{-5}$ or $5\,\times\,10^{-5}$ already can ensure satisfactory noise levels, $4.0$ and $3.9\,$~nK$_{CMB}$, respectively,
and thus could be preferred for the actual experiment optimization.
We will however use hereafter the threshold of $10^{-4}$  as it is more useful for demonstration purposes.

\begin{figure*}
	\begin{center}
	 \includegraphics[width=7cm]{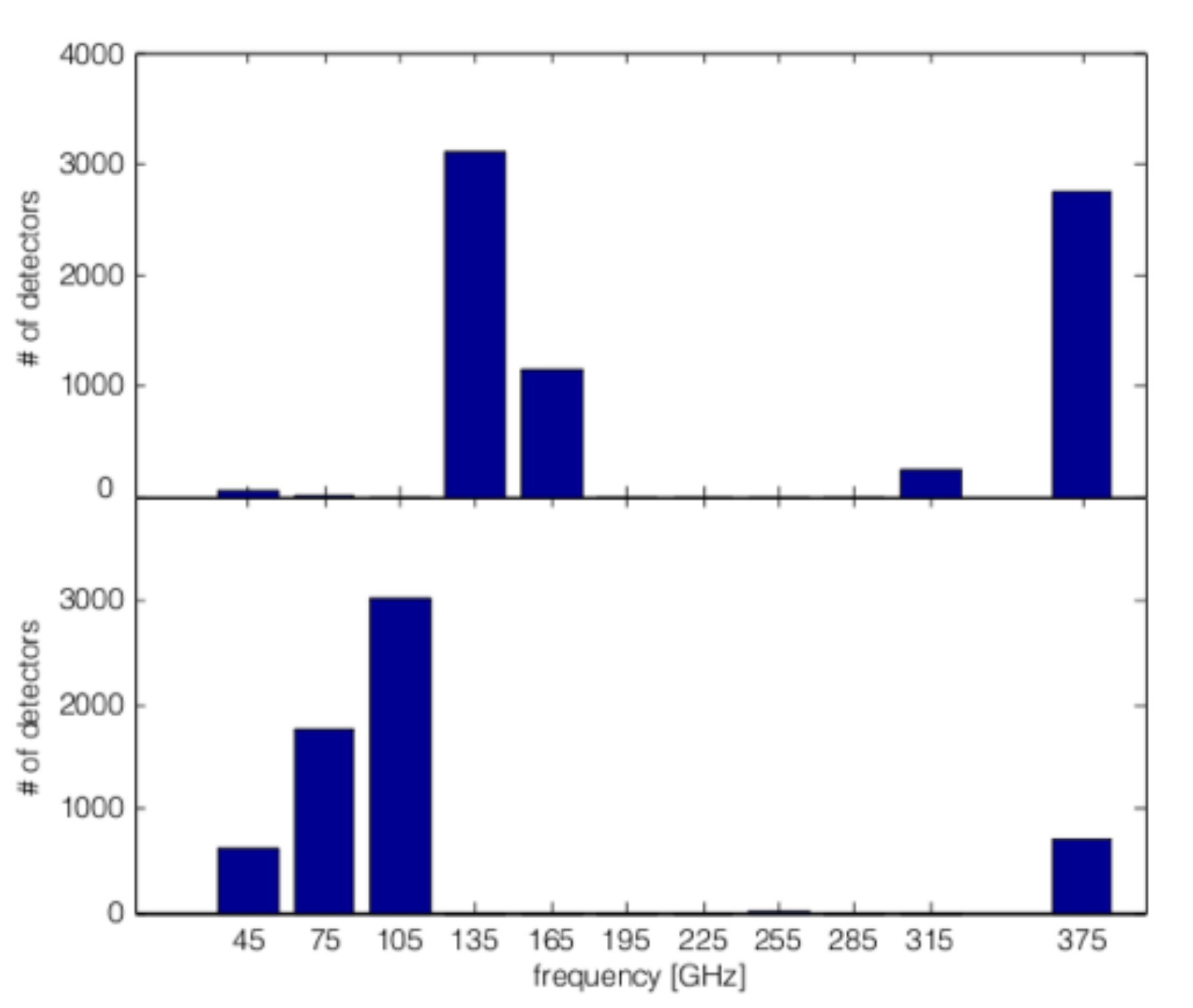}
	\includegraphics[width=6.5cm]{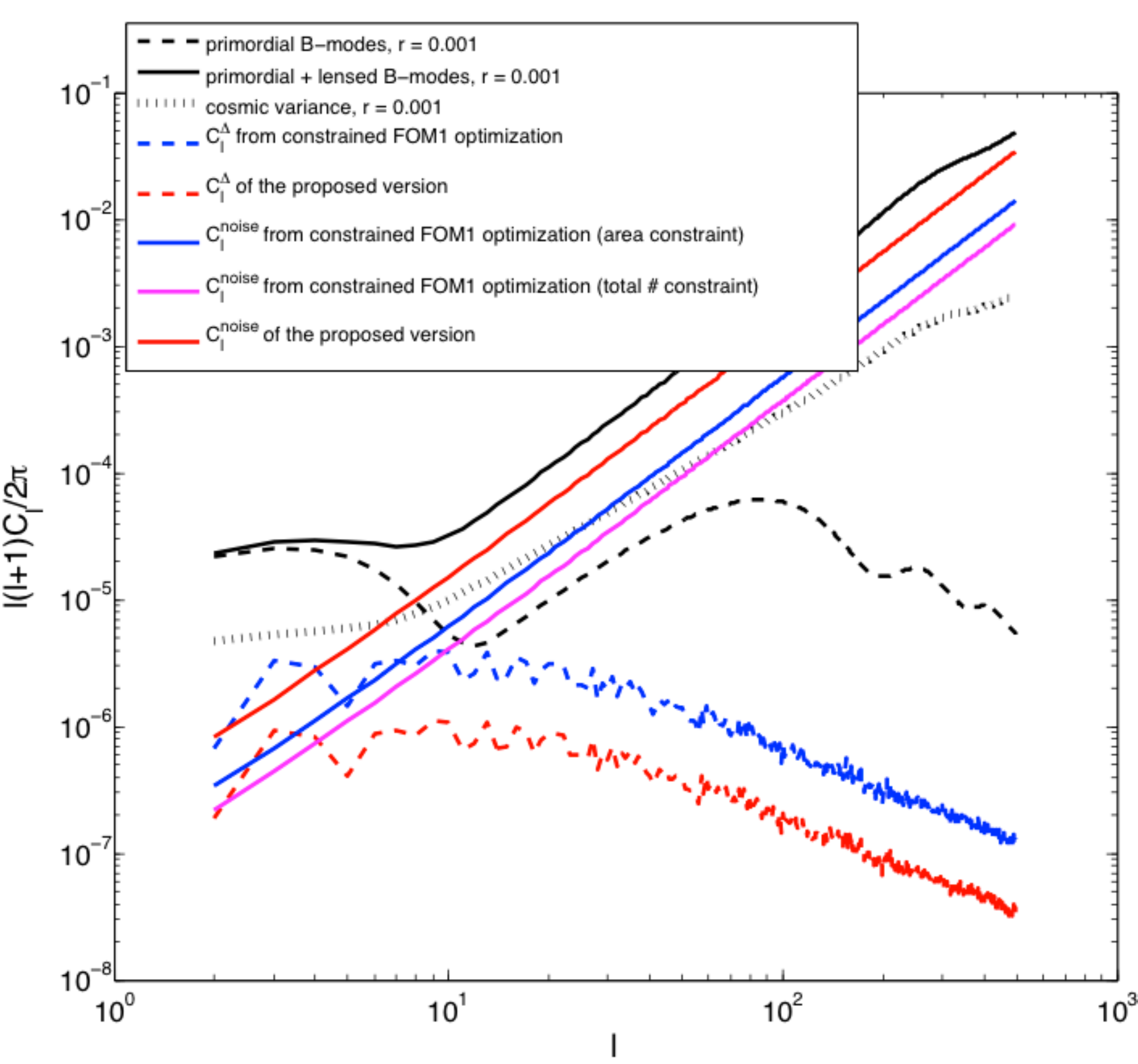}
	\caption{\textit{Left panel}: Results of the FOM\#1-based optimization derived in the case of the COrE experiment with a constraint on FOM\#2 ($<10^{-4}$), and using the P06 mask and channels with frequencies below $400$ GHz. \textit{Right panel}: Comparison of the power spectra corresponding to the proposed and optimized versions of the COrE experiment as listed in Table~\ref{table:FOM1_vs_FOM2_summary_core} and visualized in the left panel. The spectra in blue (mid-level noise spectrum and highest residuals, these latter being depicted with dashed lines) correspond to the cases with the total area constraint. On the other hand, the spectra in magenta (lowest noise level, same residuals as previously) correspond to the cases with the detector number constraint. The foreground residual spectra in both of these cases overlap perfectly in the figure with the magenta curve being invisible.} 
	 \label{fig:best_config_core_400_P06}
\end{center}
\end{figure*}

The resulting configuration is shown in Fig.~\ref{fig:best_config_core_400_P06} and summarized in Table~\ref{table:FOM1_vs_FOM2_summary_core}, where we show the results obtained for the two hardware constraints. The spectra of the noise and residuals are also displayed in the right panel of the Figure. We conclude that the detector distribution indeed resembles a hybrid between two solutions obtained
earlier  as a result of the optimization of FOMs: \#1 and \#2 separately with a respective hardware constraint, Figs.~\ref{fig:opt_results_core_area_constr} and~\ref{fig:opt_results_core_num_det_constr}. 
As anticipated above the overall level of the foreground residual spectrum is  rather high as compared to both
the $B$-mode spectrum and its respective variance due to the noise and the sky. However, as intended, the noise level has been successfully suppressed to the levels close to those computed for FOM\#3 optimized
configurations.\\

\begin{table*}
{\tiny
\begin{tabular}{|c|c||c||c|c||c||c|}
\hline 
& channels & F1-optimized  &   \multicolumn{2}{c||}{no $255$GHz channel cases} & extra channels & original \\
& (GHz)  & + constraint  & no optimization &  F1-optimized + & F1 optimized + &  version \\
&   & F2 $\leq\, 10^{-4}$  &  &   F2 $\leq\, 1.5 \times 10^{-4}$ &  F2 $\leq\, 10^{-4}$ & \cite{2011arXiv1102.2181T} \\
\hline
&45& 607 & 607 & 592 & 366 & 64 \\
&75&1771 & 1771 & 2112 & 47 & 300\\
&105& 3021 & 3021 & 2801 & 4551 &400 \\
&135& - & - & \bd{0} & - &550 \\
&165& - & - & \bd{0} & - &750 \\
Number of &195&- & -& \bd{0} & \bd{200} &1150\\
detectors &225& - & -& \bd{0} & - &1800 \\
&255& 17 & \bd{0} & \bd{0} & - &575\\
&285& - & - & \bd{0}& \bd{200} &375\\
&315& - & -&  \bd{0} & - &100\\
&375& 711 & 711& 623 & 764 &64\\
\hline
$\delta \beta_{d}\,[10^{-3}]$&& 0.74  & 0.95 & 0.91 & 0.35 & 0.28\\
$\delta \beta_{s}\,[10^{-3}]_{{}_{\,}}$&& 3.5  & 4.3 & 4.1 & 8.1& 3.4\\
$\frac{\delta \beta_{d}\delta \beta_{s}}{\delta \beta_{d}\times \delta \beta_{s}}_{{}_{\,}}$&& -0.88 & -0.92 & -0.92 & -0.66& -0.67\\
\hline
F1 $[10^{-3}]$&& 0.21 & 0.21 & 0.21 & 0.21 & 0.28\\
F2 $[10^{-3}]$&& 0.10  & 0.16 & 0.15 & 0.10&  0.028\\
F3 $\l[{\rm nK}_{{\rm cmb}}\r]_{{}_{\,}}$&& 3.6  &  3.6 & 3.6 & 3.6 &14\\
\hline
\end{tabular}
}
\caption{Comparison of performance of  the variants of the COrE setups considered in Sec.~\ref{subsect:postProcApp}.  All the optimization runs have been performed while keeping the total \# of detectors constant, used the P06 mask and only the channels below $400$GHz. The configurations in the Table include, from left to right,  (1) a result of the optimization procedure with respect to FOM\#1 with a constraint on FOM\#2 of $\le 10^{-4}$, (2) the same configuration but with the $255$GHz channel 
suppressed, (3) a configuration with the same frequency channels as in (2), but with numbers of detectors re-derived via an optimization with respect to FOM\#1 and a constraint FOM\#2 $\le 1.5\,\times\,10^{-4}$, and (4) a re-optimized configuration with the channels as before plus two extra ones with a fixed number of detectors ($=200$ each). The last column shows the original COrE configuration for comparison. Numbers in bold correspond to parameters forced to be at a given value.}
\label{table:FOM1_vs_FOM2_summary_core}
\end{table*}

\subsubsection{Post-processing}

 \label{subsect:postProcApp}

For definiteness in this Section we focus on a single, specific configuration, and choose for it  the optimized COrE setup obtained from the optimization of the FOM\#1 value, while
constraining the corresponding value of FOM\#2 to be no more than $10^{-4}$ and keeping the total number of detectors fixed, as discussed at the end of the previous Section. 
The details of this configuration are listed in the fourth column of Table~\ref{table:FOM1_vs_FOM2_summary_core} together with the respective FOMs values.

\begin{figure*}
	\begin{center}
	 \includegraphics[width=5.9cm]{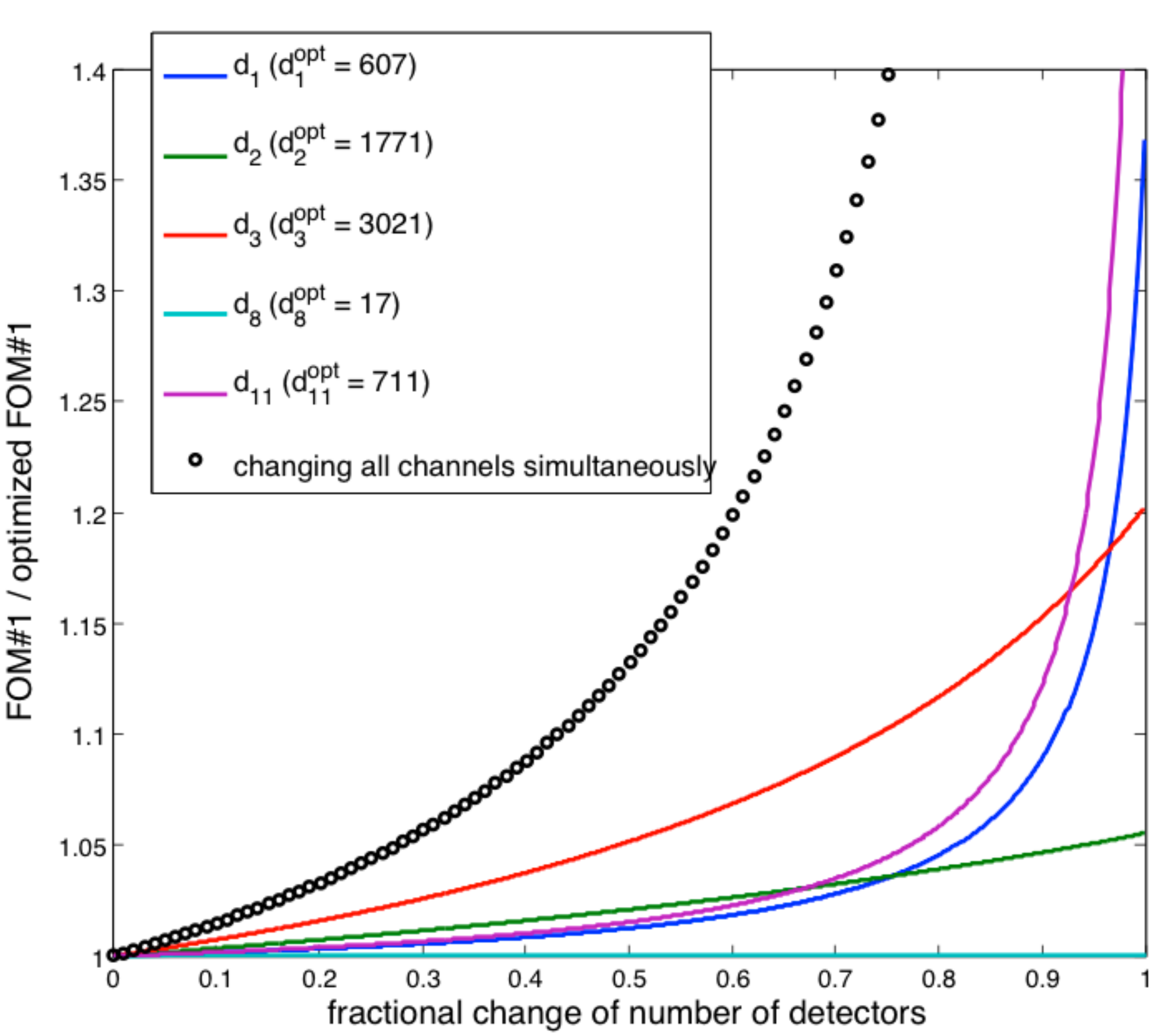}
	\includegraphics[width=5.8cm]{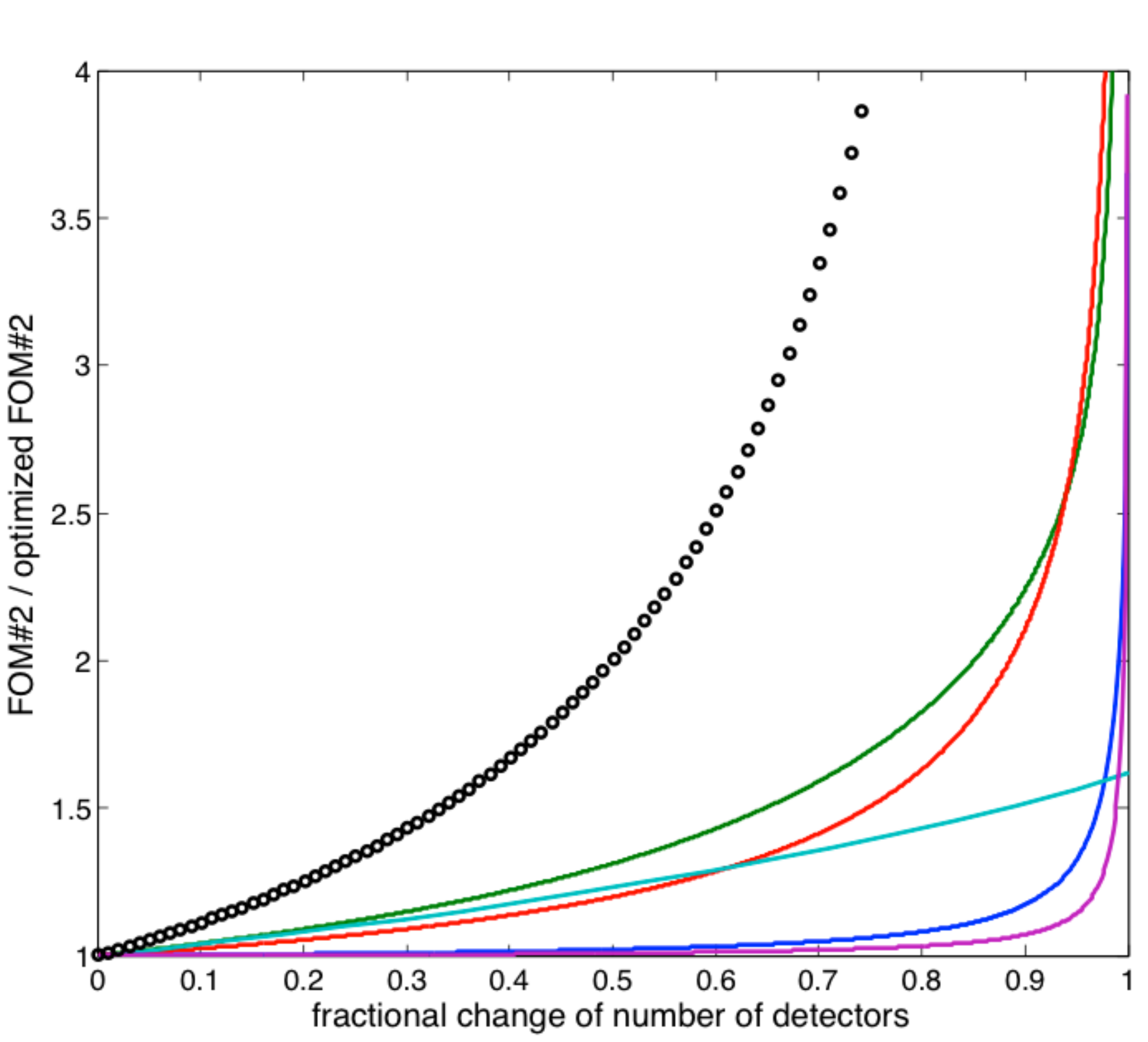}
	\includegraphics[width=5.8cm]{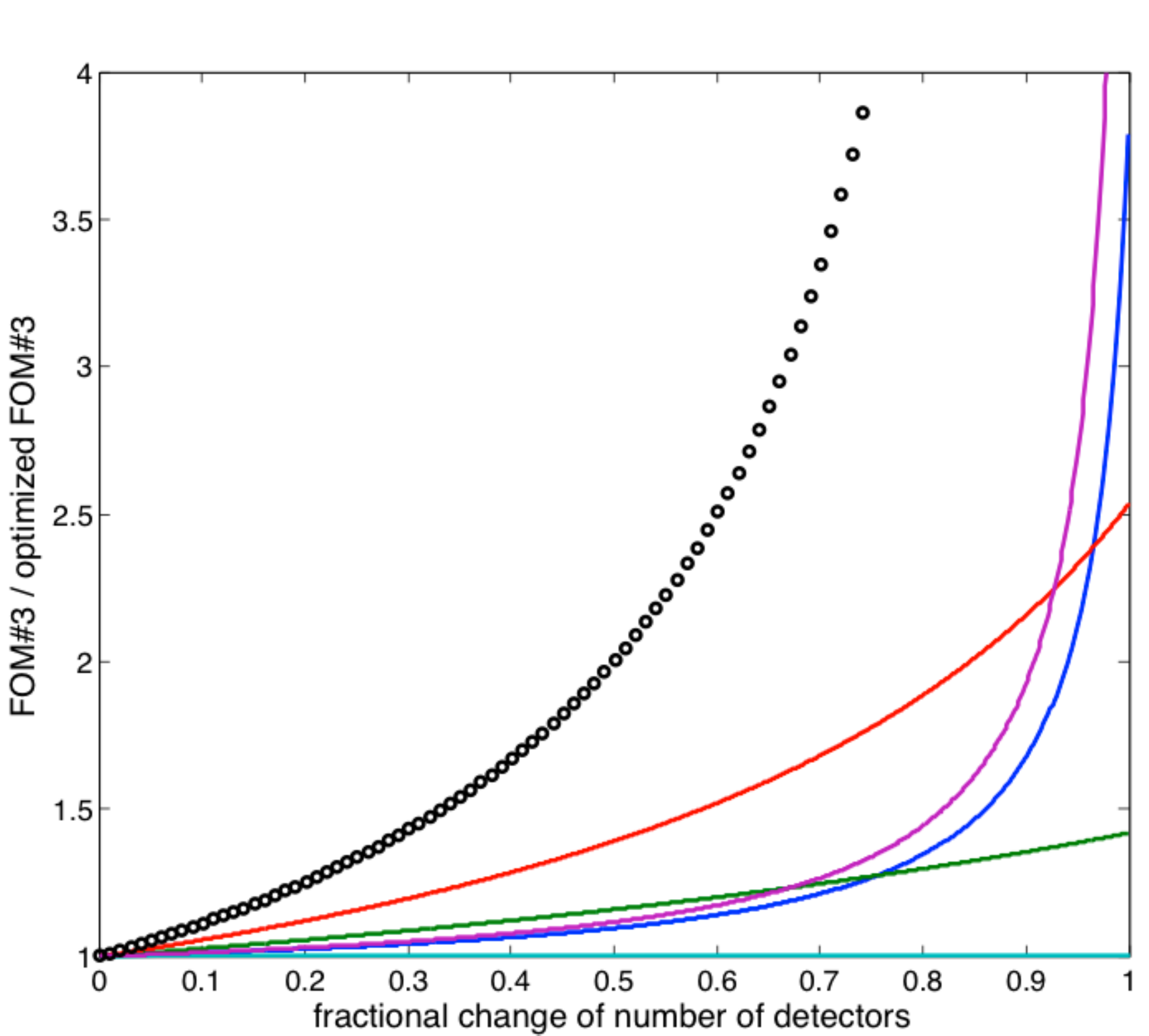}	
	 \caption{Dependence of the values of FOM\#1 (left), FOM\#2 (middle), and FOM\#3 (right), on a fractional change of a number of detectors in the hardware configuration as detailed in the fourth column of Table~\ref{table:FOM1_vs_FOM2_summary_core}. The solid lines show cases with a number of detectors in only one selected channel being gradually decreased (left to right) and all the others being kept fixed at their optimized values. The circles show the case with a number of detectors in all channels decreasing by the same fraction simultaneously. The color schemes for the lines are the same in all  the panels and described in the legend.} 
	 \label{fig:postProcResCOrE}
\end{center}
\end{figure*}
The procedure employed in this Section follows the steps outlined in Sec.~\ref{subsect:postProc}. In Fig.~\ref{fig:postProcResCOrE} we show an impact of a fractional change of a number of
detectors in one channel at the time on the values of the FOMs.  The latter are given relative to their optimized values and therefore all the curves shown in the figure are expected to start from the unity for the fractional change equal to zero, as the latter corresponds to the optimized configuration, and then grow typically monotonically with an increasing value of the fractional change. In addition, for reference we also show how the FOMs values
would change if numbers of detectors in all the channels are decreased by the same fraction. We note that at least for the two of the FOMs, i.e., FOM\#2 and \#3, the latter dependence
can be straightforwardly predicted using Eqs.~\eref{eqn:secDervSat}, \eref{eqn:resSpec0}, and~\eref{eqn:noiseProp} and shown to be inversely proportional to an actual number of detectors in the 
corresponding configurations and thus
inversely proportional to ($1-$fractional change of detectors). This indeed is adhered to by our numerical results.

The most striking features of some of the results are their apparent flatness extending on occasions to a rather high values of the fractional change. At face value that suggests that one is at liberty to change a number of detectors in some of the channels rather drastically but without noticeably penalizing the performance of the instrument. However, though some freedom indeed exists, it has to be exploited carefully. In particular, significantly changing a number of detectors in one selected channel, will usually have an effect of removing any freedom in adjusting the number of detectors in the remaining channels. Therefore if one's goal is to round-up the optimization results in a way to make them more amenable to an actual implementation
that may not be the right way to go. Below we showcase some of these issues in the specific case at hand.

Probably most conspicuous thing about the configuration considered here is the presence of a channel centered at $255$ GHz, to which are assigned only $17$ detectors, as opposed to a few thousands 
in some of the other channels. A natural question to ask is therefore whether this channel is needed at all. In fact, the two outermost panels of Fig.~\ref{fig:postProcResCOrE} seem to confirm 
our feeling that this channel is in practice irrelevant as both the FOMs \#1 and \#3 effectively do not depend on its being present. This is not so however for the FOM\#2 as shown in the middle
panel. In this case removing this channel altogether will boost the value of this FOM, and thus the level of the foreground residual by a factor of $\sim 1.5$. Though not overwhelmingly large it 
is substantial enough to justify holding on to this channel (unless of course the hardware cost of having the extra channel tips the balance the other way). These expectations are confirmed by
direct calculations, results of which as shown a 5th column in Table~\ref{table:FOM1_vs_FOM2_summary_core}. (We note that an attempt to re-optimize the resulting 4-channel system \textit{a posteriori} does not bring much improvement either; see Table~\ref{table:FOM1_vs_FOM2_summary_core}, column 6). We note that trying to keep the level of residuals
down in this case can be of particular importance given that already in its original, optimized version (Table~\ref{table:FOM1_vs_FOM2_summary_core}) the resulting values of $r_{min}$ and $r_{eff}$ are close enough to each other that this is probably
the latter, i.e, the level of residuals, which would drive the actual limit on a detectable $r$ value for this setup, rather than the statistical estimate provided by FOM\#1. Letting $r_{eff}$ grow any further
would therefore directly affect our science goals. Instead we can therefore try to trim a number of detectors in either $45$ or $375$ GHz channel. We see that we can potentially reject up to $\sim70$\%
of the detectors in the former or $\sim80$\% in the latter, without affecting the residuals level (FOM\#2) in any appreciable manner. This would have an effect of increasing FOM\#1 value by no more 
than $\sim 5$\% and FOM\#3 by no more than $\sim 50$\%, both of which may therefore look perfectly acceptable. Whichever option we opt for, we can then reuse the spare detectors by distributing 
them to some of the existing channels or creating some additional ones, say at $165$ GHz, in order to be better equipped to face some potential surprises (Sect,~\ref{subsect:postProc}). 
However a special care then has to be taken if
a number of detectors in some other channels needs to be concurrently decreased. This is because, as illustrated by lines marked with circles in Fig.~\ref{fig:postProcResCOrE}, not all directions in the parameter space are similarly flat. 

\begin{figure}[htbp] 
   \centering
   \includegraphics[width=7truecm]{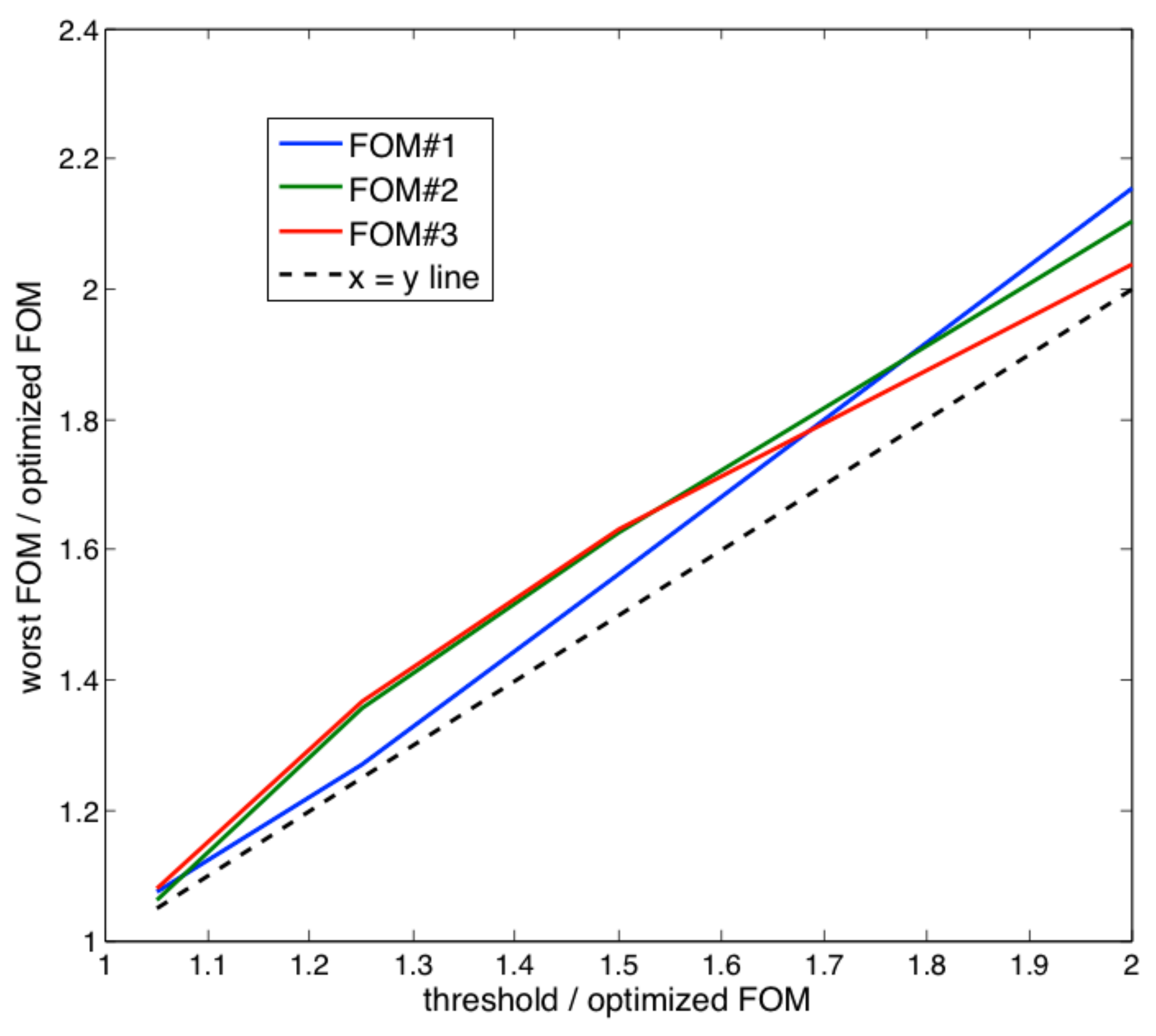} 
   \caption{The worst values of each FOM, $\tilde{v}$, computed for each of the concentric hyperellipsoids, Eq.~\ref{eqn:hyper_ellipses}, defined by the threshold values, $v$, as shown on the horizontal axis. The dotted line shows $\tilde{v} = v$ case. Clearly, $\tilde{v} \simeq v$ in all shown cases, where the latter approximate equality holds to within $10$\%. The values of $\tilde{v}$ and $v$ given here are relative to the optimized values of the respective FOMs.}
   \label{fig:v_vs_vt}
\end{figure}

If our aim is to just round-up the detector numbers we can proceed as outlined in Sec.~\ref{subsect:postProc}. We first postulate a set of fractional changes from the optimized values. In our case
these could be $\l[v_k\r] = \l[ 1.025, 1.05., 1.1, 1.15\r]$  for FOM\#1 and $\l[v_k\r] = \l[ 1.05, 1.25., 1.5, 2.0\r]$ otherwise,
and then use Fig.~\ref{fig:postProcResCOrE} to read off the corresponding values of the fractional change for each channel and each FOM. These are values denoted $\sigma$ in Sec.~\ref{subsect:postProc}. In our case for FOM\#1 they read
\begin{eqnarray}
\l\{\sigma_{j}^{\l(k\r)}\r\} =
\l\{
\begin{array}{r r r r}
     409 &     496  &      555 &    577 \\
    1017 &  1664  &     1771 & \infty \\
      880 &   1477  &     2236 &  2697 \\
      \infty &  \infty   & \infty & \infty \\
      442 &  549  & 624  &    654 \\
\end{array}
\r\},
\label{eqn:sigmasFOM1}
\end{eqnarray}
 where $k$-th column corresponds to the $k$-th value of $v_k$ and thus gives values of $\sigma$ for each of the five channels with nonzero number of detectors in the optimized configuration (see second column of Table~\ref{table:FOM1_vs_FOM2_summary_core}).
 We can use these values to define, Eq.~\eref{eqn:hyper_ellipses}, hyperellipsoidal volumes, ${\cal V}_k$, in the parameter space centered on the optimized configuration. We note  that the infinity sign marks the cases, where the desired value of $v_k$ could not have been reached due to the parameter space boundary. For instance, the values in the fourth row of Eq.~\eref{eqn:sigmasFOM1} are all infinite as in the neighborhood of the optimized configuration the value of FOM\#1 does not depend on a number of detectors in this channel as can be seen in Fig.~\ref{fig:postProcResCOrE}.

To find the worst case value of the FOM for a $k$-th hyperellipsoid, $\tilde{v}_k$, we use random sampling of first an entire volume of the ellipsoid followed by that of only its surface.
The latter requires fewer samples to ensure proper sampling density and is more efficient if we have some expectation of the FOM values monotonically deteriorating away from the 
 optimized configuration. As anticipated in Sec.~\ref{subsect:postProc} the corrected values, $\tilde{v}_k$, and initial ones, $v_k$, are indeed found to be quite close, typically within $20$\% of each other
 as illustrated in Fig.~\ref{fig:v_vs_vt}.
 
 The series of the concentric hyperellipsoids constructed here gives us a quick, though approximate, way to estimate the performance of some proposed configurations derived from the optimized one via 
 small changes of all or some optimization parameters. As an example, consider a configuration with $\l[d_j\r] = \l[600, 1700, 3000, 17, 700\r]$ detectors in each of the five channels considered here.
 Given that  for FOM\#1,
 \begin{eqnarray}
 \sum_k \frac{\l(d_j-d^{opt}_j\r)^2}{{\sigma_{j}^{\l(k\r)}}^2} \le 1
 \end{eqnarray}
 is fulfilled for any $k$, we conclude that the respective value of FOM\#1 for this case will not be larger than by a factor $\tilde{v}_{k=1} \simlt 1.025$ than the optimized value.
  Indeed a direct calculation renders a value $1.002$ times higher than the optimized one in agreement with our quick estimation.
 Similarly, we can deduce the performance of this configuration as expressed by 
the two other FOMs. These are more sensitive at least to changes in some of the channels however we find that for this specific configuration we can lose no more
than a factor of $1.05$ for both of them.  These could be compared to the actual values of $1.01$ and $1.02$, respectively, all relative to the corresponding optimized values.

In this case overall the loss of performance seems rather benign and acceptable. Moreover, as a result of rounding-down the detector numbers we have gained around 100 of those,
which we can arbitrarily assign to any of the existing channels or even create a new one to saturate the constraint on the total number of detectors. Whatever decision we make
we will not compromise any of the performance figures derived earlier.

To illustrate a process of adding some \textit{ad hoc} channels at this time we start from a configuration more drastically stripped-down than the one discussed above. Let that be for instance
$\l[d_k\r] = \l[ 500, 1500, 3000, 0, 600\r]$, where we not only reduced numbers of detectors per channel more substantially but also removed the fourth channel altogether. Using the hyperellipsoid formalism we get quickly a helpful insight into how much we have lost as a result of choosing this configuration. As we already discussed, the biggest loss is found with regard to the value of FOM\#2, which is boosted by more than $50$\% (but less than $100$\%) with FOM\#1 and FOM\#3 changing by $\simlt 1.05$ and $\sim 1.1$ respectively. (The actual values being $1.01, 1.81$ and $1.09$ for FOMs: \#1, \#2, and \#3.) 
However we have also gained as many as $400$ detectors, which can be distributed at our discretion to fill the constraint. Let us 
do so by introducing two extra channels at $195$ and $285$ GHz with $200$ detectors each. This improves the performance of the considered configuration, an improvement which we can ameliorate
even further by performing the optimization with respect to the detector numbers in the four original channels and keeping the detector numbers of the new channels fixed to $200$. We indeed find that the new setup
performs nearly as well as the initial optimized one (Table~\ref{table:FOM1_vs_FOM2_summary_core}, column 3 vs 7) but possesses a more uniform frequency coverage. If we now want 
to perform a controlled detector number rounding and analyze its impact on the configuration performance we would need to restart the entire procedure  described above.

\subsubsection{Robustness tests}

\begin{figure}[htbp]
   \centering
   \includegraphics[width=8truecm]{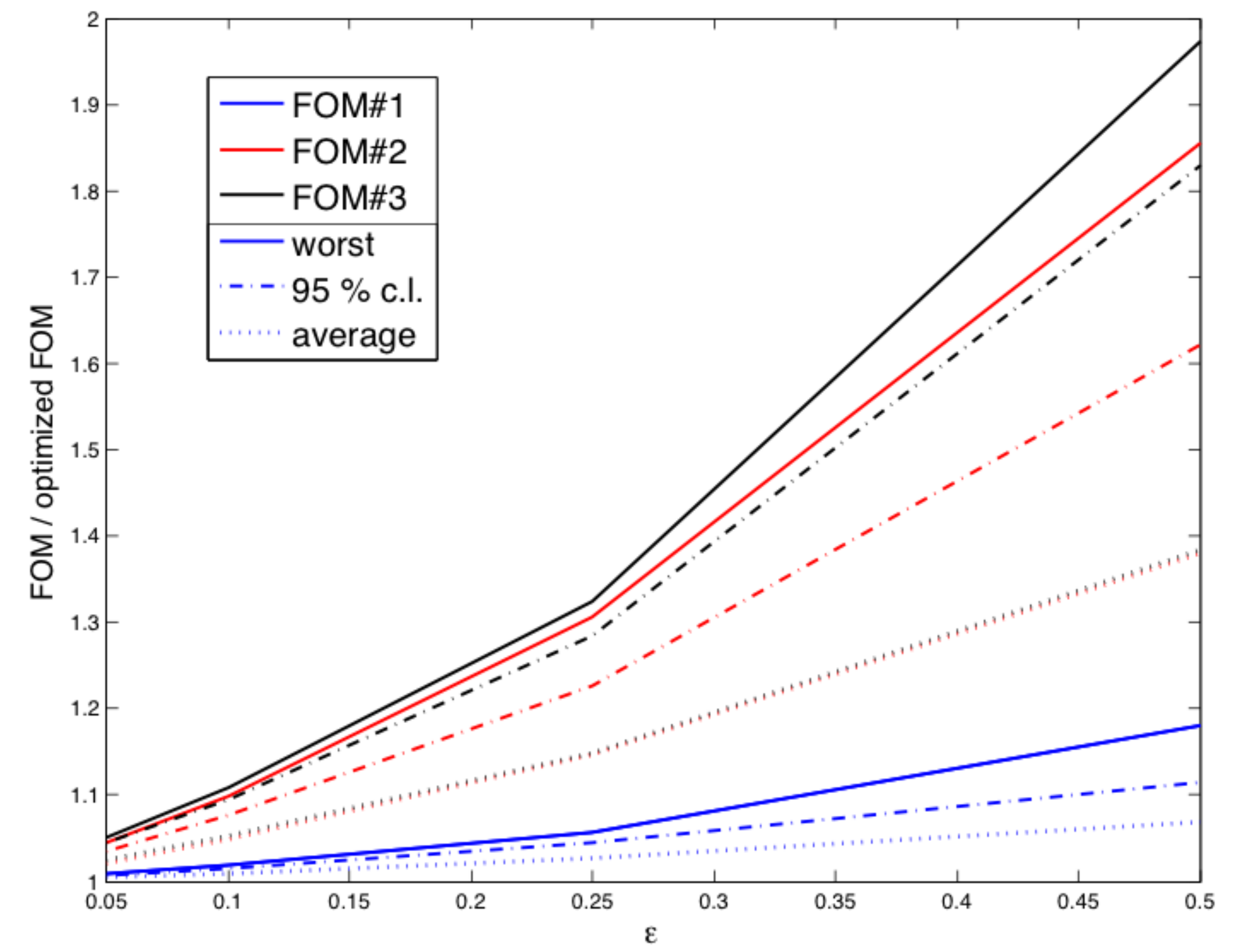} 
   \caption{Summary of our robustness tests applied to the COrE configuration obtained via the optimization of FOM\#1 with constraints of FOM\#2 $\le 10^{-4}$ and a fixed number of detectors.
   The lines of different colors correspond to different FOMs and different lines show: average (dotted), $95$\% confidence limit, (dot-dashed), and the worst value (solid). }
   \label{fig:robustnessCore}
\end{figure}

As explained in Sec.~\ref{subsect:robustness}, for each FOM, we start from the optimized configurations, as determined earlier and check how the values of the FOMs depend on a random suppression of a number of detectors in each channel by some fraction. Specifically, we assume here that the distribution of the anticipated detector failures is Gaussian with the dispersion equal to  $\varepsilon$ of which is the same
for each of the considered channel and taken to  change from $5$\%, $10$\%,  $25$\%, and  $50$ \%. We randomly draw some large number of samples, here $10^4$, and histogram the results for each of
the FOMs. We then compute the most likely value of the FOMs, $95$\%-confidence limit, and the worst drawn value. In the case of the COrE configuration studied in the previous Section we collect the results in
Fig.~\ref{fig:robustnessCore}. We conclude, as probably could have been anticipated from the results of the previous Section, that for a failure rate as large as $30$\% we will not compromise on the FOM values by more than $50$\% with respect to the optimized ones, while a failure rate of $10$\% will result in their $10$\% increase. 
These result affirm the practical soundness of the derived configuration.

\subsubsection{Robustness with respect to the foreground modeling}

\label{subsect:robustmessForeModels}

Results of the optimization procedures including thus the procedure considered here are usually only as good as the foreground models used in their course. In the specific case studied here 
we expect that  our results are fairly robust as far as foreground morphology is concerned. Our estimates are driven by two compact description of those, the foreground correlation matrix, $\bd{\hat{F}}$, and the 
foreground power spectra, which are not expected to be wildly different than what we have assumed here. We note  in particular that an increasing amplitude of the foregrounds leading to an increase
of both the elements of the matrix, $\bd{\hat{F}}$, and overall normalization of the foreground power spectra would decrease the errors on the spectral parameters and result in the amplitude of the residuals being virtually unchanged. These expectations are confirmed by the results obtained here for the three different masks.

It is more difficult to assess, though potentially more crucial, the impact of increasing a number of spectral parameters. This could be either due to more complicated spectral dependences of true foreground components, or as a result of a spatial dependence of spectral parameters. The former problem is inherent to all parametric component separation approaches including the one assumed here. In general, a wrong parametrization or frequency scaling laws assumed in such approaches may invalidate separation results. In practice, the effects are more subtle but arising biases can affect an interpretation of the results. It is therefore important that the scaling laws assumed in the optimization continue to be improved, reflecting any relevant, new observational data and more detailed, theoretical models of the foreground physics as they become avail- able. In a case of some doubts, a rather conservative approach can be fruitful, restricting channel frequencies to a range for which the scaling laws are known to provide at least good approximations to the actual ones. This is in fact an approach we used in this work by selecting  a parametric model for the dust signal with a single parameter and reduced the frequency range to those lower than $400$ GHz.

\begin{figure*}
\begin{center}
	 \includegraphics[width=8cm]{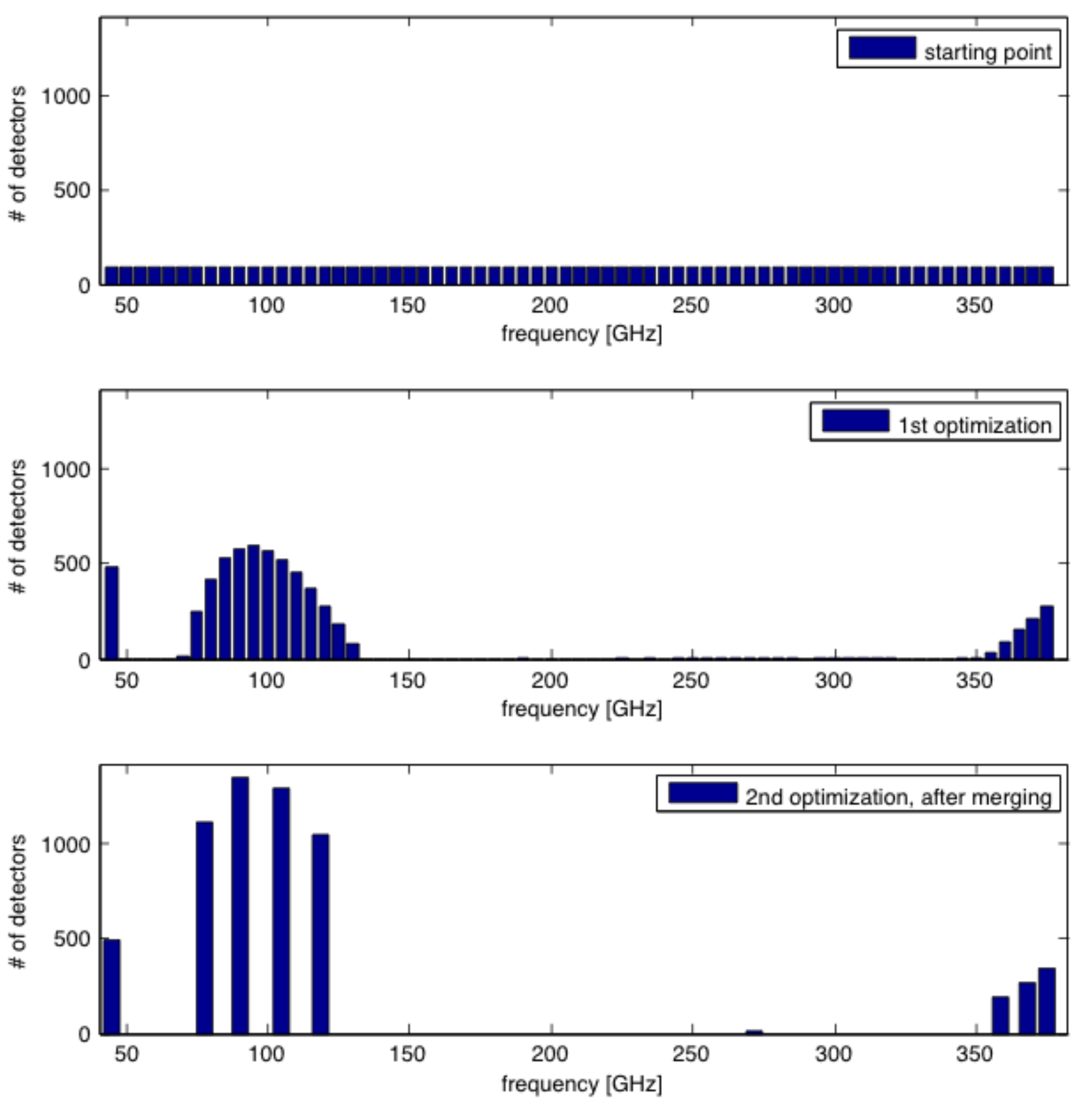}
 	 \includegraphics[width=9.5cm]{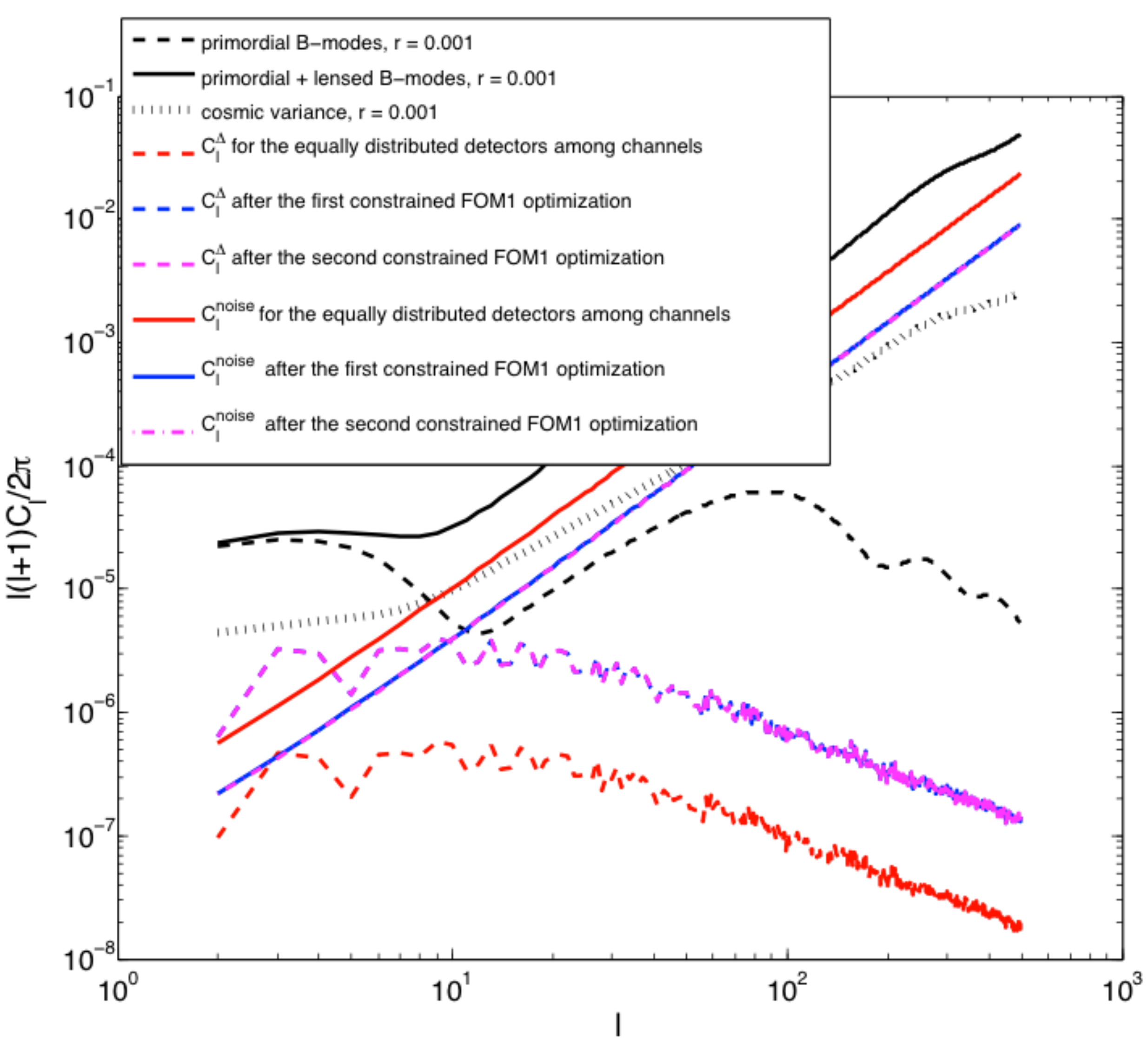}
	 \caption{Demonstration of the optimization results derived with respect to a variable number of channels, numbers of detectors per channel, and their central frequencies, while constraining the 
	 total number of detectors ($=6128$ as in the proposed COrE version). {\it Upper left panel} shows, from top to bottom,  (1) the starting configuration with all the detectors evenly distributed among a fine-grid of channels; (2) a configuration after the first optimization of FOM\#1 constrained to ensure that  ${\rm FOM}\#2 \leq 10^{-4}$; and (3) the reoptimization of configuration (2) restricted only to channels with a number of detectors larger than five and after adjacent channels merging and recentering as described in Sec.~\ref{subsect:varying_number_channels}. \textit{Right panel} shows power spectra corresponding to these configurations contrasted against the expected CMB signals. }
	 \label{fig:channles_opt_core_400_P06}
\end{center}
\end{figure*}

A spatial dependence of the scaling parameters can be treated more directly. We will implement that by dividing the observed sky into a multiple disjoint regions and introduce one set of parameters for each of those. To abstract from details of the regions shape and position, we assume that they are defined in such a way that the errors on spectral parameters are the same for each of the regions, i.e., that the differences of the overall magnitude of the matrix $\bd{\hat{F}}$ are compensated by a respective number of pixels in each area. In general this assumption would imply that more, though smaller by area, regions are defined in high-contrast foreground sky areas. This indeed could well be the case as the high-contrast 
foreground regions are expected to be more complex and may require more parameters to ensure sufficient accuracy.

For demonstration purposes we assume that we have $10$ regions with the corresponding errors on spectral parameters being $\sqrt{10}$ times larger than in the single region case as studied before. We note that cutting the sky into regions will unavoidably affect the foregrounds and thus residual power spectrum on scales larger than a typical size of the region. We will ignore this effect here, motivated by the fact that our earlier results did not find any strong dependence on the shape of the power spectrum. We also neglect here all practical difficulties such as matching the results on the map level coming from the different regions and which will have to be addressed in any actual application of the discussed method.
We limit here ourselves to the COrE-like configuration as defined earlier, calculate the FOMs as before, and optimize the configuration following the steps outlined before. As expected we find that the optimal configurations this time are not very different from the ones obtained earlier. This is because FOM\#1 and FOM\#3 are mostly trying to optimize the overall
noise level, which is the same now as before, and though the value of FOM\#2 increased by a factor $10$ due to increase of the spectral index errors this is the same configuration, which ensures
its minimum. As a consequence the new value of $r_{eff}$ is now higher than that of $r_{min}$. This clearly does not invalidate results of the optimization procedure as such, however care has to be 
exercised, while interpreting 
the obtained values of $r_{min}$, which may not be taken directly as the performance forecasts for the setup as far as detecting $r$ is concerned.

\subsection{Varying the number of channels and their frequencies}
\label{subsect:varying_number_channels}

We present here some results based on an implementation of the scheme proposed in Sec.~\ref{subsect:minProc}. We start from $\sim 70$ channels evenly spaced between $45$ and $375$ GHz every $5$ GHz, with $\sim 6000$ detectors  (total number of the COrE proposed version) equally distributed among those, as shown in Fig.~\ref{fig:channles_opt_core_400_P06}.
Then we perform the optimization with respect to FOM\#1, while keeping  ${\rm FOM}\#2 \leq 10^{-4}$ and the total number of detectors fixed. As a result we obtain a highly clustered distribution of detectors in between the initial channels, with many of these being empty. 
We therefore combine detectors of neighboring channels and replace them by a new channel with the central frequency set as a weighted, by a number of detectors, mean of the optimized distribution. The new channels are
defined to ensure proper spacing between them. Once the new channels are determined we perform a second round of the optimization, this time invoking only the new channels and aiming at optimization of the detector distribution between them.
The result is shown in the left bottom panel of Fig.~\ref{fig:channles_opt_core_400_P06}. We note that the procedure not only improved the values of the FOMs with respect to the starting (original) configuration, i.e., FOM\#1 has been decreased
by $\sim17$\% ($\sim25$\%), while the noise by a factor $\sim4$ ($\sim3$), but also, and arguably most importantly, it resulted in a configuration significantly simpler than the initial one with the number of channels reduced from $70$ down to $9$.

We note that maybe somewhat surprisingly both the configurations derived here, the final one as well as the intermediate one obtained after the first optimization step, show
only a minor, $\sim$ few percent, gain over the five-channel configuration we have considered earlier; see, e.g., the first column of Table~\ref{table:FOM1_vs_FOM2_summary_core}.
This is due to our setting the threshold for FOM\#2 rather high, while the main advantage of the significantly larger set of the initial channels used here is that it permits 
finding in principle more satisfactory compromises between the three FOMs, characterized by values of FOM\#2 lower than what could be achieved with more modest setups
discussed earlier.

\section{Conclusions}

\label{sect:conclusions}
	
In this work we have proposed a general scheme for a performance optimization and  forecasting of the CMB $B$-mode experiments in the presence of
astrophysical foregrounds. Our approach is based on a maximum likelihood parametric technique for component separation, for which we have derived
Fisher-like error estimates for spectral parameters. We use the latter to calculate the residual level of the foregrounds in cleaned CMB maps given assumed,
instrument characteristics and foreground model. We then optimize the former by minimizing a set of proposed figure of merit indicators, which reflect
our science goals. Subsequently we have applied this approach to two specific cases of recently proposed CMB $B$-mode satellites: American 
CMBpol~\cite{2009arXiv0903.0902A} and European COrE~\cite{2011arXiv1102.2181T}. 
We have discussed in detail the choices and trade-offs inevitable in such an optimization process. We have demonstrated how such a procedure
can help to simplify the resulting hardware design, while ensuring the same (or nearly the same) science outcome.

We emphasize that results of such a procedure can be only as reliable as the foreground models that are applied. This underlines the importance of developing better understanding of the polarized foregrounds, in particular, and characteristically of the parametric methods, as far as the functional form of the foreground component scaling laws is concerned. However, our approach is expected to be relatively robust as far as
other details of the foreground signals are concerned, such as, spatial distribution or spatial variability of the spectral parameters, with the latter playing a major role in determining the scientific reach of the experiment but not affecting its configuration.

We also note here that the presented framework could be extended to work with any component separation method, which implements the separation
by first estimating the mixing matrix, in a parametric or nonparametric way, and which is capable of producing estimates for the errors of the spectral parameters for any hardware configuration. 
One could, and ideally would,  therefore use the formalism proposed here to define configurations, which would ensure that many of the available component separation methods perform well.
Though the component separation methods usually conform with the first requirement, the second is more demanding and typically can be done only via computationally-heavy Monte Carlo simulations. 
Those may be often impractical for the optimization purposes, making an implementation of such a program difficult.  A related, but simpler to address, problem is whether the configurations optimized 
with one method will work for satisfactorily with the others. We will leave an investigation of both these issues to future work.

As we point out in the introduction, FOMs required for the optimization procedure are also suitable for the performance forecasting. This also clearly applies to the FOMs proposed here and in 
particular FOM\#1 and FOM\#2 seem relevant to the primordial signal detection producing values of $r_{min}$ and $r_{eff}$  on order of ${\cal O}\l(10^{-4}\r)$ for the considered optimized configurations. 
However, given that each of these two FOMs reflects a somewhat different aspect of the problem -- a statistical uncertainty in
former cases versus a systematic one  in the latter -- care has to be taken while interpreting these values. Nevertheless, as is, our results seem to support at least the contentions made elsewhere suggesting that
$r\simeq 10^{-3}$ is a realistic goal for the experiments considered here.

Finally, we point out that the science goals we have  posed for the considered CMB experiments are clearly more modest than those targeted by the original CMBpol and COrE designs.
This is responsible, at least in part, for the more complex and advanced instrumental configurations as proposed in the original proposal. More diverse science goals can, and should, be
studied in the presented framework. We leave this as well as considerations of other possible extensions, e.g., an inclusion of some of the instrumental effects, to future work.
\\
\\
\centerline{{\bf Acknowledgments}}

We acknowledge the use of the \textsc{healpix}~\cite{2005ApJ...622..759G}~\footnote{http://healpix.jpl.nasa.gov/} and \textsc{s2hat}~\cite{2011arXiv1106.0159S}~\footnote{http://www.apc.univ-paris7.fr/$\sim$radek/s2hat.html} software packages.
This work has been supported in part by French National Research Agency (ANR)
through COSINUS program (project MIDAS no. ANR-09-COSI-009) and it used HPC resources from GENCI--IDRIS (Grant No. 2011-066647) as well as from the National Energy Research Scientific Computing Center, which is supported by the Office of Science of the U.S. Department of Energy under Contract No.\ DE-AC02-05CH11231.

\bibliographystyle{apsrev}
\bibliography{optim}

\appendix

\begin{widetext}

\section{Spectral likelihood derivatives.}

\label{sect:algebra}

We present here some details of the derivation of Eqs.~\eref{eqn:firstDerAver} and~\eref{eqn:secDervSat}.

First from Eq.~\eref{eqn:profileLikeDef} we have
\begin{eqnarray}
{\partial \ln {\cal L}\over\partial\beta} & = &  \sum_p\,\l(\bd{A}_{,\beta}\,\bd{s}_p\r)^t \, \bd{N}^{-1}\,\l(\bd{d}_p-\bd{A} \,\bd{s}_p\r) 
\label{eqn:firstDer}
\end{eqnarray}
from which the second derivatives of the spectral likelihood follow as
\begin{eqnarray}
{\partial^2 \ln {\cal L} \over\partial\beta\,\partial\beta'} & = & 
 \sum_{p}\, \l\{  \l(\bd{A}_{,\beta\beta'}\,\bd{s}_p + \bd{A}_{,\beta}\,\bd{s}_{p, \beta'}\r)^t \, \bd{N}^{-1}\,\l(\bd{d}_p-\bd{A} \,\bd{s}_p\r)  
 \, -  \, \l(\bd{A}_{,\beta}\,\bd{s}_p\r)^t \, \bd{N}^{-1}\,\l(\bd{A}_{,\beta'} \,\bd{s}_p + \bd{A}\,\bd{s}_{p, \beta'}\r) 
 \r\}
\end{eqnarray}
And the noise ensemble average reads,
\begin{eqnarray}
\l\langle{\partial^2 \ln {\cal L}\over\partial\beta\,\partial\beta'}\r\rangle_{noise} & = & \sum_p\,\l\{
{\rm tr}\,\l[\bd{A}_{,\beta\beta'}^t \, \bd{N}^{-1}\,\l\langle \l(\bd{d}-\bd{A} \,\bd{s}_p\r)\,\bd{s}_p^t\r\rangle_{noise}\r]  \, -  \, {\rm tr}\,\l[\bd{A}_{,\beta}^t \, \bd{N}^{-1}\,\bd{A}_{,\beta'} \,\l\langle \bd{s}_p \, \bd{s}_p^t\r\rangle_{noise}\r]\r.\\
& + & \,\l.\phantom{\sum_p\  \ } {\rm tr} \, \l[\bd{A}_{,\beta}^t\,\bd{N}^{-1}\,\langle\l(\bd{d}_p-\bd{A} \,\bd{s}_p\r) \, \bd{s}_{p, \beta'}^t\rangle_{noise}\r] \, - \, {\rm tr} \, \l[\bd{A}_{,\beta}^t \, \bd{N}^{-1}\,\bd{A}\,\langle\bd{s}_{p, \beta'}\bd{s}_p^t\rangle_{noise}\r]\r\}.\nonumber
\end{eqnarray}

From Eqs.~\eref{eqn:dataModelFull} and\ \eref{eqn:mlMapDef} we now have
\begin{eqnarray}
\l\langle \bd{s}_p\,\bd{s}_p^t\r\rangle_{noise} & = & \bd{\bar{s}}_p\,\bd{\bar{s}}_p^t \, + \, \l( \bd{A}^t\,\bd{N}^{-1}\,\bd{A}\r)^{-1},\\
\l\langle \bd{s}_p\,\bd{s}_{p , \beta'}^t \r\rangle_{noise} \,
& = & 
- \,  \bd{\bar{s}_p}\,\bd{\bar{s}_p}^t \, \l(\bd{A}_{,\beta'}^t\bd{N}^{-1}\bd{A}+\bd{A}^t\bd{N}^{-1}\bd{A}_{,\beta'}\r)\,\l(\bd{A}^t\bd{N}^{-1}\bd{A}\r)^{-1}
\,+\, \bd{\bar{s}}_{p} \,\bd{\bar{q}}_{p\,\l(\beta'\r)}^t \,  \nonumber\\
 & - & \l(\bd{A}^t\bd{N}^{-1}\bd{A}\r)^{-1} \, \l(\bd{A}_{,\beta'}^t\bd{N}^{-1}\bd{A}\r)\,\l(\bd{A}^t\bd{N}^{-1}\bd{A}\r)^{-1}\\
\langle\l(\bd{d}_p-\bd{A} \,\bd{s}_p\r) \, \bd{s}_{p}^t\rangle_{noise} & = & 
\l(\hat{\bd{A}}\,\hat{\bd{s}}\,-\,\bd{A}\bd{\bar{s}_p}\r)\,\bd{\bar{s}_p}^t \nonumber\\
\langle\l(\bd{d}_p-\bd{A} \,\bd{s}_p\r) \, \bd{s}_{p, \beta'}^t\rangle_{noise} & = & 
- \, \l(\hat{\bd{A}}\,\hat{\bd{s}}\,-\,\bd{A}\bd{\bar{s}_p}\r)\,\bd{\bar{s}_p}^t \, \l(\bd{A}_{,\beta'}^t\bd{N}^{-1}\bd{A}+\bd{A}^t\bd{N}^{-1}\bd{A}_{,\beta'}\r)\,\l(\bd{A}^t\bd{N}^{-1}\bd{A}\r)^{-1}
\nonumber\\ 
& \phantom{=}  & \, + \, \l(\hat{\bd{A}}\,\hat{\bd{s}}\,-\,\bd{A}\bd{\bar{s}_p}\r)\,\bd{\bar{q}_{p\, \l(\beta'\r)}}^t \, + \,  \bd{A}_{,\beta'}\,\l(\bd{A}^t\bd{N}^{-1}\bd{A}\r)^{-1}\nonumber\\
& \phantom{=} &\,  + \, \bd{A}\, \l(\bd{A}^t\bd{N}^{-1}\bd{A}\r)^{-1} \, \l(\bd{A}^t\bd{N}^{-1}\bd{A}_{,\beta'}\r)\,\l(\bd{A}^t\bd{N}^{-1}\bd{A}\r)^{-1},
\end{eqnarray}

where $\hat{\bd{A}}$ and $\hat{\bd{s}}$ are the true mixing matrix and sky components, respectively, $\bd{\bar{s}}$ is a component estimate in a case
of noiseless experiment, i.e., 
\begin{eqnarray}
\bd{\bar{s}}_p & \equiv& \l( \bd{A}^t\,\bd{N}^{-1}\,\bd{A}\r)^{-1}\,\bd{A}^{t}\,\bd{N}^{-1}\,\hat{\bd{A}}\,\hat{\bd{s}}_p \label{eqn:mpDef}
\end{eqnarray}
and $\bar{\bd{q}}_{\l(\beta\r)}$ is defined as,
\begin{eqnarray}
\bar{\bd{q}}_{p\,\l(\beta'\r)} & \equiv& \l( \bd{A}^t\,\bd{N}^{-1}\,\bd{A}\r)^{-1}\,\bd{A}^{t}_{,\beta'}\,\bd{N}^{-1}\,\hat{\bd{A}}\,\hat{\bd{s}}_p.
\end{eqnarray}

Hence,
\begin{eqnarray}
\l\langle{\partial^2 \ln {\cal L}\over\partial\beta\,\partial\beta'}\r\rangle_{noise} = & -  & \,\sum_p\,\l\{
\, \l(\bd{A}_{,\beta\beta'}\,\bd{\bar{s}}_p\r)^t \, \bd{N}^{-1}\,\l(\bd{A} \,\bd{\bar{s}}_p \, - \, \hat{\bd{A}}\,\hat{\bd{s}}_p\r)  \, + \, \l(\bd{A}_{,\beta}\,\bd{\bar{s}}_p\r)^t \, \bd{N}^{-1} \,  \l(\bd{A}_{,\beta'}\,\bd{\bar{s}}_p\r)\r. \,\nonumber \\
& + & \,{\rm tr} \, \l[\bd{A}_{,\beta}^t\,\bd{N}^{-1}\, \l(\hat{\bd{A}}\,\hat{\bd{s}}_p \, - \, \bd{A}\,\bd{\bar{s}}_p\r)\,\bd{\bar{s}}_p^t\,
\l(\bd{A}_{,\beta'}^t\bd{N}^{-1}\bd{A}+\bd{A}^t\bd{N}^{-1}\bd{A}_{,\beta'}\r)\,\l(\bd{A}^t\bd{N}^{-1}\bd{A}\r)^{-1}\r] \nonumber\\
&- & \, {\rm tr}\,\l[ \l( \bd{A}_{,\beta} \, \bd{\bar{q}}_{p, \l(\beta'\r)}\r)^t \, \bd{N}^{-1}\,  \l(\hat{\bd{A}}\,\hat{\bd{s}}_p \, - \,  \bd{A}\,\bd{\bar{s}}_p\r)\r]\nonumber\\
&-& \,
{\rm tr}\,\l[\bd{A}_{,\beta}^t\, \bd{N}^{-1}\,  \bd{A}\, \l(\bd{A}^t\bd{N}^{-1}\bd{A}\r)^{-1} \,  \l(\bd{A}_{,\beta'}^t\bd{N}^{-1}\bd{A}+\bd{A}^t\bd{N}^{-1}\bd{A}_{,\beta'}\r)\, \bd{\bar{s}_p}\,\bd{\bar{s}_p}^t \r]\nonumber\\
& + & \l.\, {\rm tr}\,\l[\bd{A}_{,\beta}^t\, \bd{N}^{-1}\,  \bd{A}\, \bd{\bar{q}}_{p\,\l(\beta'\r)} \, \bd{\bar{s}}_{p}^t \r] \r\}.
\label{eqn:secDervGen0}
\end{eqnarray}
 Moreover assuming now
 the true values of the spectral indices, i.e., $\bd{\beta}=\hat{\bd{\beta}}$,
\begin{eqnarray}
\l.\l\langle{\partial^2 \ln {\cal L}_{profile}\over\partial\beta\,\partial\beta'}\r\rangle_{noise}\r|_{\bd{\beta}=\hat{\bd{\beta}}}  = & - &
\, {\rm tr} \, \l[ \bd{A}_{,\beta}^t \, \bd{N}^{-1} \,  \bd{A}_{,\beta'}\,\sum_p\,\hat{\bd{s}}_p\,\hat{\bd{s}}_p^t\r]\nonumber\\
& + &  \,
 {\rm tr}\,\l[\bd{A}_{,\beta}^t\, \bd{N}^{-1}\,  \bd{A}\, \l(\bd{A}^t\bd{N}^{-1}\bd{A}\r)^{-1} \,  \l(\bd{A}_{,\beta'}^t\bd{N}^{-1}\bd{A}+\bd{A}^t\bd{N}^{-1}\bd{A}_{,\beta'}\r)\, 
 \sum_p\,\hat{\bd{s}}_p\,\hat{\bd{s}}_p^t \r]\nonumber\\
& - & \, {\rm tr}\,\l[\bd{A}_{,\beta}^t\, \bd{N}^{-1}\,  \bd{A}\, \l(\bd{A}^t\,\bd{N}^{-1}\,\bd{A}\r)^{-1}\,\bd{A}^t_{,\beta'}\,\bd{N}^{-1}\,\bd{A} \, 
\sum_p\,\hat{\bd{s}}_p\,\hat{\bd{s}}_{p}^t \r] 
 \nonumber\\
=  & {\rm tr}&\l\{\l[ \bd{A}_{,\beta}^t\, \bd{N}^{-1}\,  \bd{A}\, \l(\bd{A}^t\bd{N}^{-1}\bd{A}\r)^{-1} \,  \bd{A}^t\bd{N}^{-1}\bd{A}_{,\beta'}\, 
- \,  \bd{A}_{,\beta}^t \, \bd{N}^{-1} \,  \bd{A}_{,\beta'}\,\r]
 \sum_p\,\hat{\bd{s}}_p\,\hat{\bd{s}}_p^t \r\}, \label{eqn:secDervFinal}
 \end{eqnarray}
from which Eq.~\eref{eqn:secDervSat} follows.

\end{widetext}

\section{Fisher matrix algebra.}

\label{sect:algebra_2}

The Fisher matrix can be expressed as~\cite{1996ApJ...465...34V},
\begin{equation}
	\centering
		F_{\alpha \beta} \equiv \l\langle \frac{\partial ^{2} ln \mathcal{L}}{\partial \lambda_{\alpha}{\partial \lambda_{\beta} }} \r\rangle = \frac{1}{2} Tr [  \mathbf{C}_{,\alpha} \mathbf{C}^{-1} \mathbf{C}_{,\beta} \mathbf{C}^{-1} ]
		\label{eq:F_def}
\end{equation}
where $\mathbf{C}$ is the covariance matrix and $\lambda$ is some parameter. 

In our case, $\lambda_{\alpha} = \lambda_{\beta} = r$, the tensor-to-scalar ratio, while the covariance 
matrix in a harmonic space, $\mathbf{C}$,  is given by,
\begin{equation}
	\centering
		\mathbf{C} \equiv \mathbf{C}_{j j'} \equiv \langle a_{lm}^{\,}a_{l'm'}^{\dagger} \rangle,
\end{equation}
where,
\begin{eqnarray}
j & = & \ell^2+\ell+m, \nonumber \\
 \ell & =  & {\rm round}[(-1 + \sqrt{ 1 + 4 j })/2],\label{eqn:j2lm}\\
 m & = & j - \ell \l(\ell+2\r), \nonumber
\end{eqnarray}
 and thus $j$ goes from $0$ to $(\ell_{max}+1)^2-1$. The function ${round}$ rounds a real number to a {\em closest} integer.
 The Fisher matrix expression can be now specialized as,
\begin{equation}
F_{rr} = \frac{1}{2}\,\sum_{j,j'} \, \frac{\partial C_{\ell}}{\partial r}  \,\left[\mathbf{C}^{-1}\right]_{j j'}^2\, \frac{\partial C_{\ell'}}{\partial r}.
\label{eqn:fishRR}
\end{equation}
where $j$ ($j'$) is related to $\ell$ ($\ell'$) as in Eqs.~\eref{eqn:j2lm}.

Because there are three uncorrelated contributions to the overall signal, which are CMB, noise, and foreground residuals, we can write,

\begin{eqnarray}
	\centering
		\mathbf{C}_{j j'} &=& \langle a_{lm}^{CMB}a_{l'm'}^{CMB,\,\dagger} \rangle +  \langle a_{lm}^{noise}a_{l'm'}^{noise, \, \dagger} \rangle 
		 +   \langle a_{lm}^{res}a_{l'm'}^{res, \, \dagger} \rangle
		\nonumber\\
		 &=& C_{l}^{CMB}\delta_{j j'} + C_{l}^{noise}\delta_{j j'}+ f_{j}^{\phantom{\dagger}}f_{j'}^{\dagger} 
		 		 \nonumber \\
		 &\equiv& D_{j j'} + f_{j}^{\phantom{\dagger}}f_{j'}^{\dagger} \ \ \ \ \ \ \ \ \ \ \ \ \ \ \ \ 
\end{eqnarray}

where  $f_j$ stands for a vector of $a_{\ell m}^{res}$ coefficients arranged according  to the $j$ index.

To compute the $\mathbf{C}^{-1}$ matrix used in equation (\ref{eq:F_def}), we can use the Sherman-Morrison formula obtaining
\begin{equation}
	\centering
		\mathbf{C}^{-1} = \mathbf{D}^{-1} -  \mathbf{D}^{-1} f (1 + f^{\dagger} \mathbf{D}^{-1}f )^{-1}f^{\dagger}\mathbf{D}^{-1},
\end{equation}
where  $(1 + f^{\dagger} \mathbf{D}^{-1}f )^{-1}$ is a number and, hence,
\begin{equation}
	\centering
		\mathbf{C}^{-1} = \mathbf{D}^{-1} -  \frac{\mathbf{D}^{-1} f f^{\dagger}\mathbf{D}^{-1}}{\displaystyle \l(1 + f^{\dagger} \mathbf{D}^{-1}f\r )},
	\label{eq:Cinv_exp}
\end{equation}
which, given that $[D^{-1}]_{j j'} =  (1/C_{\ell}) \delta_{j j'}$, becomes
\begin{eqnarray}
		\left[\mathbf{C}^{-1}\right]_{j j'} & =  &
		\frac{\delta_{j j'}}{C_{\ell}} -  \frac{ C_{\ell}^{-1}\,C_{\ell'}^{-1} \; f_{j} f_{j'}^{\dagger}}{ 1 +{\displaystyle \sum_{\ell''=0}^{\ell_{max}}\; (2\ell''+1)\,\frac{C_{\ell''}^{\Delta}}{C_{\ell''}}}},\\
	\nonumber
\end{eqnarray}
where $C^{\Delta}_\ell$ is a residuals power spectrum, Eq.~\eref{eqn:resSpec0}, defined here as
\begin{equation}
C^{\Delta}_\ell \equiv \frac{1}{2\ell+1}\,  {\displaystyle \sum_{m=\ell^2}^{\ell^2+2\ell}} |f_{m}|^{2}.
\end{equation}

So now we have finally
\begin{widetext}
\begin{eqnarray}
		\left[\mathbf{C}^{-1}\right]_{j j'}^2 & =  &  
		{\displaystyle \frac{\delta_{j j'}}{C_{\ell}^2} } \, - \,\frac{2\,f_{j} f_{j}^{\dagger}}{C_{\ell}^3\,\left( 1 +{\displaystyle \sum_{\ell''=0}^{\ell_{max}}\; (2\ell''+1)\,\frac{C_{\ell''}^{\Delta}}{C_{\ell''}}}\right)}\,\delta_{jj'}
		+ \frac{f_{j}^2 {f_{j'}^{\dagger}}^2}{C_{\ell}^2\,C_{\ell'}^2\,\left( 1 +{\displaystyle \sum_{\ell''=0}^{\ell_{max}}\; (2\ell''+1)\,\frac{C_{\ell''}^{\Delta}}{C_{\ell''}}}\right)^2},
\end{eqnarray}
\end{widetext}
which inserted into Eq.~\eref{eqn:fishRR} gives Eq.~\eref{eq:exact_fisher_result}.

\end{document}